%% file: icrc.tex
\documentclass[dvips]{article}

\usepackage[dvips]{color}
\usepackage{amssymb}
\usepackage{icrc2011}
\title{KASCADE-Grande}
\newcommand{\etal}{\MakeLowercase{\textit{et al. }}} % "et al."
\shorttitle{KASCADE-Grande}

\authors{
W.D.~Apel$^{1}$,
J.C.~Arteaga-Vel\'azquez$^{2}$,
K.~Bekk$^{1}$,
M.~Bertaina$^{3}$,
J.~Bl\"umer$^{1,4}$,
H.~Bozdog$^{1}$,
I.M.~Brancus$^{5}$,
P.~Buchholz$^{6}$,
E.~Cantoni$^{3,7}$,
A.~Chiavassa$^{3}$,
F.~Cossavella$^{4,13}$,
K.~Daumiller$^{1}$,
V.~de Souza$^{8}$,
F.~Di~Pierro$^{3}$,
P.~Doll$^{1}$,
R.~Engel$^{1}$,
J.~Engler$^{1}$,
M. Finger$^{4}$, 
D.~Fuhrmann$^{9}$,
P.L.~Ghia$^{7}$, 
H.J.~Gils$^{1}$,
R.~Glasstetter$^{9}$,
C.~Grupen$^{6}$,
A.~Haungs$^{1}$,
D.~Heck$^{1}$,
J.R.~H\"orandel$^{10}$,
D.~Huber$^{4}$,
T.~Huege$^{1}$,
P.G.~Isar$^{1,14}$,
K.-H.~Kampert$^{9}$,
D.~Kang$^{4}$, 
H.O.~Klages$^{1}$,
K.~Link$^{4}$, 
P.~{\L}uczak$^{11}$,
M.~Ludwig$^{4}$,
H.J.~Mathes$^{1}$,
H.J.~Mayer$^{1}$,
M.~Melissas$^{4}$,
J.~Milke$^{1}$,
B.~Mitrica$^{5}$,
C.~Morello$^{7}$,
G.~Navarra$^{3,15}$,
J.~Oehlschl\"ager$^{1}$,
S.~Ostapchenko$^{1,16}$,
S.~Over$^{6}$,
N.~Palmieri$^{4}$,
M.~Petcu$^{5}$,
T.~Pierog$^{1}$,
H.~Rebel$^{1}$,
M.~Roth$^{1}$,
H.~Schieler$^{1}$,
F.G.~Schr\"oder$^{1}$,
O.~Sima$^{12}$,
G.~Toma$^{5}$,
G.C.~Trinchero$^{7}$,
H.~Ulrich$^{1}$,
A.~Weindl$^{1}$,
J.~Wochele$^{1}$,
M.~Wommer$^{1}$,
J.~Zabierowski$^{11}$
}
\afiliations{
$^1$ Institut f\"ur Kernphysik, KIT - Karlsruher Institut f\"ur Technologie, Germany\\
$^2$ Universidad Michoacana, Instituto de F\'{\i}sica y Matem\'aticas, Morelia, Mexico\\
$^3$ Dipartimento di Fisica Generale dell' Universit\`a Torino, Italy\\
$^4$ Institut f\"ur Experimentelle Kernphysik, KIT - Karlsruher Institut f\"ur Technologie, Germany\\
$^5$ National Institute of Physics and Nuclear Engineering, Bucharest, Romania\\
$^6$ Fachbereich Physik, Universit\"at Siegen, Germany\\
$^7$ Istituto di Fisica dello Spazio Interplanetario, INAF Torino, Italy\\
$^8$ Universidade S$\tilde{a}$o Paulo, Instituto de F\'{\i}sica de S\~ao Carlos, Brasil\\
$^9$ Fachbereich Physik, Universit\"at Wuppertal, Germany\\
$^{10}$ Dept. of Astrophysics, Radboud University Nijmegen, The Netherlands\\
$^{11}$ Soltan Institute for Nuclear Studies, Lodz, Poland\\
$^{12}$ Department of Physics, University of Bucharest, Bucharest, Romania\\
\scriptsize{
$^{13}$ now at: Max-Planck-Institut Physik, M\"unchen, Germany; 
$^{14}$ now at: Institute Space Sciences, Bucharest, Romania; 
$^{15}$ deceased; 
$^{16}$ now at: Univ Trondheim, Norway
}
}
\email{haungs@kit.edu}

\abstract{Contributions of the KASCADE-Grande Collaboration to the 32nd International Cosmic Ray Conference, Beijing, August, 2011.
}

\begin{document}
\maketitle

%Begin the section.
{\bf Contents} \\
\begin{enumerate}
\item page 3: \hspace{0.3cm} Cosmic Ray Measurements with KASCADE-Grande
\item page 7: \hspace{0.3cm} A study of the mass composition of cosmic rays based on an event-by-event
assignment with KASCADE-Grande data
\item page 11: \hspace{0.3cm} Study of the ratio muon size to shower size as a mass sensitive parameter of
KASCADE-Grande
\item page 15: \hspace{0.3cm} The cosmic ray elemental composition based on measurement of the
$N_\mu/N_{ch}$ ratio with KASCADE-Grande
\item page 19: \hspace{0.3cm} KASCADE-Grande measurements of energy spectra for elemental groups
of cosmic rays
\item page 23: \hspace{0.3cm} Primary energy reconstruction from the S(500) observable recorded with
the KASCADE-Grande Array
\item page 27: \hspace{0.3cm} Tests of hadronic interaction models with the KASCADE-Grande muon
data
\item page 31: \hspace{0.3cm} A direct measurement of the muon component of air showers by the
KASCADE-Grande Experiment
\item page 35: \hspace{0.3cm} On the primary mass sensitivity of muon pseudorapidities measured with
KASCADE-Grande
\item page 39: \hspace{0.3cm} Gamma-Ray Source Studies using a Muon Tracking Detector (MTD)
\end{enumerate}

\clearpage
\newpage

\hspace{3cm}
\vspace{3cm}

\clearpage
\newpage
\setcounter{section}{0}
\setcounter{figure}{0}
\setcounter{table}{0}
\setcounter{equation}{0}
 \input{icrc0677.tex}

\newpage
\normalsize
\setcounter{section}{0}
\setcounter{figure}{0}
\setcounter{table}{0}
\setcounter{equation}{0}
 \input{icrc0312.tex}

\newpage
\normalsize
\setcounter{section}{0}
\setcounter{figure}{0}
\setcounter{table}{0}
\setcounter{equation}{0}
 \input{icrc0739.tex}

\newpage
\normalsize
\setcounter{section}{0}
\setcounter{figure}{0}
\setcounter{table}{0}
\setcounter{equation}{0}
 \input{icrc0504.tex}

\newpage
\normalsize
\setcounter{section}{0}
\setcounter{figure}{0}
\setcounter{table}{0}
\setcounter{equation}{0}
 \input{icrc0280.tex}
\newpage
\normalsize
\setcounter{section}{0}
\setcounter{figure}{0}
\setcounter{table}{0}
\setcounter{equation}{0}
 \input{icrc0405.tex}

\newpage
\normalsize
\setcounter{section}{0}
\setcounter{figure}{0}
\setcounter{table}{0}
\setcounter{equation}{0}
 \input{icrc0740.tex}

\newpage
\normalsize
\setcounter{section}{0}
\setcounter{figure}{0}
\setcounter{table}{0}
\setcounter{equation}{0}
 \input{icrc0953.tex}
\newpage
\normalsize
\setcounter{section}{0}
\setcounter{figure}{0}
\setcounter{table}{0}
\setcounter{equation}{0}
 \input{icrc0273.tex}

\newpage
\normalsize
\setcounter{section}{0}
\setcounter{figure}{0}
\setcounter{table}{0}
\setcounter{equation}{0}
 \input{icrc0274.tex}

\end{document}

%% file: icrc0677.tex
%%
% 32nd International Cosmic Ray Conference 2011 Beijing China

%Class Required
%%% for classical LaTeX
%andy\documentclass[dvips]{article}

%\usepackage{icrc2011}

%The paper title
\title{Cosmic Ray Measurements with KASCADE-Grande}
%The short title will appear at the header of the even pages.

%%\newcommand{\etal}{\MakeLowercase{\textit{et al. }}} % "et al."
\shorttitle{A.~Haungs \etal KASCADE-Grande}

%All paper authors
\authors{
A.~Haungs$^{1}$,
W.D.~Apel$^{1}$,
J.C.~Arteaga-Vel\'azquez$^{2}$,
K.~Bekk$^{1}$,
M.~Bertaina$^{3}$,
J.~Bl\"umer$^{1,4}$,
H.~Bozdog$^{1}$,
I.M.~Brancus$^{5}$,
P.~Buchholz$^{6}$,
E.~Cantoni$^{3,7}$,
A.~Chiavassa$^{3}$,
F.~Cossavella$^{4,13}$,
K.~Daumiller$^{1}$,
V.~de Souza$^{8}$,
F.~Di~Pierro$^{3}$,
P.~Doll$^{1}$,
R.~Engel$^{1}$,
J.~Engler$^{1}$,
M. Finger$^{4}$, 
D.~Fuhrmann$^{9}$,
P.L.~Ghia$^{7}$, 
H.J.~Gils$^{1}$,
R.~Glasstetter$^{9}$,
C.~Grupen$^{6}$,
D.~Heck$^{1}$,
J.R.~H\"orandel$^{10}$,
D.~Huber$^{4}$,
T.~Huege$^{1}$,
P.G.~Isar$^{1,14}$,
K.-H.~Kampert$^{9}$,
D.~Kang$^{4}$, 
H.O.~Klages$^{1}$,
K.~Link$^{4}$, 
P.~{\L}uczak$^{11}$,
M.~Ludwig$^{4}$,
H.J.~Mathes$^{1}$,
H.J.~Mayer$^{1}$,
M.~Melissas$^{4}$,
J.~Milke$^{1}$,
B.~Mitrica$^{5}$,
C.~Morello$^{7}$,
G.~Navarra$^{3,15}$,
J.~Oehlschl\"ager$^{1}$,
S.~Ostapchenko$^{1,16}$,
S.~Over$^{6}$,
N.~Palmieri$^{4}$,
M.~Petcu$^{5}$,
T.~Pierog$^{1}$,
H.~Rebel$^{1}$,
M.~Roth$^{1}$,
H.~Schieler$^{1}$,
F.G.~Schr\"oder$^{1}$,
O.~Sima$^{12}$,
G.~Toma$^{5}$,
G.C.~Trinchero$^{7}$,
H.~Ulrich$^{1}$,
A.~Weindl$^{1}$,
J.~Wochele$^{1}$,
M.~Wommer$^{1}$,
J.~Zabierowski$^{11}$
}
%All the affiliations.
\afiliations{
$^1$ Institut f\"ur Kernphysik, KIT - Karlsruher Institut f\"ur Technologie, Germany\\
$^2$ Universidad Michoacana, Instituto de F\'{\i}sica y Matem\'aticas, Morelia, Mexico\\
$^3$ Dipartimento di Fisica Generale dell' Universit\`a Torino, Italy\\
$^4$ Institut f\"ur Experimentelle Kernphysik, KIT - Karlsruher Institut f\"ur Technologie, Germany\\
$^5$ National Institute of Physics and Nuclear Engineering, Bucharest, Romania\\
$^6$ Fachbereich Physik, Universit\"at Siegen, Germany\\
$^7$ Istituto di Fisica dello Spazio Interplanetario, INAF Torino, Italy\\
$^8$ Universidade S$\tilde{a}$o Paulo, Instituto de F\'{\i}sica de S\~ao Carlos, Brasil\\
$^9$ Fachbereich Physik, Universit\"at Wuppertal, Germany\\
$^{10}$ Dept. of Astrophysics, Radboud University Nijmegen, The Netherlands\\
$^{11}$ Soltan Institute for Nuclear Studies, Lodz, Poland\\
$^{12}$ Department of Physics, University of Bucharest, Bucharest, Romania\\
\scriptsize{
$^{13}$ now at: Max-Planck-Institut Physik, M\"unchen, Germany; 
$^{14}$ now at: Institute Space Sciences, Bucharest, Romania; 
$^{15}$ deceased; 
$^{16}$ now at: Univ Trondheim, Norway
}
}
%email address of the contact person
\email{haungs@kit.edu}

%The abstract.
\abstract{
The detection of high-energy cosmic rays above a few hundred TeV is realized by the
observation of extensive air-showers. By using the multi-detector setup of
KASCADE-Grande, energy spectrum, elemental composition, and anisotropies of
high-energy cosmic rays in the energy range from below the knee up to 1 EeV are
investigated. The most distinct feature of the spectrum, the 'knee', is thought
to be the beginning of the end of the galactic origin of cosmic rays. As the highest
energies (above the 'ankle') are most probably of extragalactic origin, between
10 PeV to 1 EeV one expects the transition of galactic to extragalactic origin.
KASCADE-Grande is dedicated to explore this transition region. The estimation of
energy and mass of the high-energy primary particles is based on the combined
investigation of the charged particle, the electron, and the muon components measured
by the detector arrays of Grande and KASCADE.
The latest analysis results have shown that a knee-like structure in the all-particle 
energy spectrum at 83 PeV is due to a decrease of flux of the heavy mass component.
In this contribution an overview is given on various different analysis methods 
to verify this finding.}
%The keywords

% B E G I N   D O C U M E N T
%\begin{document}
\maketitle

%Begin the section.
{\bf KASCADE-Grande:} 
Main parts of the experiment are the Grande array spread over an area of $700 \times 700\,$m$^2$, 
the original KASCADE array covering $200 \times 200\,$m$^2$ with unshielded and shielded 
detectors, and additional muon tracking devices. This multi-detector system allows us to 
investigate the energy spectrum, composition, and anisotropies of cosmic rays in the energy 
range up to $1\,$EeV. The estimation of energy and mass of the primary particles is based 
on the combined investigation of the charged particle, the electron, and the muon components 
measured by the detector arrays of Grande and KASCADE. 

\begin{figure}[!t]
  \vspace{5mm}
  \centering
  \includegraphics[width=2.6in]{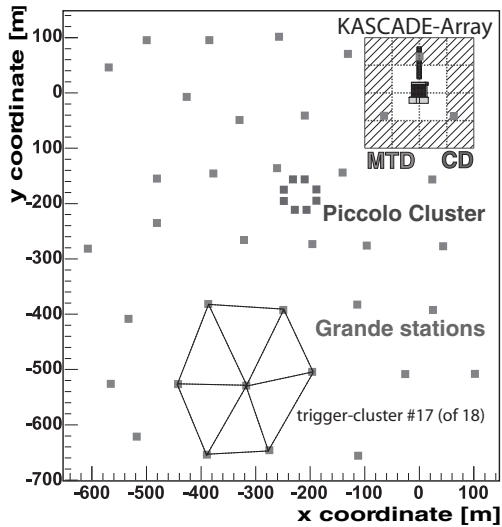}
	\caption{Layout of the KASCADE-Grande experiment: The original KASCADE (with Array, Muon Tracking Detector 
	and Central Detector), the distribution of 
	the 37 stations of the Grande array, and the small Piccolo cluster for fast trigger 
	purposes are shown. The outer 12 clusters of the KASCADE array consist of $\mu$- and $e/\gamma$-detectors, 
	the inner 4 clusters of $e/\gamma$-detectors, only. }
	\label{fig1AHa}
 \end{figure}

The multi-detector experiment KASCADE~\cite{kascadeA}
(located at 49.1$^\circ$n, 8.4$^\circ$e, 110$\,$m$\,$a.s.l.)
was extended to KASCADE-Grande 
in 2003 by installing a large array of 37 stations consisting 
of 10$\,$m$^2$ scintillation detectors each (fig.~\ref{fig1AHa}).  
KASCADE-Grande~\cite{kg-NIM10A} provides an area of 0.5$\,$km$^2$
and operates jointly with the existing KASCADE detectors.
The joint measurements with the KASCADE muon tracking devices are 
ensured by an additional cluster (Piccolo) 
close to the center of KASCADE-Grande for fast trigger purposes. 
For results of the muon tracking devices see references~\cite{icrc-doll,icrc-zabi}.
While the Grande detectors are sensitive to charged particles, 
the KASCADE array detectors measure the electromagnetic 
component and the muonic component separately. 
These muon detectors enable to reconstruct 
the total number of muons on an event-by-event basis
also for Grande triggered events. 

Basic shower observables like the core position, angle-of-incidence, 
and total number of charged particles are provided by 
the measurements of the Grande stations. 
A core position resolution
of $\approx5\,$m, a direction resolution of 
$\approx0.7^\circ$, and a resolution of the total particle number 
in the showers of $\approx15$\% is achieved.  
The total number of muons ($N_\mu$ resolution $\approx25$\%) 
is calculated using the core position determined by the Grande 
array and the muon densities measured by the KASCADE muon 
array detectors.
Full efficiency for triggering and reconstruction of air-showers is reached 
at primary energy of $\approx\:10^{16}\,$eV, slightly varying on
the cuts needed for the reconstruction of the different observables
~\cite{kg-NIM10A}. 

The strategy of the KASCADE-Grande data analysis to reconstruct the energy spectrum and 
elemental composition of cosmic rays is to use the multi-detector set-up of the experiment and 
to apply different analysis methods to the same data sample. This has advantages in various 
aspects: One would expect the same results by all 
methods when the measurements are accurate enough, when the reconstructions work 
without failures, and when the Monte-Carlo simulations describe correctly and consistently the 
shower development and detector response. 

\begin{figure}[!t]
  \vspace{5mm}
  \centering
  \includegraphics[width=3.in]{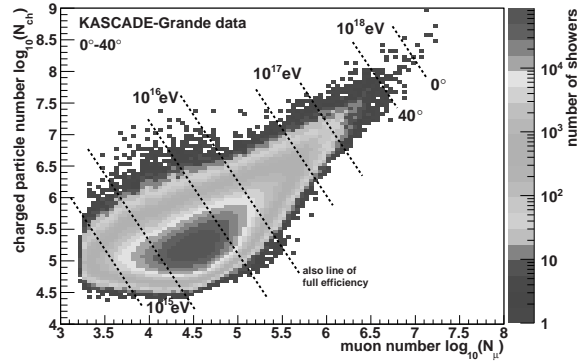}
	\caption{Two-dimensional distribution of the shower sizes charged particle number and total muon 
	number as measured by KASCADE-Grande. All quality cuts are applied, i.e. these data are the basis 
	for the mass composition studies.}
	\label{2dim}
 \end{figure}

The main air-shower observables of KASCADE-Grande, shower size and total number of muons,
could be reconstructed with high precision and low systematic uncertainties and are used 
in the following for the data analysis (fig.~\ref{2dim}).

{\bf The all-particle energy spectrum:}
In a first step of the analysis, we reconstructed the all-particle energy spectrum.
Applying various reconstruction methods to the KASCADE-Grande data  
the obtained all-particle energy spectra are compared for cross-checks of the reconstruction, 
for studies of systematic uncertainties and for testing the validity of the underlying 
hadronic interaction models. 
By combining both observables and using the hadronic interaction model QGSJet-II, a composition
independent all-particle energy spectrum of cosmic rays is reconstructed in the energy range 
of $10^{16}\,$eV to $10^{18}\,$eV within a total uncertainty in flux of 10-15\%.

\begin{figure}[!t]
  \vspace{5mm}
  \centering
  \includegraphics[width=3.1in]{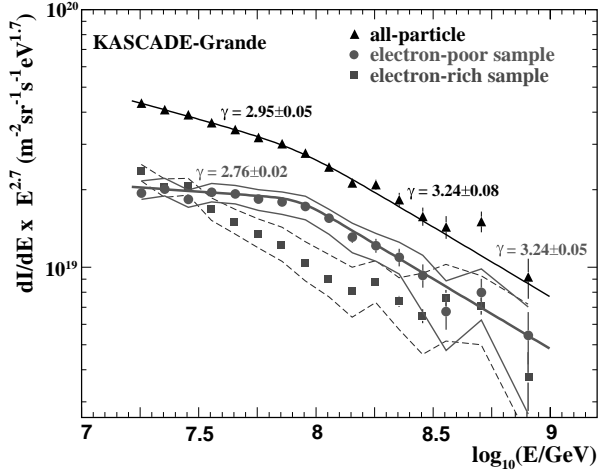}
	\caption{Reconstructed all-particle energy spectrum together with the spectra of the 
	electron-poor and electron-rich components. 
  Fits on the spectra and resulting slopes are also indicated.}
	\label{fig2AHa}
 \end{figure}

Despite the overall smooth power law behavior of the resulting all-particle spectrum, 
there are some structures observed, which do not allow 
to describe the spectrum with a single slope index~\cite{Espec}. 
There is a feature in the spectrum showing a small break at around $8 \cdot 10^{16}\,$eV. 
The power law index of $\gamma=-2.95 \pm 0.05$ is obtained by fitting the range before this break. 
Applying a second power law above the break an index of $\gamma=-3.24 \pm 0.08$ 
is obtained (see figure ~\ref{fig2AHa}). 
With a statistical significance of more than 2 sigma the two power laws are incompatible 
with each other. 
This slight slope change occurs at an energy where the rigidity dependent knee 
of the iron component would be expected (KASCADE QGSJet based analysis assigns the proton 
knee to an energy of $2-4 \cdot 10^{15}\,$eV). 
Despite the fact, that the discussed spectrum is based on the QGSJet-II hadronic interaction model, 
there is confidence that the found structures of the energy spectrum remain stable. 
Tests with EPOS as well as investigations of the spectra of the pure (and independently obtained) 
observables shower size and total muon number, confirmed the structures~\cite{Espec}. 

{\bf Composition:}
A conclusion on the origin of the found structures in the all-particle spectrum is 
not possible without investigating the composition in detail in this energy range. 
The basic goal of the KASCADE-Grande experiment is the determination of the chemical 
composition in the primary energy range $10^{16} - 10^{18}\,$eV. 
Like for the reconstruction of the energy, again several methods using different observables 
are applied to the registered data in order to study systematic uncertainties. 
However, the influence of predictions of the hadronic interaction models has a much larger influence 
on the composition than on the primary energy.  
As it is well known from KASCADE data analysis~\cite{kas-unfAHa} that the relative abundances of the 
individual elements or elemental groups are very dependent on the hadronic interaction model 
underlying the analyses
the strategy is to derive the energy spectra of the individual mass groups. 
The structure or characteristics of these spectra are found to be much less affected by the differences of 
the various hadronic interaction models than the relative abundance.
The present goal is to verify the structure found in the all-particle energy spectrum at around 100 PeV 
in the individual mass group spectra and to assign it to a particular mass. 
 \begin{figure}[!t]
  \vspace{5mm}
  \centering
  \includegraphics[width=2.6in]{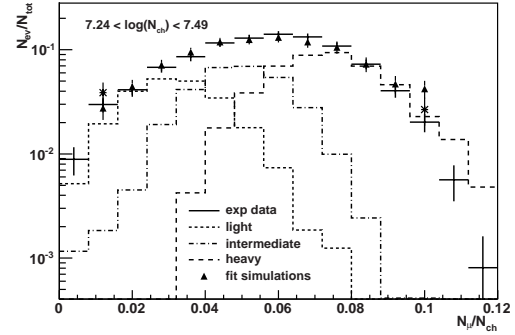}
	\caption{Shower size ratio distributions for a certain bin in charged particle number. 
	Shown is the measured distribution as well as the simulated ones for three primary mass 
	groups and the resulting sum.}
	\label{fig3AHa}
 \end{figure}

The main observables taken into account for composition studies at KASCADE-Grande are the shower size 
($N_{ch}$, or the subsequently derived electron number $N_e$) and the muon shower size ($N_\mu$). 
Figure~\ref{2dim} displays the correlation of these two observables, i.e.~this distribution is the 
basis of the composition analysis with KASCADE-Grande data.
For all the methods it is crucial to verify the sensitivity of the observables to different 
primary particles and the reproducibility of the measurements with the hadronic interaction model 
in use as a function of sizes and 
the atmospheric depth. So far, in the composition analysis we concentrate on interpreting the 
data with the hadronic interaction model QGSJet-II (and FLUKA as low energy interaction model).

Four methods of composition studies at KASCADE-Grande are discussed in the following, all showing 
the same result: The knee-like structure in the all-particle spectrum is due to the 
decrease of the flux of the heavy component of primary cosmic particles.  
\begin{itemize}
\item 
Charged particle -- muon number ratio~\cite{elena}:
The total number of charged particles $N_{ch}$ and the total 
number of muons $N_\mu$ of each recorded event are considered and the distribution of 
$N_\mu / N_{ch}$ is studied in different intervals of 
$N_{ch}$ (corresponding to different energy intervals) and zenith angle.
The experimental distribution of the observable $N_\mu / N_{ch}$ is taken into account and fitted with a 
linear combination of elemental contributions from simulations, where we distinguish three groups: 
light, medium and heavy primaries.
By this way the means and the widths of the distributions as two mass sensitive observables 
are taken into account (see as an example figure~\ref{fig3AHa}).
The width of the data distribution is in all ranges of $N_{ch}$ so large that always three mass 
groups are needed to describe them. 
Using the Monte Carlo simulations the corresponding energy of each mass group and shower size bin is 
assigned to obtain the spectra of the individual mass groups. Investigating the spectra of these individual 
mass groups, it is obvious that the break seen in the all particle spectrum is due to heavy primary masses.  
\item 
The Y-cut method~\cite{ycut}:
Here, the shower ratio $Y_{CIC} = \log{N_\mu} / \log{N_{ch}}$ between the muon and the 
charged particle numbers, 
both corrected for atmospheric attenuation by the Constant Intensity Cut method (CIC), is used as  
parameter to separate the KASCADE-Grande data into different mass groups. 
MC simulations performed with CORSIKA on the framework
of FLUKA/QGSJET-II are employed to obtain the expected $Y_{CIC}$ distributions as a function 
of the energy for different cosmic ray primaries as a basis for the separation. 
Then the $Y_{CIC}$-parameter is used to divide the KASCADE-Grande
data into electron-rich and electron-poor events, i.e.~generated by light and heavy primaries. 
The results confirm the findings above.
 \begin{figure}[!t]
  \vspace{5mm}
  \centering
  \includegraphics[width=3.3in]{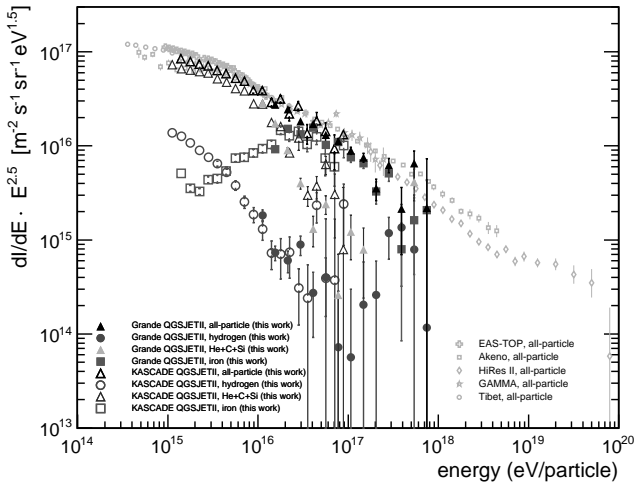}
	\caption{KASCADE and KASCADE-Grande reconstructed energy spectra of individual mass groups.}
	\label{unf}
 \end{figure}
\item 
The k-parameter method~\cite{marioA}:
Using the above mentioned reconstruction of the energy spectrum by correlating the size of the 
charged particles $N_{ch}$ and muons $N_{\mu}$ on an event-by-event basis, the mass 
sensitivity is minimized by means of the parameter $k(N_{ch},N_{\mu})$. 
On the other hand, the evolution of $k$ as a function of energy keeps track of the evolution of 
the composition, and allows an event-by-event separation between light and heavy primaries. 
Using $k$ as separation parameter for different mass groups, where the exact values of $k$ have to be 
determined with help of simulations, directly the energy spectra of the mass groups are obtained.
Figure~\ref{fig2AHa} shows the resulting spectra. As the significance of the break of the all-particle 
spectrum  increases significantly when enhancing heavy primaries, it is obvious that this 
structure is introduced by heavy primaries.   
\item 
The Gold-unfolding method~\cite{unfol}:
This method is based on the unfolding of the two-dimensional shower size spectrum (fig.~\ref{2dim}) 
in a similar way as it was developed for the KASCADE data analysis~\cite{kas-unfAHa}.
Due to the fact that the accuracies in reconstructing the shower sizes for KASCADE-Grande are not as 
high as in case of KASCADE, for comparing both results, primary masses are combined to light (H), 
medium (He+C+Si) and heavy (Fe) spectra. 
The resulting individual mass group spectra can be combined to provide a solution 
for the entire energy range from 1 PeV to 1 EeV. 
This analysis was performed for the combination of $N_{ch}$ with $N_\mu$~\cite{unfol} as well as for
the combination $N_e$ with $N_\mu$. Figure~\ref{unf} shows the results for the latter case~\cite{finger}.
The obtained spectra are compared with the results using KASCADE data in the lower energy range, 
both using the hadronic interaction model QGSJet~II with FLUKA. 
The agreement in the overlapping energy range of the spectra is remarkable.
The knee-like feature in the iron component
is found at an energy which is about 26 times higher than the proton knee in case of KASCADE.  
\end{itemize}

Summarizing, by these first composition studies it is seen that the knee-like feature in the energy spectrum 
at about $8 \cdot 10^{16}\,$eV is due to a kink in the spectrum of the heavy component of 
primary cosmic particles. In addition, at least three primary elements are needed to
describe the experimental data over the entire energy range accessible by KASCADE-Grande, i.e. up to 1 EeV. 
Furtheron, it was found that QGSJet-II, the hadronic interaction model in use, can fairly well 
reproduce the data and, in particular, provides a consistent solution on the elemental composition, 
independent of the method in use. 
One has to remark, that using another hadronic interaction model would probably lead to significant 
changes in the relative abundances of the elemental groups as different models predict different 
shower sizes for a certain energy and mass of the primary cosmic ray.
But, we are confident that the obtained spectral form for the heavy and light component 
of the cosmic ray spectrum will remain unchanged. 

\vspace*{0.01cm} \footnotesize{
{\bf Acknowledgement:} KASCADE-Grande is supported by the BMBF of Germany, the MIUR and INAF of Italy, the
Polish Ministry of Science and Higher Education and the Romanian Authority for Scientific Research.
}
\vspace*{-0.25cm} 
%\vspace{\baselineskip}
%%\

\clearpage

%\end{document}

%% file: icrc0312.tex
%%
% 32nd International Cosmic Ray Conference 2011 Beijing China

%Class Required
%%% for classical LaTeX
%The paper title
\title{A study of the mass composition of cosmic rays based on an
event-by-event assignment with KASCADE-Grande data}
%The short title will appear at the header of the even pages.

\shorttitle{M. Bertaina \etal A study of the mass composition ... 
with KASCADE-Grande}

%All paper authors
\authors{
M.~Bertaina$^{1}$,
W.D.~Apel$^{2}$,
J.C.~Arteaga-Vel\'azquez$^{3}$,
K.~Bekk$^{2}$,
J.~Bl\"umer$^{2,4}$,
H.~Bozdog$^{2}$,
I.M.~Brancus$^{5}$,
P.~Buchholz$^{6}$,
E.~Cantoni$^{1,7}$,
A.~Chiavassa$^{1}$,
F.~Cossavella$^{4,13}$,
K.~Daumiller$^{2}$,
V.~de Souza$^{8}$,
F.~Di~Pierro$^{1}$,
P.~Doll$^{2}$,
R.~Engel$^{2}$,
J.~Engler$^{2}$,
M. Finger$^{4}$, 
D.~Fuhrmann$^{9}$,
P.L.~Ghia$^{7}$, 
H.J.~Gils$^{2}$,
R.~Glasstetter$^{9}$,
C.~Grupen$^{6}$,
A.~Haungs$^{2}$,
D.~Heck$^{2}$,
J.R.~H\"orandel$^{10}$,
D.~Huber$^{4}$,
T.~Huege$^{2}$,
P.G.~Isar$^{2,14}$,
K.-H.~Kampert$^{9}$,
D.~Kang$^{4}$, 
H.O.~Klages$^{2}$,
K.~Link$^{4}$, 
P.~{\L}uczak$^{11}$,
M.~Ludwig$^{4}$,
H.J.~Mathes$^{2}$,
H.J.~Mayer$^{2}$,
M.~Melissas$^{4}$,
J.~Milke$^{2}$,
B.~Mitrica$^{5}$,
C.~Morello$^{7}$,
G.~Navarra$^{1,15}$,
J.~Oehlschl\"ager$^{2}$,
S.~Ostapchenko$^{2,16}$,
S.~Over$^{6}$,
N.~Palmieri$^{4}$,
M.~Petcu$^{5}$,
T.~Pierog$^{2}$,
H.~Rebel$^{2}$,
M.~Roth$^{2}$,
H.~Schieler$^{2}$,
F.G.~Schr\"oder$^{2}$,
O.~Sima$^{12}$,
G.~Toma$^{5}$,
G.C.~Trinchero$^{7}$,
H.~Ulrich$^{2}$,
A.~Weindl$^{2}$,
J.~Wochele$^{2}$,
M.~Wommer$^{2}$,
J.~Zabierowski$^{11}$
}
%All the affiliations.
\afiliations{
$^1$ Dipartimento di Fisica Generale dell' Universit\`a Torino, Italy\\
$^2$ Institut f\"ur Kernphysik, KIT - Karlsruher Institut f\"ur Technologie, Germany\\
$^3$ Universidad Michoacana, Instituto de F\'{\i}sica y Matem\'aticas, Morelia, Mexico\\
$^4$ Institut f\"ur Experimentelle Kernphysik, KIT - Karlsruher Institut f\"ur Technologie, Germany\\
$^5$ National Institute of Physics and Nuclear Engineering, Bucharest, Romania\\
$^6$ Fachbereich Physik, Universit\"at Siegen, Germany\\
$^7$ Istituto di Fisica dello Spazio Interplanetario, INAF Torino, Italy\\
$^8$ Universidade S$\tilde{a}$o Paulo, Instituto de F\'{\i}sica de S\~ao Carlos, Brasil\\
$^9$ Fachbereich Physik, Universit\"at Wuppertal, Germany\\
$^{10}$ Dept. of Astrophysics, Radboud University Nijmegen, The Netherlands\\
$^{11}$ Soltan Institute for Nuclear Studies, Lodz, Poland\\
$^{12}$ Department of Physics, University of Bucharest, Bucharest, Romania\\
\scriptsize{
$^{13}$ now at: Max-Planck-Institut Physik, M\"unchen, Germany; 
$^{14}$ now at: Institute Space Sciences, Bucharest, Romania; 
$^{15}$ deceased; 
$^{16}$ now at: Univ Trondheim, Norway
}
}
%email address of the contact person
\email{bertaina@to.infn.it}

%The abstract.
\abstract{The cosmic ray energy spectrum between 10$^{16}$ eV and
10$^{18}$ eV is reconstructed in KASCADE-Grande by correlating the size
of the charged particles (N$_{ch}$) and muons (N$_{\mu}$) on an 
event-by-event basis. In the energy assignment, the mass sensitivity is
minimized by means of a parameter $k(N_{ch},N_{\mu})$. On the other hand,
the evolution of $k$ as a function of energy keeps track of the evolution
of the composition, and allows an event-by-event separation between 
electron rich and electron poor primaries. A first result on the evolution of 
$k$ and its connection
with the shape of the energy spectrum, is presented in the framework
of the QGSJet~II-03 interaction model. 
}
%The keywords

\maketitle

%Begin the section.
\section{Introduction}
The study of the energy spectrum and of the chemical composition of
cosmic rays are fundamental tools to understand origin, acceleration
and propagation of cosmic rays. The energy range between 10$^{16}$ eV and 
10$^{18}$ eV is quite important from astrophysical point of view because 
it is expected that in this energy range the transition between
galactic and extra-galactic origin of cosmic rays will occur. The results
obtained at lower energies by KASCADE \cite{kas-unfB} and EAS-TOP 
\cite{eas-com} as well as by other experiments suggest that
the knee in the primary energy spectrum around 3 - 4 $\times$ 10$^{15}$ eV
is due to the break in the spectra of light elements. Moreover, KASCADE
results seem to indicate that a rigidity dependent mechanism is 
responsible for such knees. Therefore, a knee of the heaviest components 
would be expected in the range of 10$^{16}$ eV to 10$^{18}$ eV.
Various theories with different assumptions try to explain the rather 
smooth behavior of the cosmic ray energy spectrum in this energy range
(i.e. \cite{berezinsky,hillasB}). In order to discriminate between the 
different models, a very precise measurement of the possible structures 
of the energy spectrum and of the evolution of the composition is needed.

\section{The Technique}
The technique employed to derive the all-particle energy spectrum 
and the abundance of electron-rich (e.r.) and electron-poor (e.p.) primaries 
is based on the
correlation between the size of the charged particles (N$_{ch}$) and 
muons (N$_{\mu}$) on an event-by-event basis. The method itself has been
described in detail in \cite{bertainaB}.\\
A sample of Monte Carlo data was simulated including the full air shower
development in the atmosphere, the response of the detector and its 
electronics as well as their uncertainties. In this way, the parameters
reconstructed from simulation are obtained in the same way as for real data.
The EAS events were generated with an isotropic distribution with spectral
index $\gamma = -3$ and were simulated with CORSIKA \cite{corsB} and the
hadronic Monte Carlo generators FLUKA \cite{flukaB} and QGSJet~II-03 
\cite{qgsB}.
Sets of simulated events were produced in the energy range from
$10^{15}\,$eV to $10^{18}\,$eV with high statistics and for five
elements: H, He, C, Si and Fe, representative for
different mass groups ($\approx 353.000$ events 
per primary). Few events up to $3 \cdot 10^{18}\,$eV  were also generated 
in order to cross-check the reconstruction behavior at the highest 
energies.\\
Grande stations \cite{KG-NIM10B} are used to provide core position and 
angle-of-incidence, 
as well as the total number of charged particles in the shower, by means 
of a maximum likelihood procedure comparing the measured number of 
particles with the one expected from a modified NKG lateral distribution 
function \cite{hajo} of charged particles in the EAS.\\
The total number of muons is calculated using the core position
determined by the Grande array and the muon densities measured by the
KASCADE muon array detectors.
The total number of muons $N_\mu$ in the shower disk (above the energy
threshold of 230~MeV) is derived from a maximum likelihood estimation 
assuming a lateral distribution function based 
on the one proposed by Lagutin and Raikin \cite{Lagutin:2001nc}. 
The reconstruction procedures and obtained accuracies of KASCADE-Grande
observables are described in detail in reference \cite{KG-NIM10B}.\\
For the reconstructed events, we restricted ourselves to events with 
zenith angles lower than $40^\circ$.
Additionally, only air showers with cores located in a central 
area on KASCADE-Grande were selected. With this cut on
the fiducial area, border effects are discarded and possible under- and
overestimations of the muon number for events close to and far away from 
the center of the KASCADE array are reduced. 
All of these cuts were applied also to the Monte Carlo simulations to 
study the effects and to optimize the cuts. Full efficiency for triggering
 and reconstruction of air-showers is reached at primary energy of
$\approx 10^{16}$eV, slightly depending on the cuts needed for the
reconstruction of the different observables \cite{KG-NIM10B}.\\
The analysis presented here is finally based on 1173 days of data and the 
cuts on the sensitive central area and zenith angle correspond to a total 
acceptance of $ A = 1.976 \cdot 10^{9}$ cm$^2 \cdot$ sr, and an exposure of
$N = 2.003 \cdot 10^{17}$ cm$^2 \cdot$ s$ \cdot $sr, respectively.\\
 \begin{figure}[th!]
   \vspace{-5mm}
  \centering
  \includegraphics[width=2.7in]{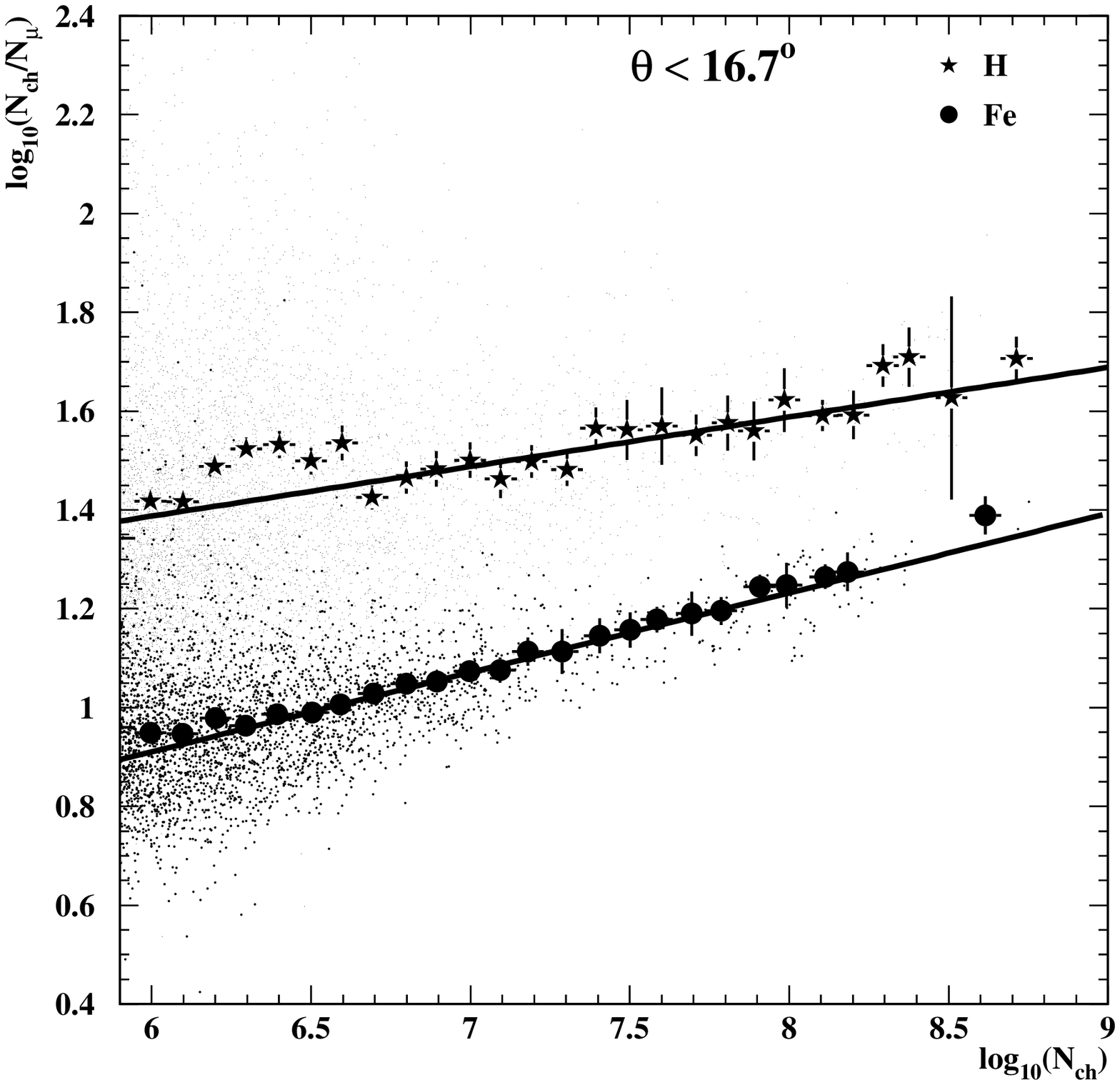}
   \vspace{-5mm}
  \caption{Scatter plot of the reconstructed $N_{ch}/N_{\mu}$ vs. 
$N_{ch}$ for H and Fe primaries for the first angular bin.
The full dots and error bars indicate the mean and its statistical error of 
the distribution of the individual events (small dots).
The fits result in parameters c and d of expression ~\ref{equn3}.}
  \label{fig1}
 \end{figure}
 \begin{figure}[th!]
   \vspace{-5mm}
  \centering
  \includegraphics[width=2.7in]{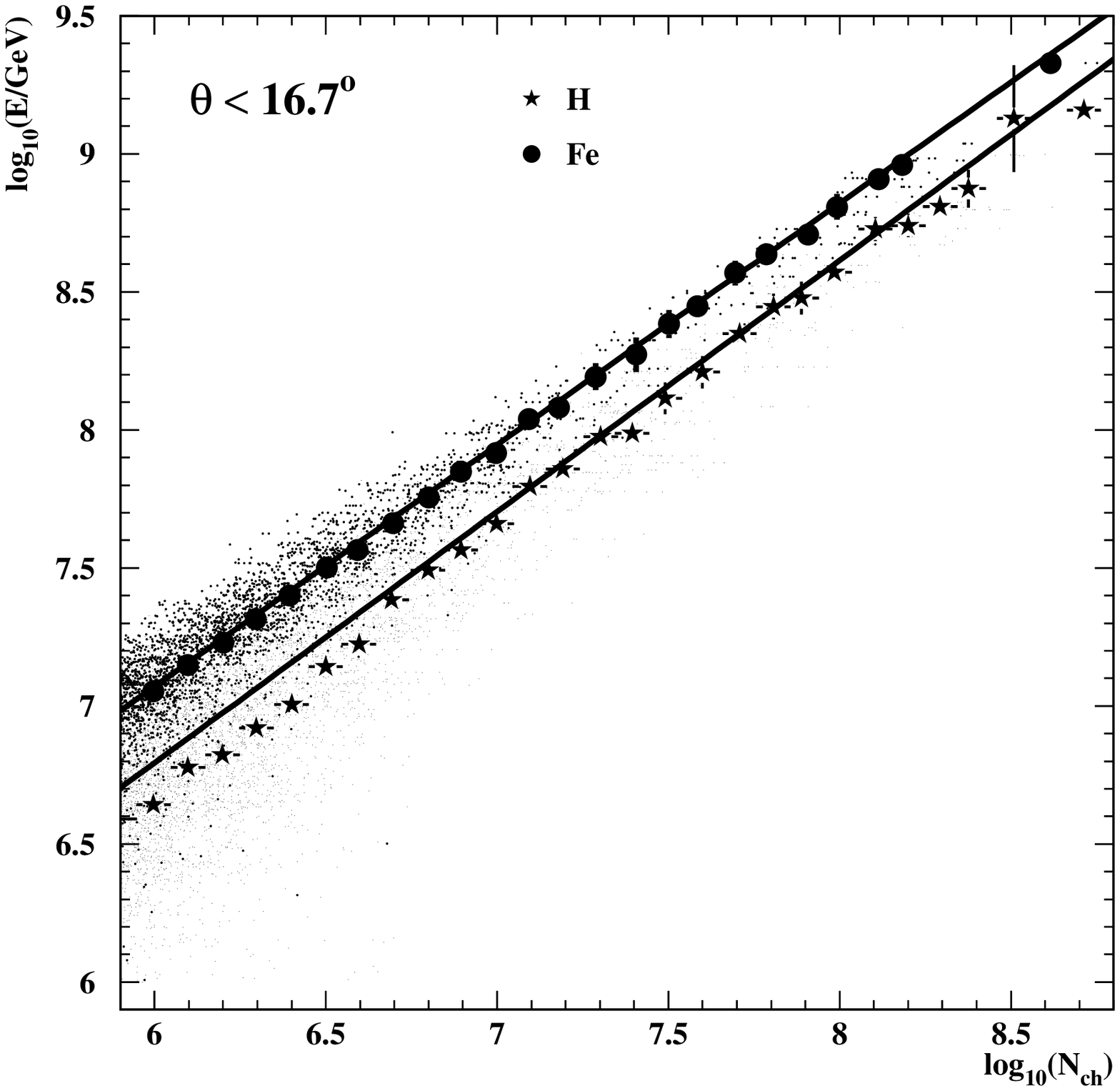}
   \vspace{-5mm}
  \caption{Scatter plot of $E$ vs. $N_{ch}$ for Fe and H primaries.
The fits result in parameters a and b of expression ~\ref{equn1}.}
  \label{fig2}
 \end{figure}
With Monte Carlo simulations a formula is obtained to calculate the 
primary energy per individual shower on the basis of $N_{ch}$ and $N_\mu$.
The formula takes into account the mass sensitivity in order to minimize 
the composition dependence in the energy assignment, and at the same time,
provides an event-by-event separation between e.r. and e.p. 
candidates. 
The formula is defined for 5 different zenith
angle intervals 
($\theta$ $<$ 16.7, 16.7 $\le$ $\theta$ $<$ 24.0,
24.0 $\le$ $\theta$ $<$ 29.9, 29.9 $\le$ $\theta$ $<$ 35.1,
35.1 $\le$ $\theta$ $<$ 40.0)
independently, to take into account the shower attenuation
in atmosphere. Data are combined only at the very last stage to obtain
a unique power law spectrum and mass composition.
 \begin{table}[th!]
  \caption{Parameters of the calibration functions.}
   {\scriptsize
  \centering
  \begin{tabular}{|l|c|c|c|c|c|c|c|c|}
  \hline
Angles[deg]&\multicolumn{2}{c|}{a}&\multicolumn{2}{c|}{b}&\multicolumn{2}{c|}{c}&\multicolumn{2}{c|}{d}\\ \hline
&H&Fe&H&Fe&H&Fe&H&Fe\\ \hline \hline
$0.0-16.7$ & 0.910 & 0.876 & 1.333 & 1.817 & 0.100 & 0.161 & 0.786 & -0.055 \\ \hline
$16.7-24.0$ & 0.894 & 0.878 & 1.495 & 1.923 & 0.081 & 0.179 & 0.884 & -0.254 \\ \hline
$24.0-29.9$ & 0.937 & 0.889 & 1.301 & 1.935 & 0.104 & 0.156 & 0.677 & -0.170 \\ \hline
$29.9-35.1$ & 0.934 & 0.881 & 1.458 & 2.099 & 0.109 & 0.171 & 0.543 & -0.351 \\ \hline
$35.1-40.0$ & 0.919 & 0.875 & 1.748 & 2.287 & 0.105 & 0.156 & 0.412 & -0.348 \\ \hline
  \end{tabular}
  \label{Tab01}
   }
 \end{table}
The energy assignment is defined as $E = f(N_{ch}$$,k)$
(see equation 1), where N$_{ch}$ is the size of the charged particle
component and the parameter $k$ is defined through the
ratio of the sizes of the N$_{ch}$ and muon (N$_\mu$) components:
$ k = g (N_{ch}$,$N_\mu)$ (see equation 2).
The main aim of the $k$ variable is to take into account the average 
differences in the N$_{ch}/$N$_\mu$ ratio among different primaries with 
same N$_{ch}$ and the shower to shower fluctuations for events
of the same primary mass:
{ %  \setlength{\arraycolsep}{0.0em}
  \begin{eqnarray}
  log_{10}(E[GeV])&{}={}&[a_H+(a_{Fe}-a_H) \cdot k]\cdot log_{10}(N_{ch})+
\nonumber\\
  &&{+}\: b_H+(b_{Fe}-b_H) \cdot k
  \label{equn1}
  \end{eqnarray}              %
  \begin{equation}
    k = \frac{log_{10}(N_{ch}/N_{\mu})-log_{10}(N_{ch}/N_{\mu})_H}
        {log_{10}(N_{ch}/N_{\mu})_{Fe}-log_{10}(N_{ch}/N_{\mu})_H}
    \label{equn2}
   \end{equation}
  \begin{equation}
    log_{10}(N_{ch}/N_{\mu})_{H,Fe} = c_{H,Fe} \cdot log_{10} (N_{ch}) + d_{H,Fe}.
    \label{equn3}
   \end{equation}
}
The k parameter is, by definition of eq.~(\ref{equn2}), a number centered 
around 0 for H showers and 1 for Fe ones if 
expressed as a function of $N_{ch}$, while slightly shifted when reported
as a function of energy (see Fig.~\ref{fig4}). The complete list
of parameters $a$ -$d$ can be found in table \ref{Tab01}.\\
 \begin{figure}[th!]
  \vspace{-5mm}
  \centering
  \includegraphics[width=2.7in]{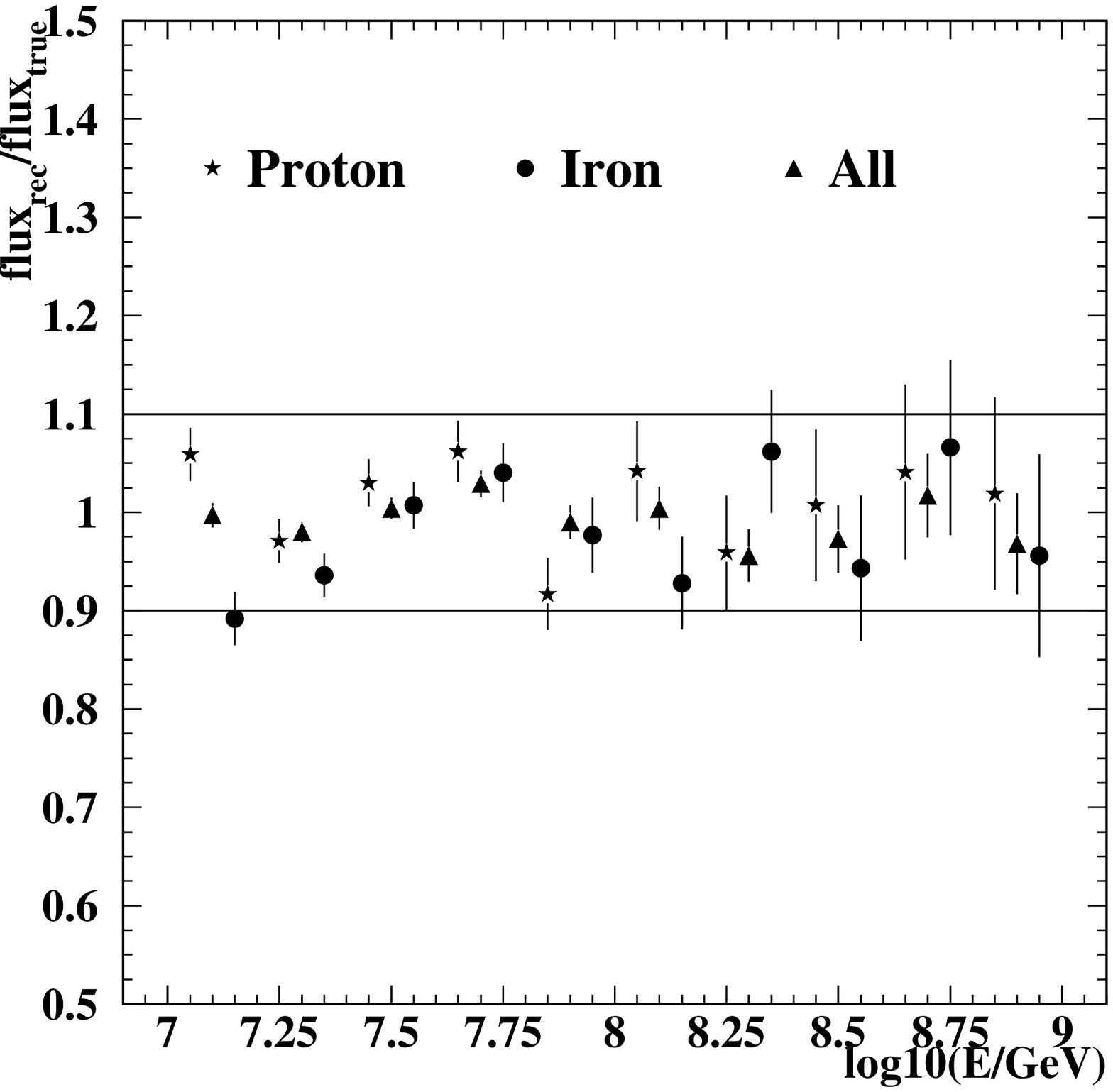}
  \vspace{-5mm}
  \caption{Ratio between the reconstructed and true simulated energy 
spectra for H, Fe and all mixed primaries summing up all angular bins.}
  \label{fig3z}
 \end{figure}
Figs.~\ref{fig1},~\ref{fig2} show the scatter plots with the parametrizations
defined for H and Fe in the first angular bin, while Fig.~\ref{fig3z}
shows the capability of reproducing simulated energy spectra. Pure 
spectra of H, Fe and a mixture of 5 different primaries with 20\% 
abundance each are shown as examples. The true flux is always reproduced 
within 10\% uncertainty.

\section{Results}
\begin{figure}[t!]
   \vspace{-5mm}
  \hspace{10mm}
  \centering
  \includegraphics[width=2.7in]{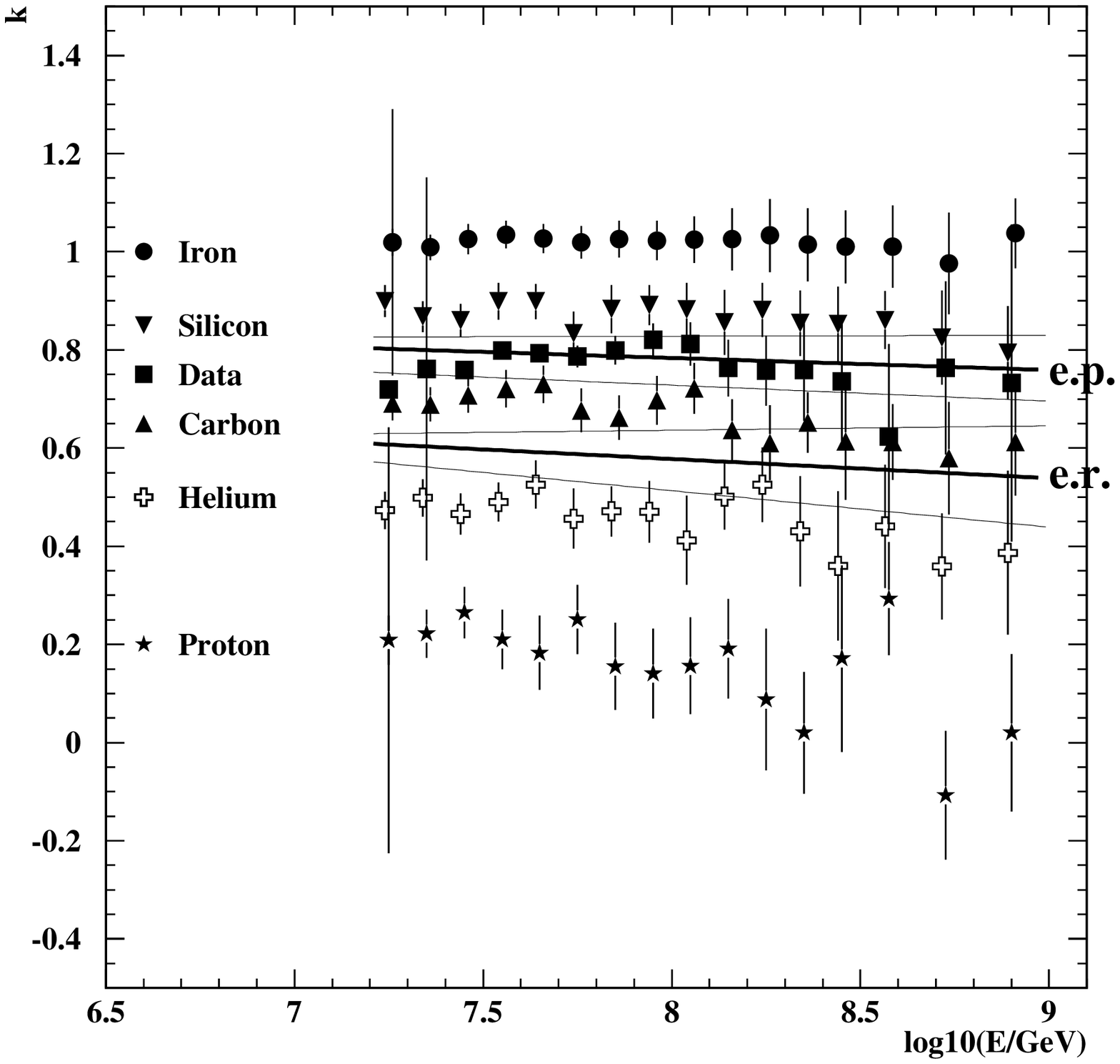}
  \vspace{-5mm}
  \caption{Evolution of the $k$ parameter as a function of the 
reconstructed energy for experimental data compared with pure primary
spectra for the angular range 0-24$^\circ$. The error bars assign statistical
as well as recostruction uncertainties of $k$.}
  \label{fig4}
 \end{figure}

\begin{figure}[th!]
   \vspace{-5mm}
  \centering
  \includegraphics[width=2.7in]{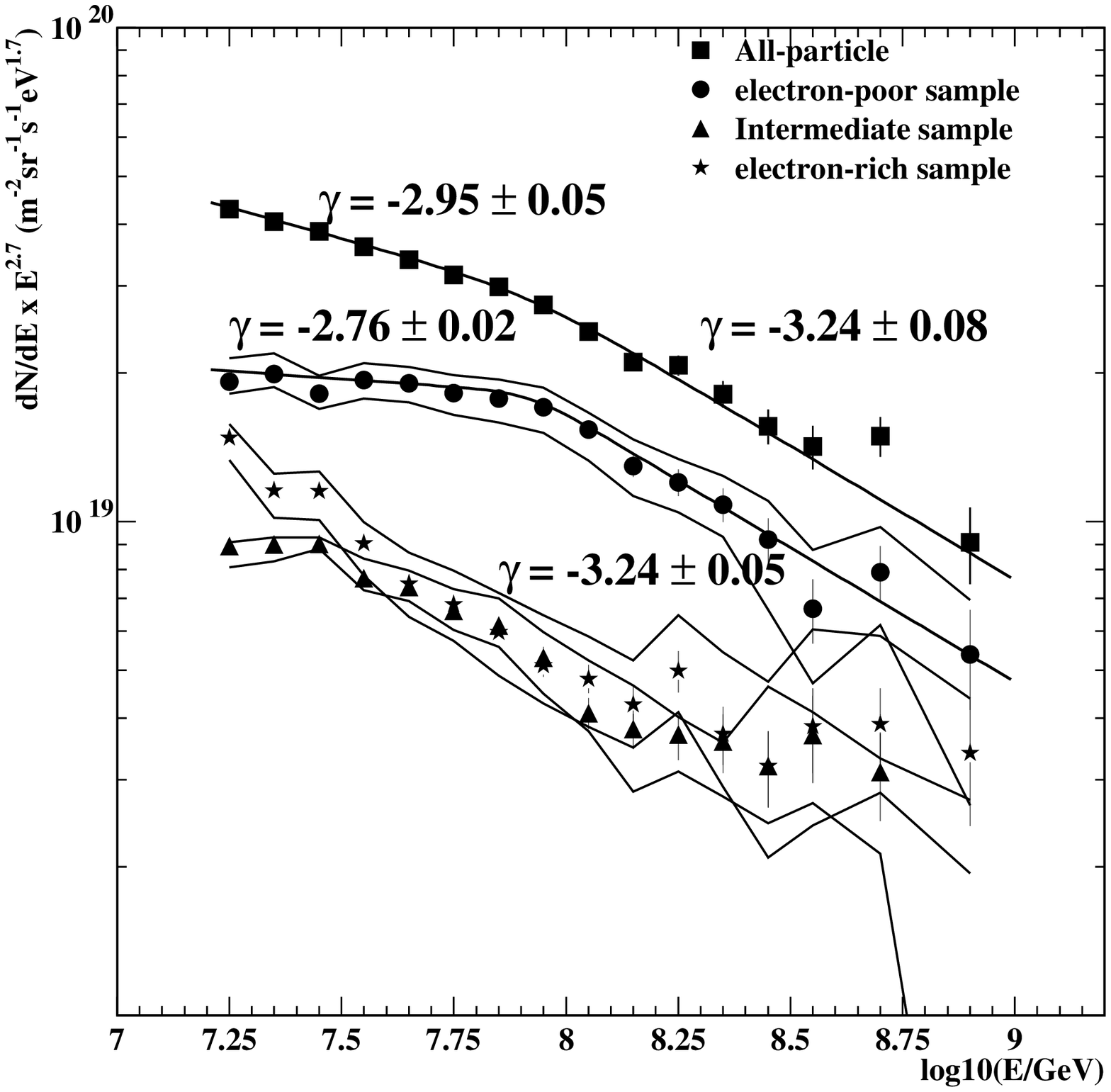}
   \vspace{-5mm}
  \caption{Reconstructed energy spectrum of the e.r., e.p. and
intermediate samples as classified in the text together with the all-particle
spectrum.}
  \label{fig5}
 \end{figure}

\begin{figure}[th!]
   \vspace{-5mm}
  \centering
  \includegraphics[width=2.7in]{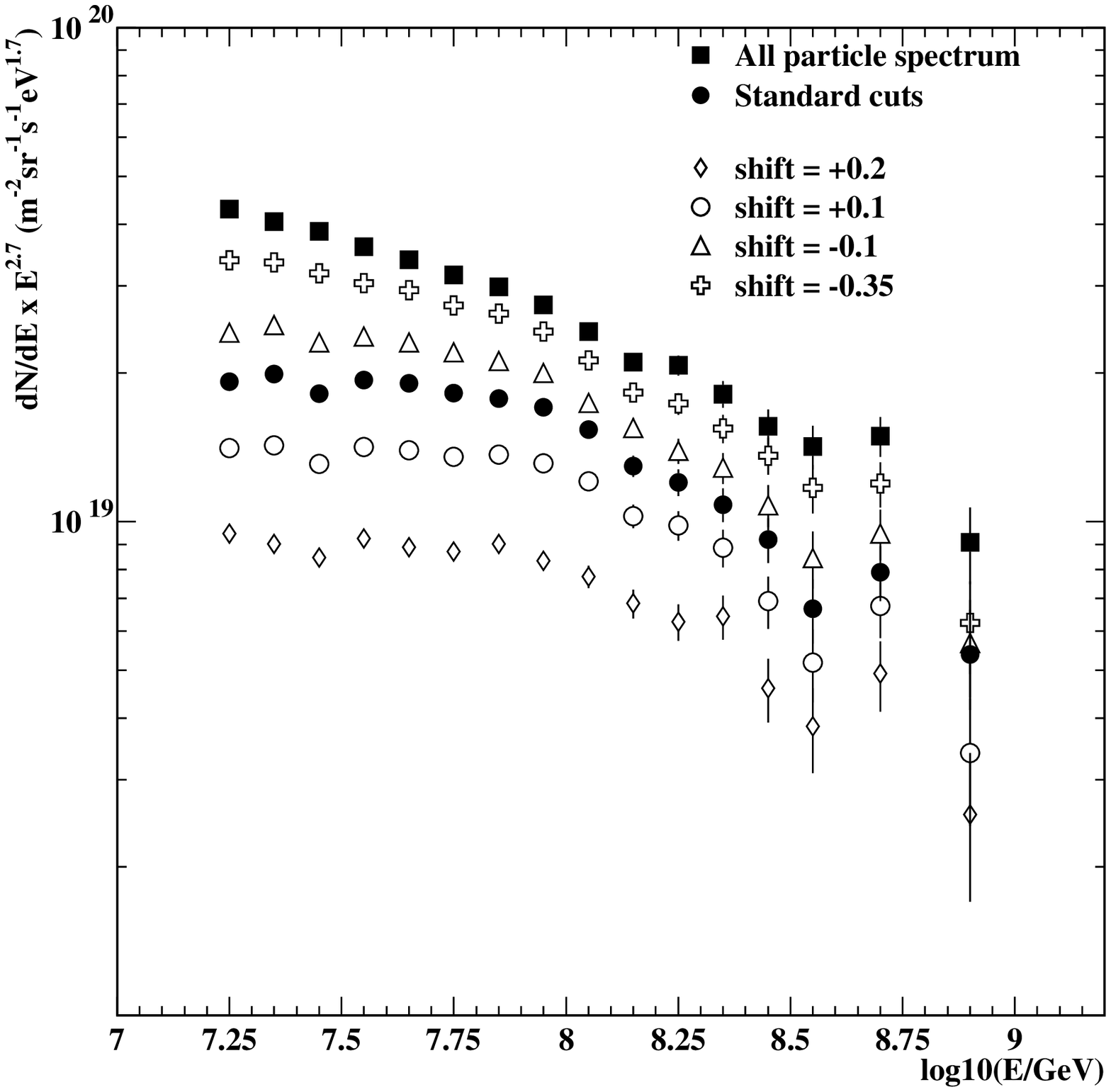}
   \vspace{-5mm}
\caption{Effect of the shift of the cut line of the e.p. component 
on its energy spectrum.}
  \label{fig6}
 \end{figure}

Fig.~\ref{fig4} shows, for the sum of the first two angular bins,
the evolution of the $k$ parameter as a function
of the reconstructed energy obtained by simulating with QGSJet~II-03 model
pure H, He, C, Si, Fe primary spectra, as well as for the experimental 
data. 
Similar behavior is obtained also for the other angular bins. 
The error bars indicate the average dispersion of the
$k$ parameter among different bins, which include statistical errors, and 
systematic uncertainty of equations~\ref{equn3} in each angular bin. 
Fig.~\ref{fig4} shows also two thick straight lines which are used to classify 
events into different mass groups. The classification is done 
for each angular bin, independently, to avoid possible systematic 
effects among the bins.
The top thick line represents
the separation between e.p. and intermediate samples and it is 
defined by fitting the $k_{e.p.}(E)=(k_{Si}(E)+k_{C}(E))/2$ points which are 
obtained by averaging the values of $k$ for Si and C components.   
In analogy, the bottom thick line represents
the separation line between intermediate and e.r. samples and it is 
defined by fitting the $k_{e.r.}(E)=(k_{C}(E)+k_{He}(E))/2$ points which are 
obtained by averaging the values of $k$ for C and He components. In the 
following, e.p. events will be defined as those having a $k$ value 
higher than the top thick line and e.r. events those with $k$ below 
the bottom thick line.
The region in between, which is dominated mainly by CNO and highly 
contaminated by Si and He, will be defined as the intermediate sample.
Such an assignment is, therefore, chosen on an event-by-event basis.
Naturally, the absolute abundances of the events in the three classes 
depend on the location of the straight lines. 
For that reason, two thin lines with different slopes bracket the range of
possible positions of each thick line. They represent the uncertainties in 
defining these energy-dependent selection cuts.\\
Fig.~\ref{fig5} shows the flux of the three components classified as previously
explained.
The reconstructed spectrum of the e.p. sample shows a distinct knee-like
feature around 8 $\times$ 10$^{16}$ eV. Applying a fit of two power laws to the
spectrum with a smooth transition \cite{plaw} results in a statistical
significance of 3.5$\sigma$ that the entire spectrum cannot be fitted with a
single power law. The change of the spectral slope is $\Delta \gamma$ = -0.48
from $\gamma$ = -2.76 $\pm$ 0.02 to $\gamma$ = -3.24 $\pm$ 0.05 with the break
position at log$_{10}$(E/eV) = 16.92 $\pm$ 0.04. Applying the same function to
the all-particle spectrum results in a statistical significance of only 
2.1$\sigma$ that a fit of two power laws is needed to describe the spectrum.
Here the change of the spectral slope is from $\gamma$ = -2.95 $\pm$ 0.05 to
$\gamma$ = -3.24 $\pm$ 0.08, but with the break position again at 
log$_{10}$(E/eV) = 16.92 $\pm$ 0.10. The spectrum of the e.r. sample
is compatible with a single power law ($\gamma$ $\sim$ -3.37).\\
The error bands in the spectra show that the result is independent from the
particular choice of the selection-cut lines. However, in order to further 
validate the present result, parallel shifts of the cut-line of the 
e.p. sample have been applied. Specifically, by shifting the cut line
to higher values of $k$, the e.p. sample is enhanced, and its
flux diminuishes, while shifting the cut line towards lower values of $k$, the
e.p. sample becomes more contaminated by e.r. events
and the flux increases. Fig.~\ref{fig6} indicates that shifting up the line
cut keeps the knee-like structure unchanged, while shifts down tend to smooth
out the structure. This result confirms that the structure seen in the spectrum
is caused by the e.p. component, and that the conclusion is 
essentially independent of the particular hadronic interaction model used in 
the analysis. 

\section{Conclusions}
A first result on the evolution of the $k$ parameter, which is
sensitive to the evolution of the average mass composition has been
derived using an event-by-event approach, based on
QGSJet~II-03 model. 
It shows that the e.p. component shows a knee-like structure
at (8.3 $\pm$ 0.8) $\times$ 10$^{16}$ eV and the change of slope is about 
$\Delta \gamma$ $\sim$ 0.5 through the break. 
The change of slope of $\Delta \gamma$ $\sim$ 0.3 observed in the all-particle
spectrum is located in the same energy range where the change of slope of the 
e.p. component occurs, indicating that it reflects the knee in the
e.p. component. The e.r. component has a much steeper spectrum
compared to the e.p. one.
In the framework of the QGSJet-II-03 model, the e.p. component is the
most dominant one in the energy range 10$^{16}$ - 10$^{18}$ eV.
This result is in agreement with those obtained by other analyses of 
KASCADE-Grande data \cite{arteagaB,cantoni,fuhrmann}.

\vspace*{0.5cm} \footnotesize{
{\bf Acknowledgement:} KASCADE-Grande is supported by the BMBF of Germany, the MIUR and INAF of Italy, the
Polish Ministry of Science and Higher Education (this work in part by grant
for 2009-2011) and the Romanian Authority for Scientific Research.}

%\vspace{\baselineskip}

\clearpage

%% file: icrc0739.tex
%%
% 32nd International Cosmic Ray Conference 2011 Beijing China

%Class Required
%%% for classical LaTeX

%The paper title
\title{Study of the ratio muon size to shower size as a mass 
sensitive parameter of KASCADE-Grande}
%The short title will appear at the header of the even pages.

\shorttitle{J.C. Arteaga \etal Study of the ratio muon size to shower size
as ....}

%All paper authors
\authors{
J.C.~Arteaga-Vel\'azquez$^{1}$,
A.~Chiavassa$^{2}$,
W.D.~Apel$^{3}$,
F. Balestra$^{2}$,
K.~Bekk$^{3}$,
M.~Bertaina$^{2}$,
J.~Bl\"umer$^{3,4}$,
H.~Bozdog$^{3}$,
I.M.~Brancus$^{5}$,
P.~Buchholz$^{6}$,
E.~Cantoni$^{2,7}$,
F.~Cossavella$^{4,13}$,
K.~Daumiller$^{3}$,
V.~de Souza$^{8}$,
F.~Di~Pierro$^{2}$,
P.~Doll$^{3}$,
R.~Engel$^{3}$,
J.~Engler$^{3}$,
M. Finger$^{4}$, 
D.~Fuhrmann$^{9}$,
P.L.~Ghia$^{7}$, 
H.J.~Gils$^{3}$,
R.~Glasstetter$^{9}$,
C.~Grupen$^{6}$,
A.~Haungs$^{3}$,
D.~Heck$^{3}$,
J.R.~H\"orandel$^{10}$,
D.~Huber$^{4}$,
T.~Huege$^{3}$,
P.G.~Isar$^{3,14}$,
K.-H.~Kampert$^{9}$,
D.~Kang$^{4}$, 
H.O.~Klages$^{3}$,
K.~Link$^{4}$, 
P.~{\L}uczak$^{11}$,
M.~Ludwig$^{4}$,
H.J.~Mathes$^{3}$,
H.J.~Mayer$^{3}$,
M.~Melissas$^{4}$,
J.~Milke$^{3}$,
B.~Mitrica$^{5}$,
C.~Morello$^{7}$,
G.~Navarra$^{2,15}$,
J.~Oehlschl\"ager$^{3}$,
S.~Ostapchenko$^{3,16}$,
S.~Over$^{6}$,
N.~Palmieri$^{4}$,
M.~Petcu$^{5}$,
T.~Pierog$^{3}$,
H.~Rebel$^{3}$,
M.~Roth$^{3}$,
H.~Schieler$^{3}$,
F.G.~Schr\"oder$^{3}$,
O.~Sima$^{12}$,
G.~Toma$^{5}$,
G.C.~Trinchero$^{7}$,
H.~Ulrich$^{3}$,
A.~Weindl$^{3}$,
J.~Wochele$^{3}$,
M.~Wommer$^{3}$,
J.~Zabierowski$^{11}$
}
%All the affiliations.
\afiliations{
$^1$ Universidad Michoacana, Instituto de F\'{\i}sica y Matem\'aticas, Morelia, Mexico\\
$^2$ Dipartimento di Fisica Generale dell' Universit\`a Torino, Italy\\
$^3$ Institut f\"ur Kernphysik, KIT - Karlsruher Institut f\"ur Technologie, Germany\\
$^4$ Institut f\"ur Experimentelle Kernphysik, KIT - Karlsruher Institut f\"ur Technologie, Germany\\
$^5$ National Institute of Physics and Nuclear Engineering, Bucharest, Romania\\
$^6$ Fachbereich Physik, Universit\"at Siegen, Germany\\
$^7$ Istituto di Fisica dello Spazio Interplanetario, INAF Torino, Italy\\
$^8$ Universidade S$\tilde{a}$o Paulo, Instituto de F\'{\i}sica de S\~ao Carlos, Brasil\\
$^9$ Fachbereich Physik, Universit\"at Wuppertal, Germany\\
$^{10}$ Dept. of Astrophysics, Radboud University Nijmegen, The Netherlands\\
$^{11}$ Soltan Institute for Nuclear Studies, Lodz, Poland\\
$^{12}$ Department of Physics, University of Bucharest, Bucharest, Romania\\
\scriptsize{
$^{13}$ now at: Max-Planck-Institut Physik, M\"unchen, Germany; 
$^{14}$ now at: Institute Space Sciences, Bucharest, Romania; 
$^{15}$ deceased; 
$^{16}$ now at: Univ Trondheim, Norway
}
}
%email address of the contact person
\email{arteaga@ifm.umich.mx; achiavas@to.infn.it}

%The abstract.
\abstract
{
 In this work, the shower ratio $Y^{CIC} = \log N_{\mu}/\log N_{ch}$ between the
 muon and the charged numbers, both corrected by atmospheric attenuation with the 
 Constant Intensity Cut method (CIC), is studied as a parameter to separate
 the KASCADE-Grande data into different mass groups. MC simulations performed
 with CORSIKA on the framework of FLUKA/QGSJet II are employed to obtain the 
 expected $Y^{CIC}$ distributions as a function of the energy for different
 cosmic ray primaries as a basis for the separation. Then the $Y^{CIC}$ parameter is 
 used to divide the KASCADE-Grande data into ``electron rich'' and ``electron poor''
 events. The fraction of events generated by H (Fe) primaries and classified as 
 ``electron rich'' (``electron poor'') and its dependence on the primary energy 
 is discussed. Finally, the energy spectra of these samples are reconstructed.
 A knee-like structure is found around $10^{17}$ in the ``electron poor''
 sample associated with the heavy component of cosmic rays.
}
%The keywords

\maketitle

%Begin the section.
\section{Introduction}

 In order to understand the origin, nature, propagation and acceleration mechanism
 of galactic cosmic rays, detailed measurements with enough statistics on their 
 energy spectrum, composition and arrival directions are needed. The task is 
 complicated since at high energies ($\gtrsim 1$ PeV), cosmic rays must be
 studied indirectly by means of extensive air showers (EAS) detected with Earth-bound
 experiments. Information about composition and primary energy requires the
 simultaneous observation of several EAS parameters, such as the muon and
 the electron particle contents at ground level. By combining the informations
 from both EAS observables, the KASCADE experiment was able to separate for
 the first time the cosmic ray energy spectra of different mass groups in the 
 so called \textit{knee} energy region (around $10^{15}$ eV) \cite{kas-unf}. The picture of
 the galactic cosmic ray spectrum is still incomplete due to the lack of 
 statistics in the interval $10^{16} - 10^{18}$ eV, where a \textit{knee} in 
 the heavy component of cosmic rays is predicted by several models. To explore
 this energy regime KASCADE was upgraded to KASCADE-Grande, a multidetector-setup
 with enhanced area also designed to measure with precision the charged, muon and 
 electron contents of air showers for cosmic ray studies \cite{kg-NIM10C}.  In this 
 paper we present an analysis based on the ratio between the muon and the charged
 particle numbers in EAS. This ratio is exploited to divide the events 
 in two samples: the \textit{electron poor} and the \textit{electron rich} 
 events. The two samples can be identified with EAS generated
 by primaries belonging to different mass groups (i.e. light and heavy
 elements). The technique is illustrated by applying it to the KASCADE-Grande
 data to reconstruct the light and heavy spectra of cosmic rays.

 \section{The KASCADE-Grande experiment}

  KASCADE-Grande, located in Karlsruhe, Germany (110 m a.s.l.), is a cosmic ray detector
 designed to measure EAS within the energy interval of $10^{16} - 10^{18}$ eV.
 Two detector arrays of different sizes are the main components of
 the experiment \cite{kg-NIM10C}. The first one is composed by 37 plastic scintillator detectors and
 covers a surface of $700 \times 700 \, \mbox{m}^2$. It is aimed to provide 
 measurements of the charged particle number ($N_{ch}$), core position and arrival
 direction of air showers\cite{kg-NIM10C} . The penetrating component ($N_{\mu}$) is measured with the
 aid of a smaller array of $200 \times 200 \, \mbox{m}^2$ integrated by 
 $252$ $e/\gamma$ and shielded detectors \cite{kascadeC, kg-NIM10C}. The electron shower size is
 obtained from the difference between $N_{ch}$ and $N_{\mu}$ parameters.

 \section{Simulations and data selection}

 To perform the present analysis, MC simulations must be invoked. FLUKA
 \cite{fluka} and QGSJet II-03 \cite{qgsC} were employed to describe the hadronic 
 interactions at low and high energies, respectively. CORSIKA \cite{corsC} was 
 used to describe the EAS development and GEANT 4
 to simulate the response of the KASCADE-Grande experiment to the passage of
 the air shower. Both simulated and experimental data are saved with identical
 output formats and analyzed with the same reconstruction program. MC data
 was generated for H, He, C, Si and Fe nuclei. Events were sampled from 
 an isotropic distribution with spectral index $\gamma = -2$. The simulated 
 data was weighted to emulate a $\gamma = -3$ energy spectrum. 
 
 With the aid of Monte Carlo simulations, quality cuts were investigated.
 They were carefully selected to reduce the  systematic uncertainties on the
 muon and charged particle numbers. First, the experimental data sample 
 was built out of events successfully reconstructed with the KASCADE-Grande
 procedure and collected during stable periods of data acquisition . Only
 EAS detected with more than 35 active stations were considered in this
 work.  Then border effects were avoided by picking events with shower cores 
 located inside a central area of $1.37 \times 10^{5} \, \mbox{m}^2$  in the 
 Grande array. Finally, the analysis was restricted to EAS with arrival zenith
 angles lower than $30^\circ$, $\log N_{ch} > 6$ and $\log N_{\mu} > 5.1$. 
 The $N_\mu$ value is corrected for systematic effects by means of a proper correction 
 function parameterized out of MC simulations. With the above selection cuts,
 the effective time of observation is equivalent to 1173 days. Full efficiency 
 ($\gtrsim 95 \%$) is achieved for $\log (E/\mbox{GeV}) \geq 7.4$.

 \begin{figure}[!t]
  \centering
  \includegraphics[height=2.5in, width=3.1in]{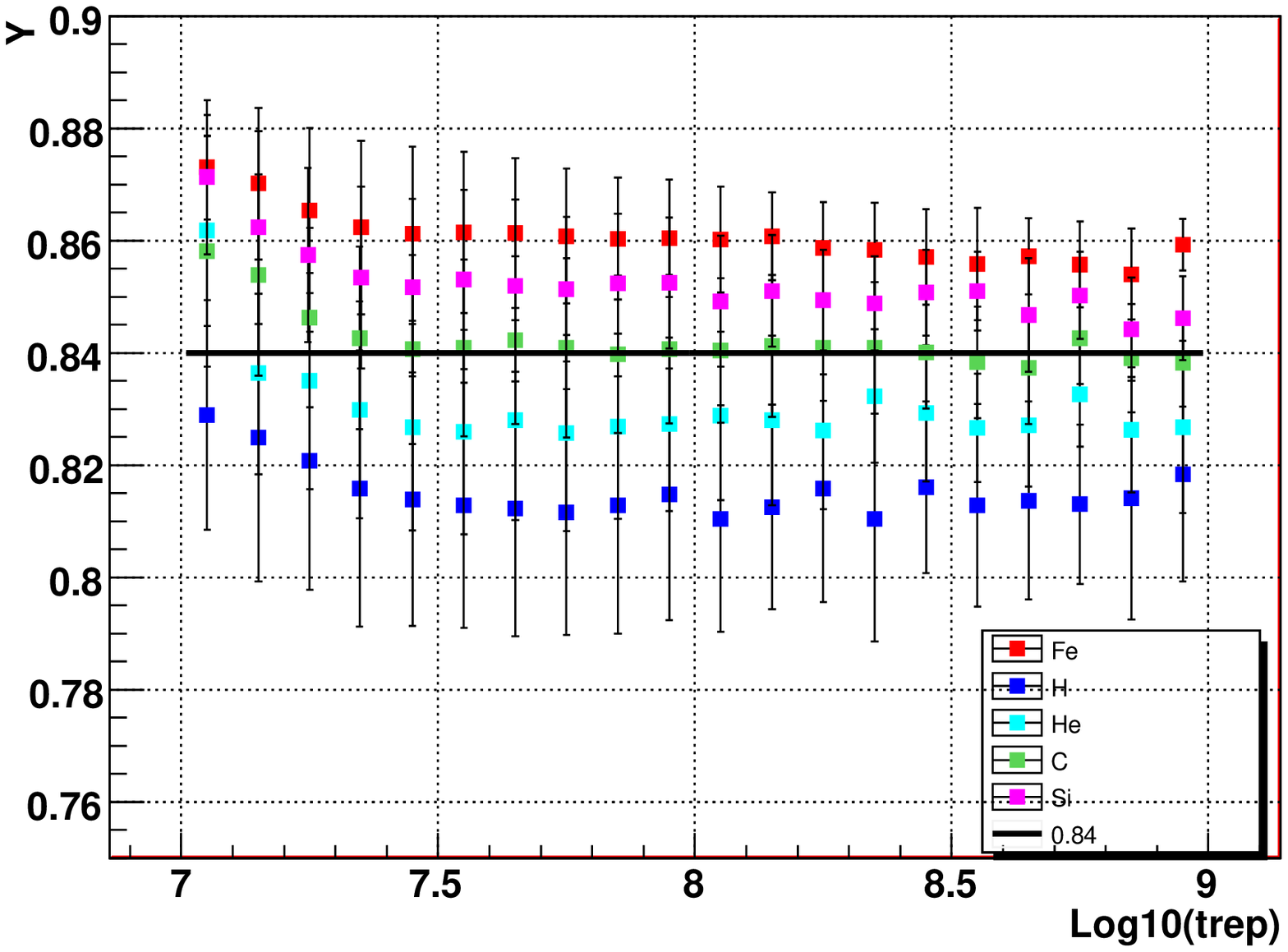}
  \caption{Estimations of the $Y^{CIC}$ ratio for five different primaries 
  shown as a function of the reconstructed energy. The results are obtained by
  means of a full EAS (based on the QGSJet II-03 high energy hadronic 
  interaction model) and detector simulation.}
  \label{icrc0739_fig01}
 \end{figure}

 \begin{figure*}[!t]
 \centering
 \includegraphics[height=2.5in, width=3.3in]{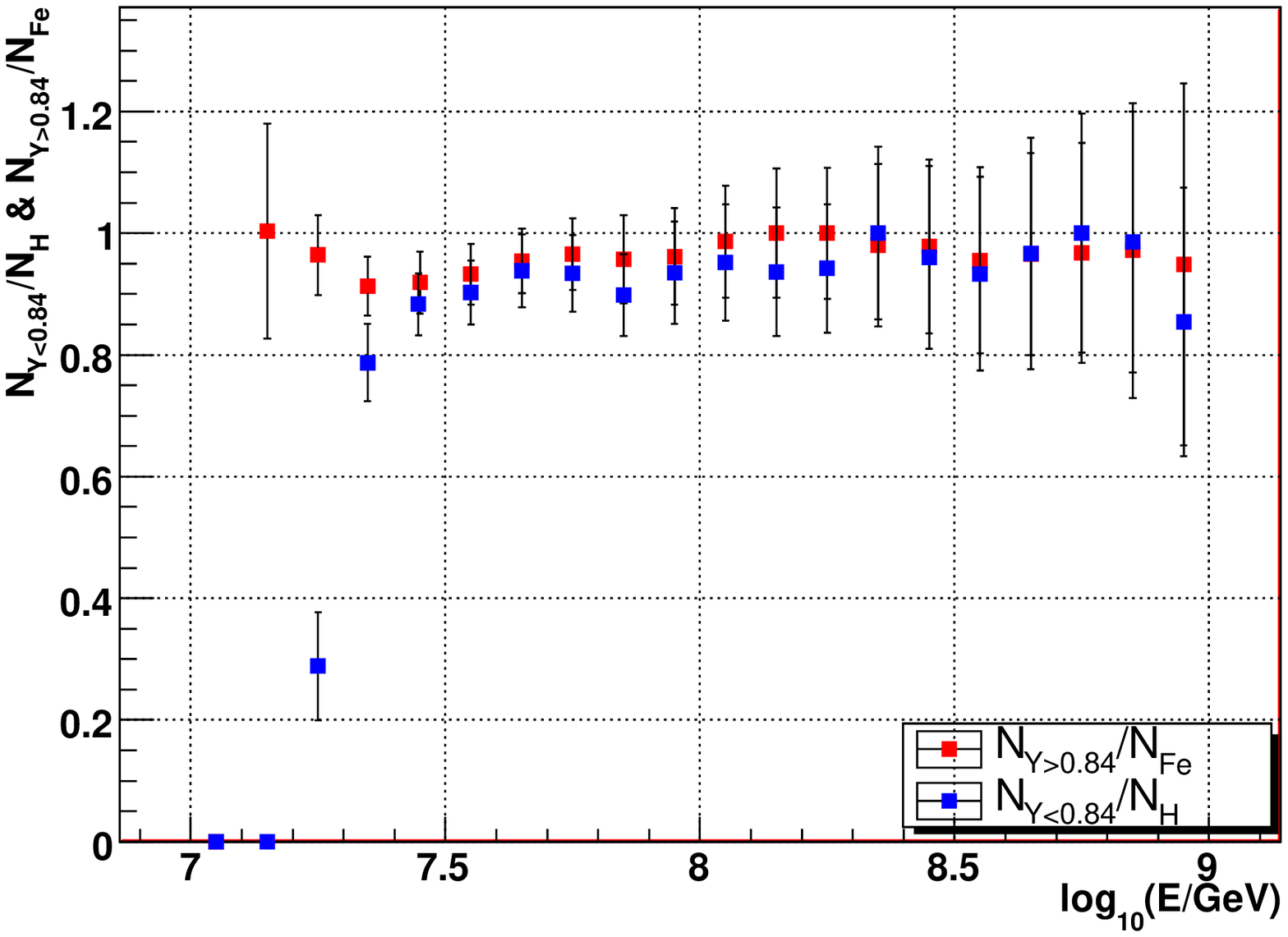}
 \includegraphics[height=2.5in, width=3.3in]{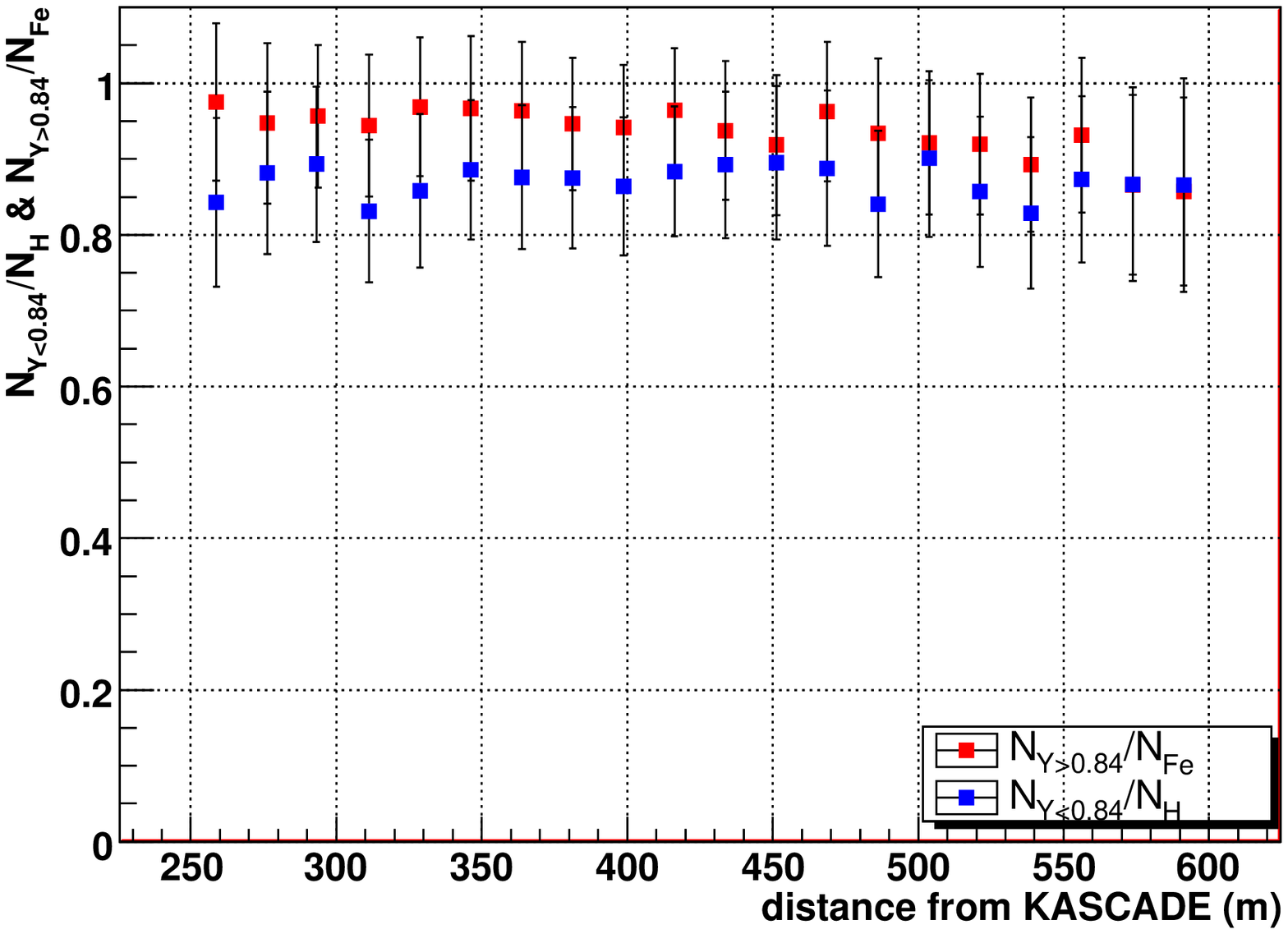}
  \caption{Fraction of hydrogen (iron) events classified as \textit{electron
  rich} (\textit{electron poor}) using the cut $Y^{CIC}_{cut} = 0.84$. The 
  fractions for protons (stars) and iron nuclei (dots) are plotted against the 
  reconstructed primary energy and the distance of the core position to the 
  center of KASCADE. These results are obtained with a full EAS simulation 
  based on the QGSJet II-03 interaction model.}
 \label{icrc0739_fig0203}
\end{figure*}

 \section{The $Y^{CIC}$ ratio}

 The KASCADE-Grande $(N_\mu, N_{ch})$ data offers a unique opportunity to
 investigate the chemical composition of galactic cosmic rays.
 Several methods can be envisaged for this enterprise.
 The one described in this contribution relies on the classification of the EAS 
 on the basis of the $Y^{CIC}$ ratio, a parameter defined as
 \begin{equation}
   Y^{CIC} = \log N_\mu(\theta_{ref})/ \log N_{ch}(\theta_{ref}).
   \label{eqn1}
 \end{equation}
 In the above formula, the EAS observables, $N_\mu$ and $N_{ch}$, have been
 corrected  event by event for attenuation effects in the atmosphere in such a
 way that they correspond to the shower sizes at a reference zenith angle,
 $\theta_{ref}$. The correction is achieved by applying the Constant Intensity 
 Cut Method (CIC) as described in references \cite{kang} and \cite{arteaga}. 
 This procedure allows to combine data collected from different zenith angles 
 in a model independent way. Here, $\theta_{ref}$ was chosen to $22^{\circ}$.

  The $Y^{CIC}$ parameter is a sensitive quantity to the composition of  
 primary cosmic rays. The point can be better appreciated by looking at 
 figure \ref{icrc0739_fig01}. There the mean values of $Y^{CIC}$ are plotted 
 as a function of the reconstructed energy for five single primaries 
 (see \cite{bertainaC} for details about the energy estimation), the error 
 bars represent the RMS of the $Y^{CIC}$
 distributions. We can notice that at a fixed energy, the 
 $Y^{CIC}$ ratio grows with the mass of the primary particle. 
 The effect can be understood as a result of the fact that the 
 physics of hadronic interactions favors the production of more secondary 
 charged pions, and hence of muons, for heavy nuclei. This implies
 a lower production of electrons and charged particles in general when
 increasing the mass number of the parent nuclei. As seen in figure
 \ref{icrc0739_fig01}, the $Y^{CIC}$ distributions are almost energy 
 independent in the region of $100 \%$ efficiency. Taking advantage of the 
 overall behavior of $Y^{CIC}$ and in a first attempt 
 to study the primary composition of cosmic rays in KASCADE-Grande by means of 
 this parameter a cut is applied on the data at $Y^{CIC}_{cut} = 0.84$.  
 Therefore the events are divided into two different sets for a subsequent 
 analysis: the \textit{electron rich} and \textit{electron poor} groups 
 corresponding to events with $Y^{CIC} < 0.84$ and $Y^{CIC} \geq 0.84$, 
 respectively. From figure \ref{icrc0739_fig01}, it is clear that the first set 
 covers the mean $Y^{CIC}$ values for light mass elements, here H and He, while 
 the second set contains the region for heavy mass nuclei, such as Si and Fe. 
 In this way the \textit{electron rich} and \textit{electron poor} samples 
 become representative of the light and heavy mass groups.
 
 We remind that the $Y^{CIC}$ distributions were obtained in the framework of the 
 QGSJet II-03 hadronic model. However other hadronic interaction models, such as 
 EPOS 1.99 \cite{eposC}, were investigated. Through these studies it was 
 observed that the mean values of $Y^{CIC}$ still
 do not depend on the primary energy, while their absolute values are modified. 
 Thus, in the case of the EPOS 1.99 hadronic interaction model, the optimal 
 separation between light and heavy mass groups is $Y^{CIC}_{cut} = 0.86$.

 The separation into different mass groups employing the $Y^{CIC}$ cut is not 
 completely clean due to the size of the fluctuations inside each energy bin. 
 Using MC simulations we have evaluated, for each element, the fraction of 
 events selected as {\it electron rich} or {\it electron poor}. For simplicity, 
 only the results obtained for hydrogen and iron nuclei are shown in figure 
 \ref{icrc0739_fig0203} as a function of the reconstructed energy (left panel) 
 and the core position (right panel).
 For energies greater than $\log (E/GeV) \geq 7.4$ (i.e. $100
 \%$ efficiency) the fraction of events properly classified through the
 $Y^{CIC}$ ratio is almost energy independent and greater than $80 \%$. This
 fraction is somewhat higher for iron primaries because of the lower shower
 fluctuations for heavy nuclei. It is not surprising to find out that their 
 corresponding curves are approximately constant (in the energy range of full 
 detection efficiency) over the whole effective area since the selection cuts on the EAS
 parameters were explicitly chosen to guarantee that measured shower parameters 
 do not depend on the experimental conditions, such as the core distance from 
 the KASCADE array center (i.e. mean distance of the muon detectors from the
 shower core).
 
 These results show that possible distortions introduced in the reconstructed 
 energy spectra of the different event samples, separated using the 
 $Y^{CIC}_{cut}$ value, are minimized. In the next section, the method is used
 to get insight into the composition of the primary cosmic ray flux.

\section{Applications: Energy spectra for two mass groups}

\begin{figure}[!t]
  \centering
  \includegraphics[height=2.5in, width=3.1in]{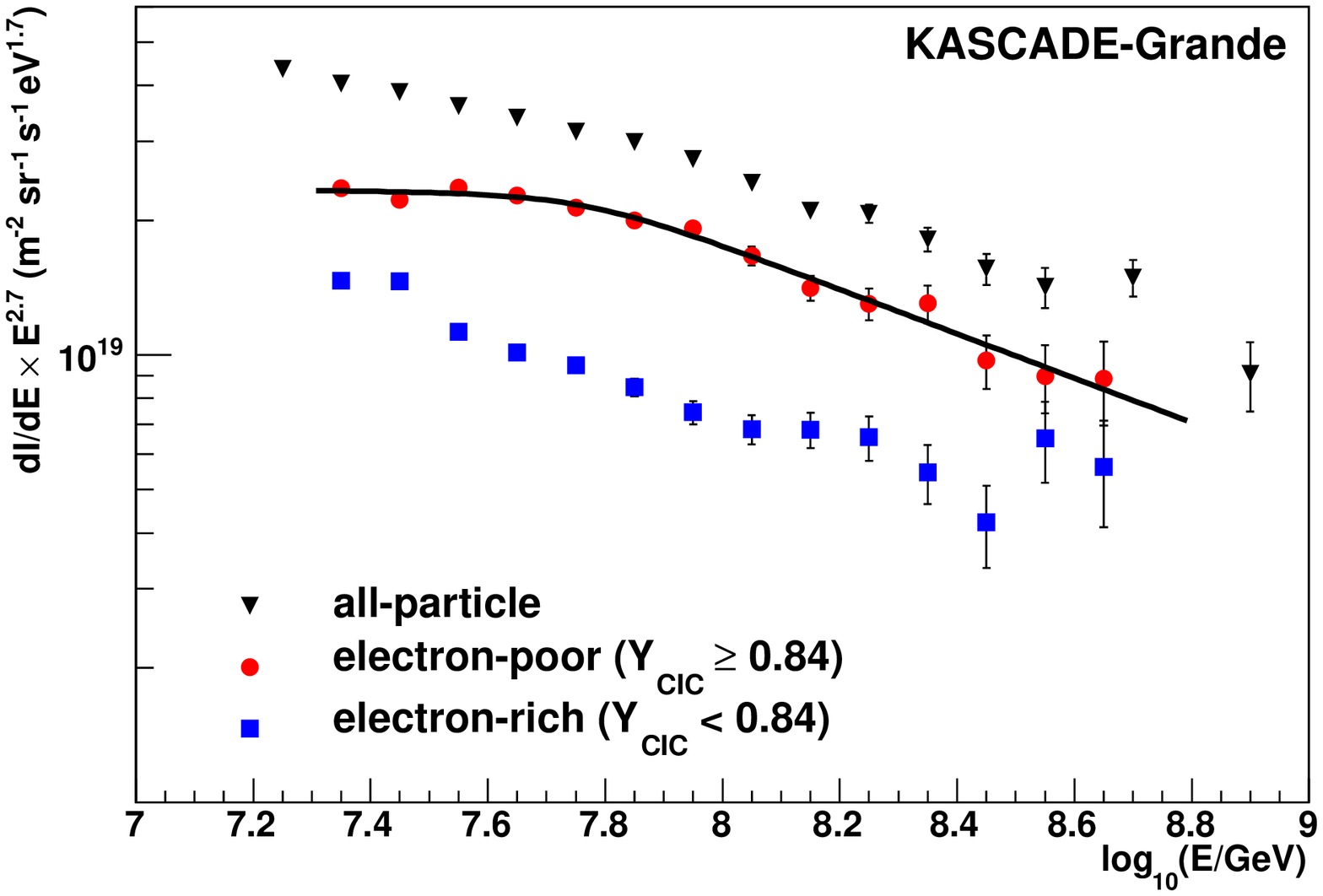}
  \caption{Light and heavy energy spectra derived from the 
   KASCADE-Grande $(N_\mu, N_{ch})$ data using the $Y_{CIC}$ ratio as a mass 
   estimator. The all-particle cosmic ray energy spectrum (squares) derived 
   by adding the individual mass group fluxes is also shown.}
  \label{icrc0739_fig04}
 \end{figure}

  Once the experimental data is divided into  \textit{electron poor} and 
  \textit{electron rich} groups using the above procedure above, the
  primary energy of each event is estimated following \cite{bertainaC}.
  After the energy derivation, the energy spectrum is reconstructed for each
  electron group leading to the energy spectra for  the light and heavy
  components of the cosmic ray primary flux, which are presented in figure 
  \ref{icrc0739_fig04}. The all-particle spectrum is also shown, and it is
  estimated as the direct sum of the two mass group spectra here estimated. 
  The plots are not corrected for migration effects.

\begin{figure}[!t]
  \centering
  \includegraphics[height=2.5in, width=3.1in]{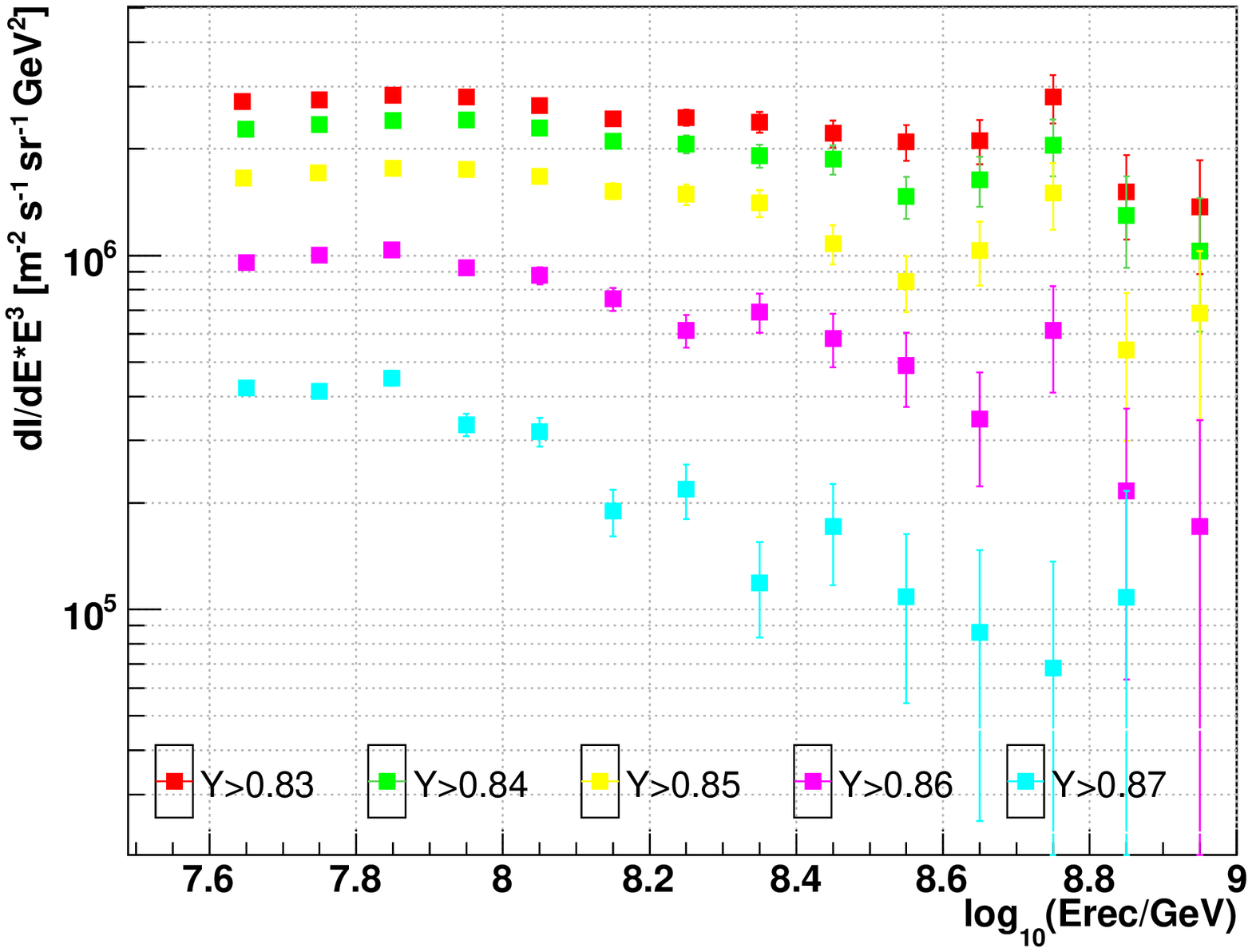}
  \caption{Energy spectra for the \textit{electron poor} sample obtained for 
   different $Y^{CIC}$ cuts.}
  \label{icrc0739_fig05}
 \end{figure}

  From figure \ref{icrc0739_fig04} an outstanding feature is immediately 
  appreciated in the heavy component of cosmic rays which corresponds to a 
  \textit{knee}-like structure. This feature also appears in more detailed  
  analyses of the KASCADE-Grande data \cite{bertaina2, daniel, elenaC}. A fit with a 
  broken power-law spectrum to the \textit{electron poor} spectrum indicates a 
  change of slope  $\Delta \gamma = -0.47$ at $\log_{10}(E/\mbox{GeV}) = 7.77 
  \pm 0.06$ from $\gamma = -2.72 \pm 0.02$ to $\gamma = -3.19
  \pm 0.02$. 
  As seen in figure \ref{icrc0739_fig05}, the knee-like feature is a structure
  which does not depend on the position of the $Y^{CIC}_{cut}$, i.e., it is 
  inherent to the cosmic ray data. 

  In figure  \ref{icrc0739_fig04}, it can be noticed that the
  all-particle energy spectrum shows also a break. From a fit
  with a broken power-law spectrum, the position of the break is found around
  $\log_{10}(E/\mbox{GeV}) = 7.92 \pm 0.10$, but the change of slope
  is smaller for this case: $\Delta \gamma = -0.29$, from 
  $\gamma = -2.95 \pm 0.05$ to $\gamma = -3.24 \pm 0.08$. It is seen that this 
  feature arises as a consequence of the knee 
  in the heavy component, but is less pronounced due to the contribution of 
  the light group, which is not zero for $\log (E/\mbox{GeV}) \geq 7.4$.

 \section{Conclusions}

  In this contribution, the potential of the $Y^{CIC}$ ratio as a mass
  sensitive parameter was studied. It was shown that the classification of 
  EAS data into \textit{electron rich} and \textit{electron poor} events by 
  applying the cut $Y^{CIC}_{cut} = 0.84$ can lead to the separation of the
  light and heavy mass groups in the cosmic ray flux. The fraction of events
  generated by hydrogen (iron) primaries and classified as \textit{electron
  rich} (\textit{electron poor}) does not depend on the primary energy and (in 
  the frame of the QGSJet II-03 interaction model) is greater than $80 \%$.
  Properly choosing the selection cuts on the EAS parameters the
  separation efficiency can be expected to be constant with energy, core
  position and zenith angle avoiding the introduction of distortions to the
  corresponding energy spectra. By applying the $Y^{CIC}$ method, the 
  energy spectra of the light and heavy mass groups were obtained. The
  results show that the heavy component presents a knee around 
  $\log_{10}(E/\mbox{GeV}) = 7.77 \pm 0.06$ and that the contribution to 
  the all-particle energy spectrum 
  from the light component is different from zero.

\vspace*{0.5cm} \footnotesize{
{\bf Acknowledgment:} KASCADE-Grande is supported by the BMBF of Germany, 
the MIUR and INAF of Italy, the Polish Ministry of Science and Higher Education 
and the Romanian Authority for Scientific Research. This study was partly 
supported by the DAAD-Proalmex program (2009-2010) and CONACYT.}

%\vspace{\baselineskip}
\

\clearpage

%% file: icrc0504.tex
%%
% 32nd International Cosmic Ray Conference 2011 Beijing China

%Class Required
%%% for classical LaTeX

%The paper title
\title{The cosmic ray elemental composition based on measurement of the 
N$_\mu$/N$_{ch}$ ratio with KASCADE-Grande}
%The short title will appear at the header of the even pages.

\shorttitle{E. Cantoni \etal The cosmic ray elemental composition......}

%All paper authors
\authors{
E.~Cantoni$^{1,7}$,
W.D.~Apel$^{2}$,
J.C.~Arteaga-Vel\'azquez$^{3}$,
K.~Bekk$^{2}$,
M.~Bertaina$^{1}$,
J.~Bl\"umer$^{2,4}$,
H.~Bozdog$^{2}$,
I.M.~Brancus$^{5}$,
P.~Buchholz$^{6}$,
A.~Chiavassa$^{1}$,
F.~Cossavella$^{4,13}$,
K.~Daumiller$^{2}$,
V.~de Souza$^{8}$,
F.~Di~Pierro$^{3}$,
P.~Doll$^{2}$,
R.~Engel$^{2}$,
J.~Engler$^{2}$,
M. Finger$^{4}$, 
D.~Fuhrmann$^{9}$,
P.L.~Ghia$^{7}$, 
H.J.~Gils$^{2}$,
R.~Glasstetter$^{9}$,
C.~Grupen$^{6}$,
A.~Haungs$^{2}$,
D.~Heck$^{2}$,
J.R.~H\"orandel$^{10}$,
D.~Huber$^{4}$,
T.~Huege$^{2}$,
P.G.~Isar$^{2,14}$,
K.-H.~Kampert$^{9}$,
D.~Kang$^{4}$, 
H.O.~Klages$^{2}$,
K.~Link$^{4}$, 
P.~{\L}uczak$^{11}$,
M.~Ludwig$^{4}$,
H.J.~Mathes$^{2}$,
H.J.~Mayer$^{2}$,
M.~Melissas$^{4}$,
J.~Milke$^{2}$,
B.~Mitrica$^{5}$,
C.~Morello$^{7}$,
G.~Navarra$^{1,15}$,
J.~Oehlschl\"ager$^{2}$,
S.~Ostapchenko$^{2,16}$,
S.~Over$^{6}$,
N.~Palmieri$^{4}$,
M.~Petcu$^{5}$,
T.~Pierog$^{2}$,
H.~Rebel$^{2}$,
M.~Roth$^{2}$,
H.~Schieler$^{2}$,
F.G.~Schr\"oder$^{2}$,
O.~Sima$^{12}$,
G.~Toma$^{5}$,
G.C.~Trinchero$^{7}$,
H.~Ulrich$^{2}$,
A.~Weindl$^{2}$,
J.~Wochele$^{2}$,
M.~Wommer$^{2}$,
J.~Zabierowski$^{11}$
}
%All the affiliations.
\afiliations{
$^1$ Dipartimento di Fisica Generale dell' Universit\`a Torino, Italy\\
$^2$ Institut f\"ur Kernphysik, KIT - Karlsruher Institut f\"ur Technologie, Germany\\
$^3$ Universidad Michoacana, Instituto de F\'{\i}sica y Matem\'aticas, Morelia, Mexico\\
$^4$ Institut f\"ur Experimentelle Kernphysik, KIT - Karlsruher Institut f\"ur Technologie, Germany\\
$^5$ National Institute of Physics and Nuclear Engineering, Bucharest, Romania\\
$^6$ Fachbereich Physik, Universit\"at Siegen, Germany\\
$^7$ Istituto di Fisica dello Spazio Interplanetario, INAF Torino, Italy\\
$^8$ Universidade S$\tilde{a}$o Paulo, Instituto de F\'{\i}sica de S\~ao Carlos, Brasil\\
$^9$ Fachbereich Physik, Universit\"at Wuppertal, Germany\\
$^{10}$ Dept. of Astrophysics, Radboud University Nijmegen, The Netherlands\\
$^{11}$ Soltan Institute for Nuclear Studies, Lodz, Poland\\
$^{12}$ Department of Physics, University of Bucharest, Bucharest, Romania\\
\scriptsize{
$^{13}$ now at: Max-Planck-Institut Physik, M\"unchen, Germany; 
$^{14}$ now at: Institute Space Sciences, Bucharest, Romania; 
$^{15}$ deceased; 
$^{16}$ now at: Univ Trondheim, Norway
}
}
%email address of the contact person
\email{cantoni@to.infn.it, achiavas@to.infn.it}

%The abstract.
\abstract{The KASCADE-Grande experiment, located at Karlsruhe Institute of 
Technology, is a multi-component Extensive Air Shower (EAS) detector studying 
primary cosmic rays in the $10^{16}-10^{18}$ eV energy range.
In this contribution a measurement of the cosmic ray chemical composition, 
based on the comparison of the experimental distributions of the ratio
between the EAS muon size and the total charged size (N$_\mu$/N$_{ch}$) 
with those expected from a complete EAS simulation based on the QGSJet II-03 
interaction model, is presented.
It has already been shown that, in the frame of this interaction model, the 
detector performances allow to separate three different mass groups: light, 
intermediate and heavy. With the employed technique, the relative abundances of the 
three mass groups are derived and their evolution as a function of the charged
particle size is observed. The differential spectra of the three mass groups are
then inferred and discussed
}
%The keywords

% B E G I N   D O C U M E N T
\maketitle

%Begin the section.
\section{Introduction}

The energy range from $10^{16}$ to $10^{18}$ eV is very important to deeply
investigate the details of the knee of the primary energy spectrum. In the last ten 
years different experiments (e.g KASCADE \cite{kascunf} and EAS-TOP \cite{et}) have 
shown that this feature can be attributed to the light component of the primaries 
(i.e. H or He). A definitive confirmation of the astrophysical origin of the knee can
only be obtained by the detection of the foreseen change of slope of the heavy 
component at a primary energy $\sim 10^{17}$ eV.

In this contribution we present a study of the primary cosmic
ray chemical composition in the $10^{16}-10^{18}$ eV energy range performed
comparing the KASCADE-Grande experiment \cite{kg-NIM10} data with the 
results of a full Extensive Air Shower simulation based on the CORSIKA \cite{cors}
code and on the QGSJet II-03 \cite{qgs} interaction model.

The KASCADE-Grande experiment is located at the Campus North of the
Karlsruhe Institute of Technolgy, 110 m a.s.l. It consists of an array of
37 plastic scintillator modules 10 m$^2$ each (Grande) spread over
an area of 700x700 m$^2$, working jointly with the co-located and formerly
present KASCADE experiment \cite{kascade}, consisting of 252 electron and muon 
scintillation detectors placed over a 200x200 m$^2$ area. All events triggering 
a sevenfold coincidence of Grande array detectors, arranged on an hexagonal 
grid (mean side lenght of ~130 m), are reconstructed. The 
charged particle size N$_{ch}$ is determined by the Grande array and the muon 
number N$_\mu$ by the KASCADE array. Both (N$_{ch}$) and (N$_\mu$) are measured 
with an accuracy $\leq 15\%$; a detailed 
description of the reconstruction procedure can be found in \cite{kg-NIM10}.

\section{The Chi Square Method}

\begin{figure}[!t]
 \vspace{5mm}
  \centering
  \includegraphics[width=3.2in]{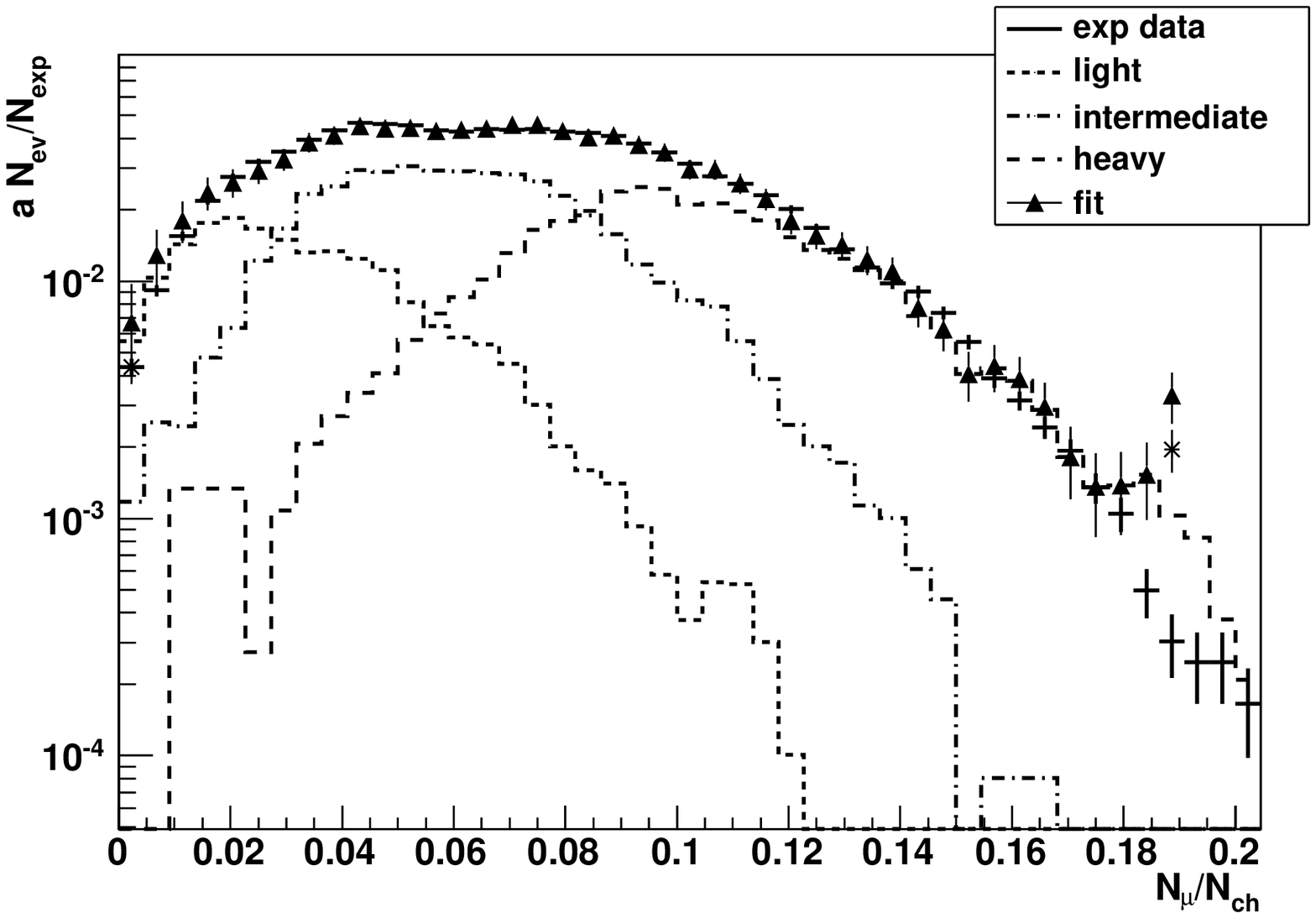}
  \caption{The experimental $N_\mu / N_{ch}$ distribution (solid line) measured 
  in the $6.24 \leq Log(N_{ch}) < 6.36$ interval for vertical events
  ($0^{\circ} \leq \theta \leq 24^{\circ}$) fitted by a combination of
  Hydrogen (thin dashed line), Helium + Carbon in a 50\% mixture (dot dashed
  line) and Silicon + Iron 
  in a 50\% mixture (thick dashed line). The triangles show the fit result. The 
  experimental plot is normalized to 1 and every simulated component is 
  normalized to its relative abundance. The tails of the 
  experimental distribution are treated summing the events (the
  corresponding value is shown by a star) in a single bin, 
  in order to have at least 5 events in  each of them. In the shown example, 
  the results for the abundances and the fit are: 
  $\alpha_{light} = 0.14 \pm 0.02$; $\alpha_{intermediate} = 0.59 \pm 0.02$; 
  $\alpha_{heavy} = 0.27 \pm 0.01$.  The chi square and the cumulative function
  values are: $\chi^{2}_{0}/\nu = 56.52/46 = 1.23$, 
  $P(\chi^{2}>\chi_{0}^{2}) = 0.14$}
 \label{fig_nchdist}
\end{figure}

\begin{figure}[!t]
 \vspace{0mm}
  \centering
  \includegraphics[width=3.2in]{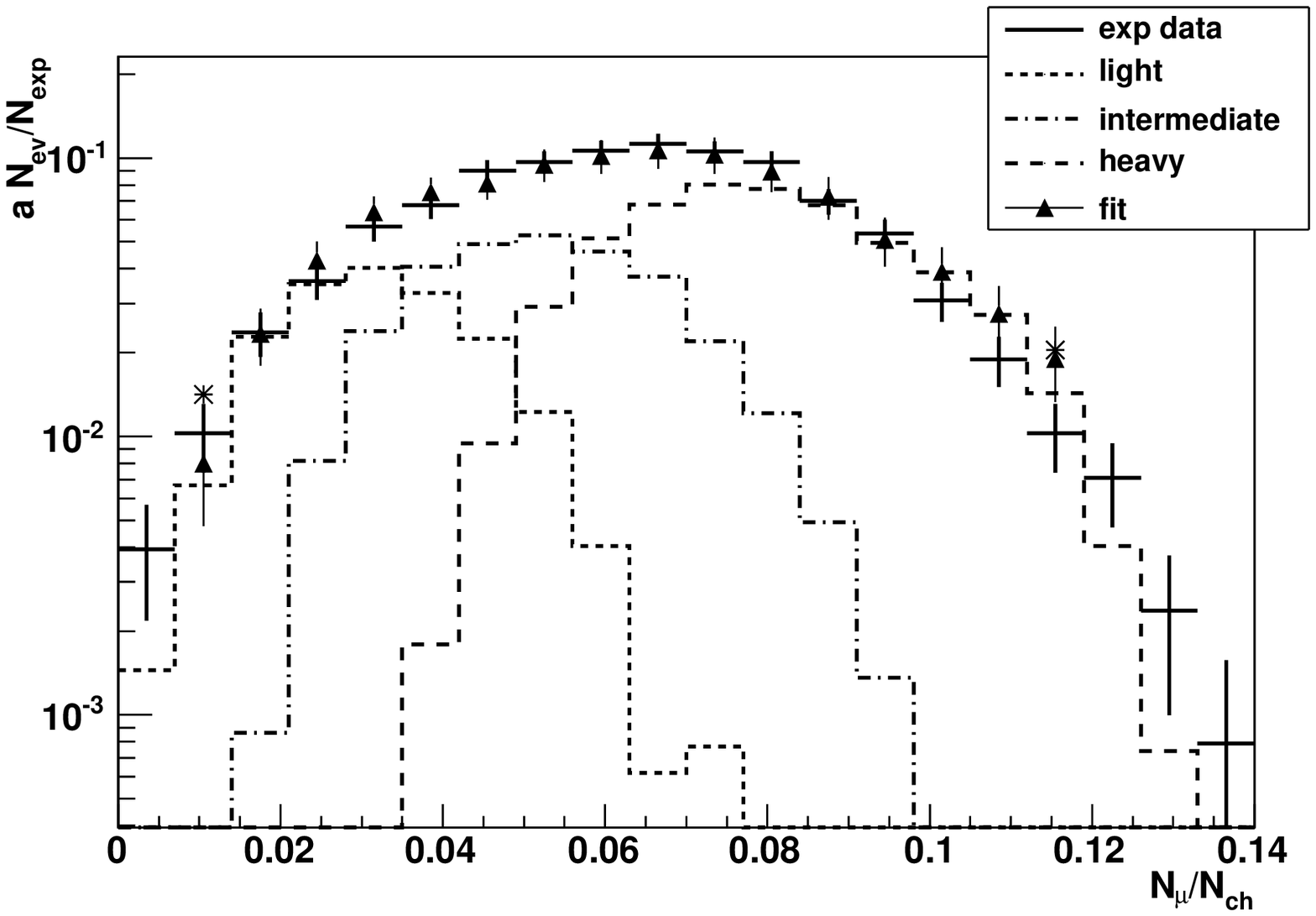}
  \caption{The experimental $N_\mu / N_{ch}$ distribution (solid line) measured 
  in the $7.15 \leq Log(N_{ch}) < 7.24$ interval for vertical events
  ($0^{\circ} \leq \theta \leq 24^{\circ}$) fitted by a combination of
  light (thin dashed line), intermediate (dot dashed line) and heavy (thick
  dashed line). The triangles show the fit results. In the shown example, the 
  results for the abundances and the fit are: 
  $\alpha_{light} = 0.16 \pm 0.02$; $\alpha_{intermediate} = 0.30 \pm 0.04$; 
  $\alpha_{heavy} = 0.52 \pm 0.03$.  The chi square and the cumulative function
  values are: $\chi^{2}_{0}/\nu = 9.66/14 = 0.69$, 
  $P(\chi^{2}>\chi_{0}^{2}) = 0.79$}
 \label{fig_nchdis2}
\end{figure}

The present analysis is performed, to minimize effects of a possible
incorrect description (in the simulation) of the EAS evolution in atmosphere, on 
vertical events ($0^{\circ} \leq \theta \leq 24^{\circ}$). Data 
are divided in twelve N$_{ch}$ intervals 
(from Log N$_{ch}$=6.0 to Log N$_{ch}$=8.0) and the N$_\mu$/N$_{ch}$ 
experimental distributions are fitted with a linear combination of those obtained by a 
complete simulation of events generated with a power law distribution in
the $10^{15}-10^{18}$ eV energy range for five different primaries (H, He, 
C, Si, Fe). We have already shown that the KASCADE-Grande experiment perfomances 
allow us to separate, with such algorithm, events into three samples originated 
by different mass groups \cite{kg-lodz09}. The combination giving the best fits 
(examples for two of the twelve N$_{ch}$ intervals are shown in figures 
\ref{fig_nchdist} and \ref{fig_nchdis2}) is the one composed by the following 
mass groups: light, 100\% hydrogen; intermediate, 50\% helium and 50\% carbon; 
heavy, 50\% silicon and 50\% iron.
The linear combination of the simulated mass groups is expressed as follows:
\begin{equation}\label{eq:sim}
F_{sim}(i) = \sum_{j}\alpha_{j}f_{sim,j}(i)
\end{equation}
where $F_{sim}(i)$ is the total theoretical fraction of simulated events falling in the 
histogram channel $i$, $f_{sim,j}(i)$ is the fraction for the single component 
$j$ (j = 1,2,3), $\Sigma_{j}(i)$ is the sum over the different components and 
$\alpha_{j}$ is the fit parameter representing the relative abundance of the component 
$j$. The fit parameters fulfill the conditions
\begin{equation}\label{eq:cond1}
0. \leq \alpha_{j} \leq 1.
\end{equation}
and
\begin{equation}\label{eq:cond2}
\sum_{j}\alpha_{j} = 1.
\end{equation}
The fit is performed minimizing the following Chi Square function:
\begin{equation}\label{eq:chi}
\chi^{2} = \sum_{i} \frac{(F_{exp}(i) - F_{sim}(i))^{2}}{\sigma(i)^{2}}
\end{equation}
where $F_{exp}(i)$ is the fraction of experimental events falling in the histogram 
channel $i$ and $\sigma(i)$ is the error on the theoretical 

expression (\ref{eq:sim}). A fit is validated and the relative abundances of the mass
groups are accepted if the cumulative function, $P(\chi^{2}>\chi_{0}^{2})$, is included in
the $0.05 < P < 0.95$ interval.

The choice of sampling events in bins of the charged particle size is made in order
to have a selection depending only on the performances of the experiment and not
being influenced by an EAS simulation (as could be in the case of reconstructed energy
bins).

\begin{figure}[!t]
 \vspace{0mm}
  \centering
  \includegraphics[width=3.2in]{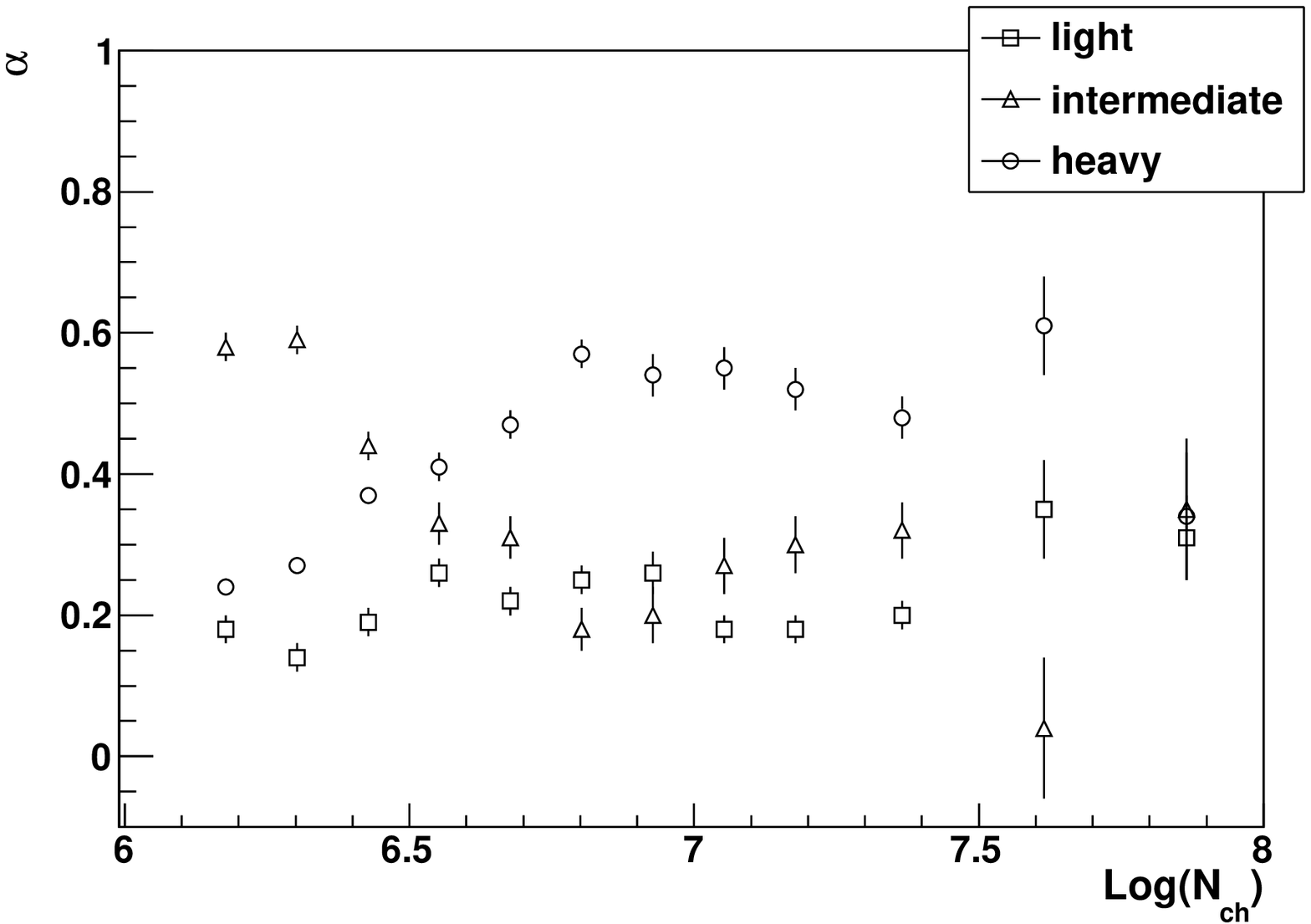}
  \caption{Relative abundances of the three mass groups versus the charged 
  particle size $N_{ch}$ measured for vertical events.}
 \label{fig_relabun}
\end{figure}

The behavior of the relative abundances versus the charged particle size is shown 
in figure \ref{fig_relabun}. The light and intermediate components show a trend with
bigger fluctuations with respect to the heavy mass group. The relative abundance of 
the light mass group is almost independent on the charged particle shower size, 
while the heavy one shows a sudden change at $\log N_{ch} \sim 6.8$. The
intermediate mass group seems to be more abundant near 
(in $N_{ch}$ bins) the threshold, decreasing for higher $N_{ch}$ and 
becoming nearly similar to those of light elements.

\section{Spectra of the single mass groups}

From the number of events measured in each $\Delta \log(N_{ch})$ interval
($N_{exp}(k), k=1,12$) and having determined the relative abundance of each 
mass group ($j$) we can calculate the number of events that are originated by 
primaries belonging to the $j^{th}$ mass group 
($N_j (k)= \alpha_{j}\cdot N_{exp}(k)$). The three mass group fluxes in each 
charged particle size bin are then derived:
\begin{equation}
\Phi_j(k) = N_j(k)/(A \Omega \Delta t) (m^{-2} sr^{-1} s^{-1})
\end{equation}
where $A=1.52\times 10^4 m^2$, $\Omega=0.52 sr$ and $\Delta t= 1.1 \times 10^8 
s$.

From a full EAS (based on the QGSJet II-03 interaction model) and detector 
simulation we derive, for each mass group, a power law relation 
($E_{0} = 10^{b}\cdot N_{ch}^{a}$) between the charged particle size and the 
primary energy (the values of the $a$ and 
$b$ parameters are reported in table \ref{tab_ab}).
\begin{table}[t]
\begin{center}
\begin{tabular}{l|ccc}
\hline
 & light & intermediate & heavy \\
\hline
$a$&0.99 $\pm$ 0.01&0.93 $\pm$ 0.004&0.91 $\pm$ 0.01\\
$b$&0.75 $\pm$ 0.06&1.30 $\pm$ 0.02&1.61 $\pm$ 0.03\\
\hline
\end{tabular}
\caption{$a$, $b$ parameters of the power law correlating, for each mass group, 
the charged particle size N$_{ch}$ and the primary energy $E_0$}
\label{tab_ab}
\end{center}
\end{table}
With these relations we convert the N$_{ch}$ intervals into the corresponding 
energy bins (that are thus not the same for the three mass groups) and from the 
measured fluxes we derive the differential flux at the central energy of 
each interval.

\begin{figure}[!t]
 \vspace{0mm}
  \centering
  \includegraphics[width=3.2in]{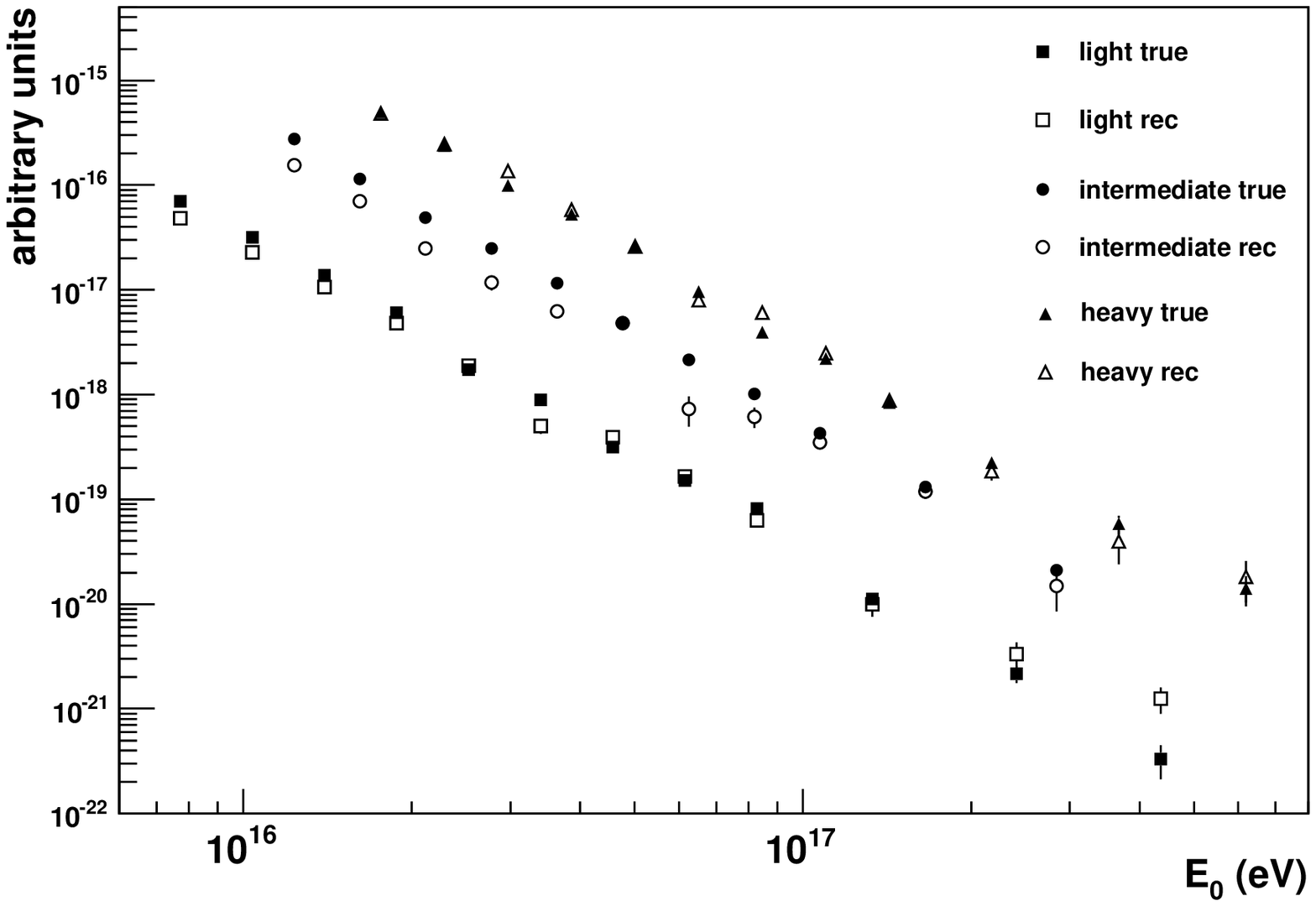}
  \caption{Comparison between test and reconstructed spectra for the three mass
  groups. In the fake spectrum all mass groups have equal abundance, in
  the plot the fluxes of light and heavy mass groups are multiplied by arbitrary
  factors for clarity.}
 \label{fig_testsp}
\end{figure}

To verify if, through the described algorithm, we are able to reproduce the
spectra of the single mass groups we have performed the same analysis on test 
spectra: half of the simulated data set is used as fake experimental data while 
the other half is used as reference for the fitting procedure. With the goal of
identifying possible spectral distortions artificially introduced by the
analysis we have chosen the test spectra with a slope $\gamma = -3$ and 
without breaks for all elements. The results are shown in figure 
\ref{fig_testsp}: the heavy mass
group shows good agreement between test and reconstructed spectra in the whole
energy range. The light and intermediate mass groups are in good agreement for 
energies greater than $4\times 10^{16}$ eV while near the threshold the 
reconstructed spectra are lower than the test ones. This result is not 
unexpected as shower and experimental fluctuations (that are bigger for low 
masses and originate an event migration from bin to bin) are not yet corrected 
for in the conversion from $N_{ch}$ to energy. This procedure is under development, 
its results will be shown soon. It is worthwhile to point out that the  
uncorrected reconstructed test spectra show, for all mass groups, no 
artificial breaks.

\begin{figure}[!t]
 \vspace{5mm}
  \centering
  \includegraphics[width=3.2in]{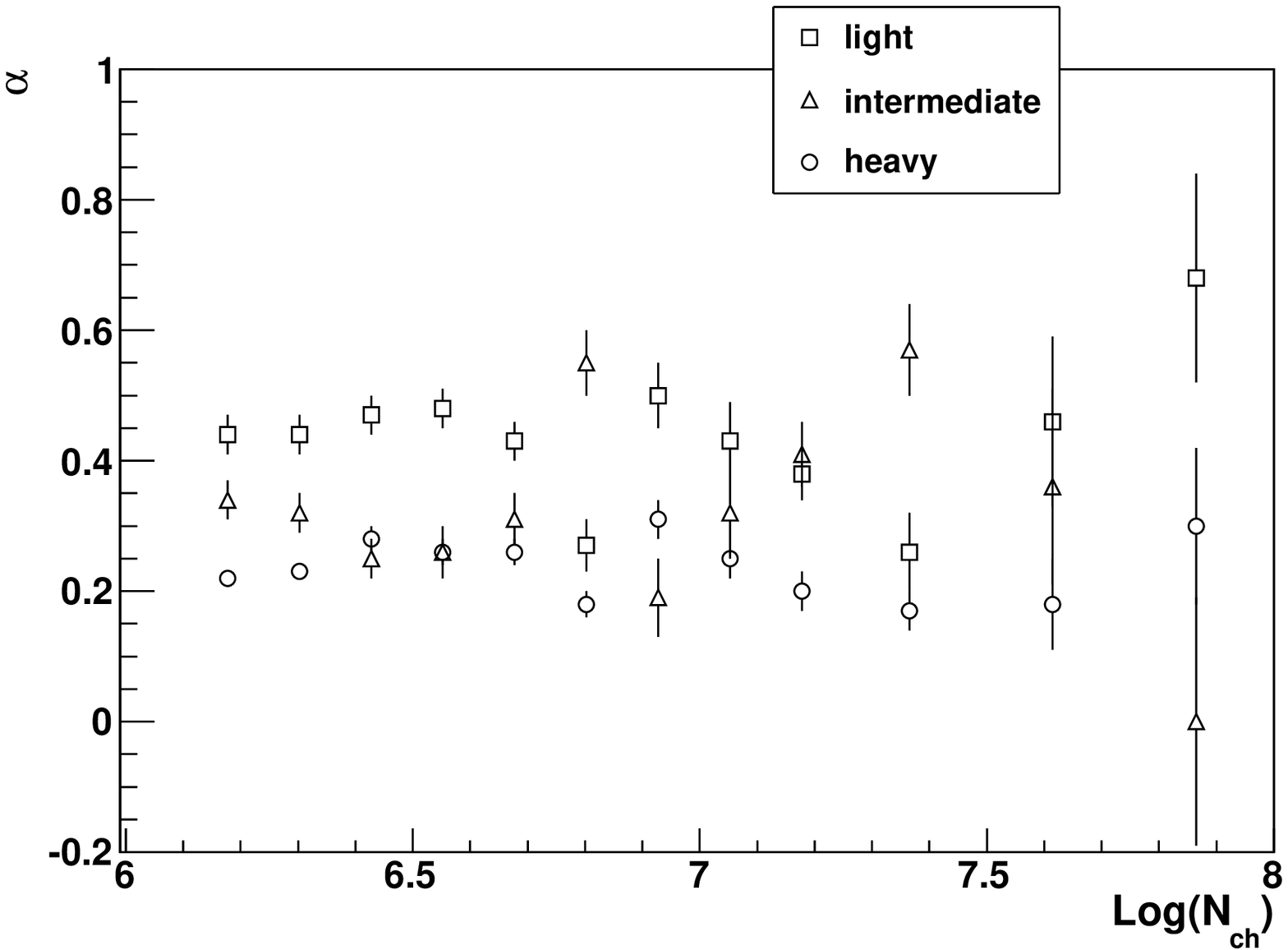}
  \caption{Reconstructed relative abundances of the three mass groups versus 
  the charged particle size $N_{ch}$ obtained analyzing test spectra with
  constant and equal slope for all elements.}
 \label{fig_relats}
\end{figure}

Another hint of the correct behavior of the described analysis algorithm can be
derived from figure \ref{fig_relats} showing the relative abundances
reconstructed for the previously described test spectra. These abundances show
no dependence on $N_{ch}$ in the whole range, even if fluctuations 
(mainly of the light and intermediate mass groups) around the 
expected value can be seen.

\section{Conclusions}

The performances of the algorithm aiming at the measurement of the evolution of
the primary chemical composition of cosmic rays are described. We have shown
how, with the KASCADE-Grande experiment performances, we are able to separate 
three mass groups and that all of them are needed to reconstruct the measured 
$N_\mu/N_{ch}$ distibutions. The measured chemical composition gets heavier as
the shower size $N_{ch}$ increases as can be inferred from the evolution of 
the relative abundances of the three mass groups with $N_{ch}$.

\begin{figure}[!t]
 \vspace{5mm}
  \centering
  \includegraphics[width=3.2in]{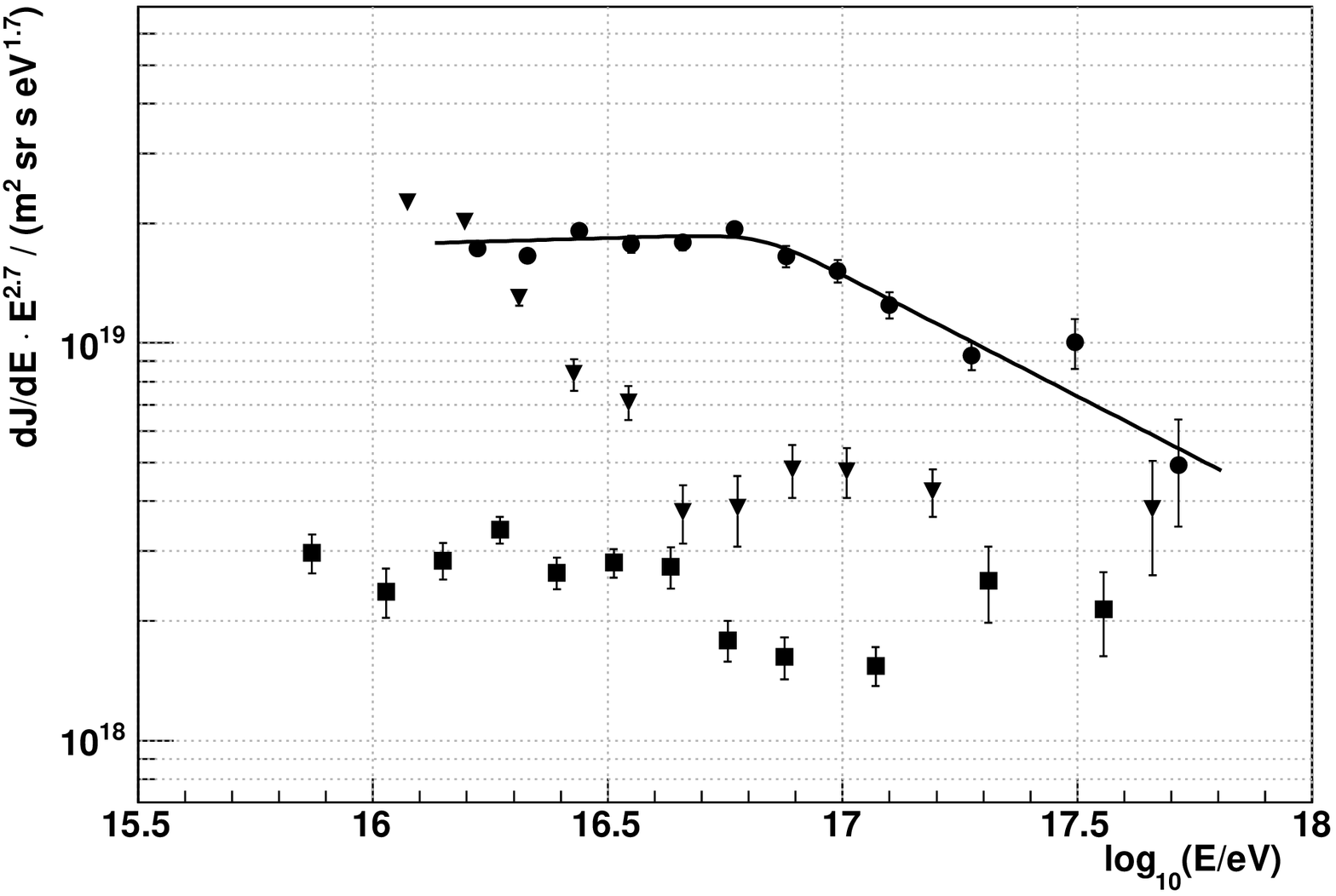}
  \caption{Differential energy spectra of the three mass groups: heavy (Si+Fe, dots),
  intermediate (He+C, triangles) and light (H, squares) elements.}
 \label{fig_expsp}
\end{figure}

The measured spectra of the three mass groups are shown in figure 
\ref{fig_expsp} as obtained in the frame of the QGSJet II-03 interaction model. The spectrum 
of the heavy mass group shows a distinct change of slope and cannot be described by a single 
power law. This reconstructed spectrum is fitted with a function \cite{kascfit} describing 
a spectral shape with two different slopes interconnected by a smooth knee at energy $E_k$. The results, 
shown by the solid line in figure \ref{fig_expsp}, are: $\gamma_1= 2.67\pm 0.02$, $\gamma_2 = 
3.29\pm 0.02$ and $\log E_k (eV) = 17.79\pm 0.04$. The statistical significance of the
change of slope is $\sim 4\sigma$. At a similar energy a (weaker) steepening 
is also observed in the all particle spectrum measured by KASCADE-Grande \cite{bertaina}.

In contrast both the intermediate and light mass groups spectra are descriable 
by a single power law due to the low event numbers. These spectra are dominated by statistical 
fluctuations, and no significant conclusion can thus be derived at the moment.

All shown spectra do not yet take into account effects of event-to event shower  
fluctations at the energy assignment: a procedure to unfold them is under development.

It is important to emphasize that the present results, and in particular the relative
abundances of the mass groups, heavily depend on the choice of 
QGSJet II-03 as interaction model. The entire analysis will be, in near 
future, repeated on basis of a complete EAS simulation based on a 
different interaction models, i.e. EPOS \cite{epos}.

\vspace*{0.5cm} \footnotesize{
{\bf Acknowledgement:} KASCADE-Grande is supported by the BMBF of Germany, 
the MIUR and INAF of Italy, the Polish Ministry of Science and Higher Education 
and the Romanian Authority for Scientific Research.}

%\vspace{\baselineskip}
%\vspace{-0.4cm}

\clearpage

%% file: icrc0280.tex
%%
% 32nd International Cosmic Ray Conference 2011 Beijing China

%Class Required
%%% for classical LaTeX
%The paper title
\title{KASCADE-Grande measurements of energy spectra for elemental groups of cosmic rays}
%The short title will appear at the header of the even pages.

\shorttitle{D. Fuhrmann \etal KASCADE-Grande measurements of energy spectra}

%All paper authors
\authors{
D.~Fuhrmann$^{9}$,
W.D.~Apel$^{1}$,
J.C.~Arteaga-Vel\'azquez$^{2}$,
K.~Bekk$^{1}$,
M.~Bertaina$^{3}$,
J.~Bl\"umer$^{1,4}$,
H.~Bozdog$^{1}$,
I.M.~Brancus$^{5}$,
P.~Buchholz$^{6}$,
E.~Cantoni$^{3,7}$,
A.~Chiavassa$^{3}$,
F.~Cossavella$^{4,13}$,
K.~Daumiller$^{1}$,
V.~de Souza$^{8}$,
F.~Di~Pierro$^{3}$,
P.~Doll$^{1}$,
R.~Engel$^{1}$,
J.~Engler$^{1}$,
M.~Finger$^{4}$, 
P.L.~Ghia$^{7}$, 
H.J.~Gils$^{1}$,
R.~Glasstetter$^{9}$,
C.~Grupen$^{6}$,
A.~Haungs$^{1}$,
D.~Heck$^{1}$,
J.R.~H\"orandel$^{10}$,
D.~Huber$^{4}$,
T.~Huege$^{1}$,
P.G.~Isar$^{1,14}$,
K.-H.~Kampert$^{9}$,
D.~Kang$^{4}$, 
H.O.~Klages$^{1}$,
K.~Link$^{4}$, 
P.~{\L}uczak$^{11}$,
M.~Ludwig$^{4}$,
H.J.~Mathes$^{1}$,
H.J.~Mayer$^{1}$,
M.~Melissas$^{4}$,
J.~Milke$^{1}$,
B.~Mitrica$^{5}$,
C.~Morello$^{7}$,
G.~Navarra$^{3,15}$,
J.~Oehlschl\"ager$^{1}$,
S.~Ostapchenko$^{1,16}$,
S.~Over$^{6}$,
N.~Palmieri$^{4}$,
M.~Petcu$^{5}$,
T.~Pierog$^{1}$,
H.~Rebel$^{1}$,
M.~Roth$^{1}$,
H.~Schieler$^{1}$,
F.G.~Schr\"oder$^{1}$,
O.~Sima$^{12}$,
G.~Toma$^{5}$,
G.C.~Trinchero$^{7}$,
H.~Ulrich$^{1}$,
A.~Weindl$^{1}$,
J.~Wochele$^{1}$,
M.~Wommer$^{1}$,
J.~Zabierowski$^{11}$
}
%All the affiliations.
\afiliations{
$^1$ Institut f\"ur Kernphysik, KIT - Karlsruher Institut f\"ur Technologie, Germany\\
$^2$ Universidad Michoacana, Instituto de F\'{\i}sica y Matem\'aticas, Morelia, Mexico\\
$^3$ Dipartimento di Fisica Generale dell' Universit\`a Torino, Italy\\
$^4$ Institut f\"ur Experimentelle Kernphysik, KIT - Karlsruher Institut f\"ur Technologie, Germany\\
$^5$ National Institute of Physics and Nuclear Engineering, Bucharest, Romania\\
$^6$ Fachbereich Physik, Universit\"at Siegen, Germany\\
$^7$ Istituto di Fisica dello Spazio Interplanetario, INAF Torino, Italy\\
$^8$ Universidade S$\tilde{a}$o Paulo, Instituto de F\'{\i}sica de S\~ao Carlos, Brasil\\
$^9$ Fachbereich Physik, Universit\"at Wuppertal, Germany\\
$^{10}$ Dept. of Astrophysics, Radboud University Nijmegen, The Netherlands\\
$^{11}$ Soltan Institute for Nuclear Studies, Lodz, Poland\\
$^{12}$ Department of Physics, University of Bucharest, Bucharest, Romania\\
\scriptsize{
$^{13}$ now at: Max-Planck-Institut Physik, M\"unchen, Germany; 
$^{14}$ now at: Institute Space Sciences, Bucharest, Romania; 
$^{15}$ deceased; 
$^{16}$ now at: Univ Trondheim, Norway
}
}
%email address of the contact person
\email{fuhrmann@physik.uni-wuppertal.de}

%The abstract.
\abstract{The KASCADE-Grande experiment, located at KIT-Karlsruhe, Germany, consists of a large scintillator array for measurements of charged particles, $N_{ch}$, and of an array of shielded scintillation counters used for muon counting, $N_{\mu}$. KASCADE-Grande is optimized for cosmic ray measurements in the primary energy range $10^{16}$~eV to $10^{18}$~eV, thereby enabling the verification of a possible second knee expected at approximately $10^{17}$~eV. Exploring the composition in this energy range is of fundamental importance for understanding the transition from galactic to extragalactic cosmic rays. Following earlier studies of elemental spectra reconstructed in the first knee energy range from KASCADE data, we shall now extend these measurements to beyond $10^{17}$~eV. By analyzing the two-dimensional shower size spectrum $N_{ch}$ vs. $N_{\mu}$, we reconstruct the energy spectra of different mass groups by means of unfolding methods. The procedure and its results, which yield a strong indication for a kink in the iron spectrum at around 80 PeV, will be presented.}
%The keywords

% B E G I N   D O C U M E N T
\maketitle

%Begin the section.
\section{Introduction}
The spectrum of cosmic rays follows a power law over many orders of magnitude in energy, overall appearing rather featureless. However, there are a few structures observable. In 1958 Kulikov and Khristiansen \cite{lit:first_knee_measurement} discovered a distinct steepening in the spectrum at around $10^{15}$~eV. Three years later Peters \cite{lit:peters} concluded that the position of this kink, also called the cosmic ray ``knee'', will depend on the atomic number of the cosmic ray particles if their acceleration is correlated to magnetic fields. Round about half a century later, the KASCADE experiment \cite{lit:kascade_allgemein_nimpaper} clarified that this change in spectral index is caused by a decrease of the so far dominating light\footnote{The description ``light'' refers to the atomic mass of the cosmic ray particles, which are primarily nuclei.} component of cosmic rays \cite{lit:kascade-unfolding}. This result was achieved by means of an unfolding analysis disentangling the manifold convoluted energy spectra of five mass groups from the measured two-dimensional shower size distribution of electrons and muons at observation level. Based on the high energy interaction model QGSJET~01~\cite{lit:qgsjet01} it was shown, that the kink in the all-particle spectrum at around $5\times10^{15}$~eV corresponds to a knee observed in the hydrogen flux.\\
Nowadays, there are numerous theories about the origin and acceleration of cosmic rays. Concerning the knee position, some of them predict in contrast to the rigidity dependence considered by Peters a correlation with the mass of the particles. Hence, it is of great interest to verify whether also the spectra of heavy components exhibit analogous structures and if, at what energies. The KASCADE-Grande experiment \cite{lit:kascade_grande_allgemeim_nimpaper} extends the accessible energy range of KASCADE to higher energies up to $10^{18}$~eV and allows by this to investigate the cosmic rays composition at regions where the possible, so-called iron knee is expected. The determination of this ``second'' knee enables the validation of the various theoretical models. Following this purpose, the KASCADE-Grande measurements have been analyzed similar to the afore-mentioned studies \cite{lit:kascade-unfolding} of the KASCADE data. The applied unfolding method will be outlined in the next section. Thereafter, the uncertainties of the analysis and the resulting elemental energy spectra will be shown and studied. A more comprehensive description can be found in \cite{lit:phd}.
\section{Outline of the analysis}\label{sec:outline}
\begin{figure}[!t]
  %\vspace{+5mm}
  \centering
  \includegraphics[clip, width=0.8\columnwidth, bb= 0 4 567 360]{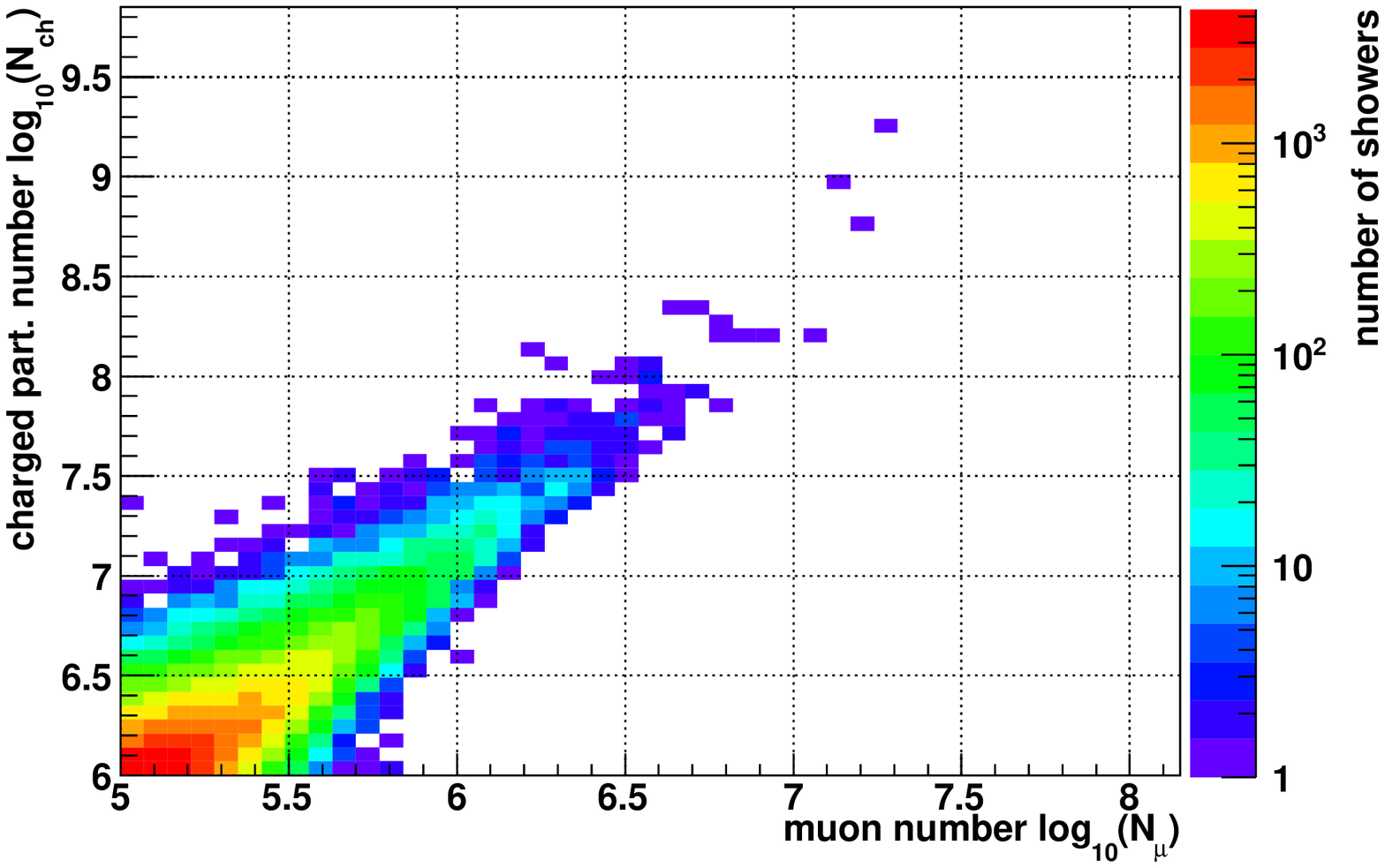}
  \caption{Measured shower size distribution.}
  \label{fig:2d_data_plane}
 \end{figure}

The analysis' objective is to compute the energy spectra of five\footnote{Due to effects of limited resolution not any number of mass groups can be treated.} cosmic ray mass groups, represented by hydrogen (H), helium (He), carbon (C), silicon (Si) and iron (Fe) nuclei, from $10^{16}$~eV beyond primary energies of $10^{17}$~eV. The two-dimensional shower size spectrum $\mathrm{lg}N_{ch}$ vs. $\mathrm{lg}N_{\mu}$ of charged particles and muons measured with KASCADE-Grande is used as starting point for the unfolding analysis (Fig.~\ref{fig:2d_data_plane}). 
All measured\footnote{Most of the cuts, e.g. the chosen zenith angle range, are also applied to the simulated air showers.} air showers have to pass certain quality cuts to ensure good reconstructed shower sizes. Particularly, only air showers with zenith angles less than $18^{\circ}$ are used exhibiting at least $10^6$ charged particles and $10^5$ muons. The measurement time covers approximately $1\,318$~days resulting in $78\, 000$ accepted events passing all quality cuts. This corresponds to an exposure of $164\, 709$~m$^2$~sr~yr.\\
The convolution of the sought-after differential fluxes $\mathrm d J_n/\mathrm d\, \mathrm{lg}E$ of the primary cosmic ray nuclei $n$ into the measured number of showers $N_i$ contributing to the cell $i$ of shower size plane, and thus to the content of this specific charged particle and muon number bin $\left( \mathrm{lg}(N_{ch}), \mathrm{lg}(N_{\mu}) \right)_{i}$, can be described by an integral equation:
\begin{equation}\label{eq:convolution_integral}
N_i=\sum\limits_{n=1}^{N_\mathrm n} \; \int\limits_{T_{\mathrm m}} \int\limits_{\mathit{\Omega}_{\mathrm{tot}}} \int\limits_{A_{\mathrm f}} \int\limits_{E}  \frac{\mathrm d J_n}{\mathrm d\, \mathrm{lg}E}\; p_n\; \mathrm d\,\mathrm{lg}E\ \cos \theta\ \mathrm dA\ \mathrm d\mathit{\Omega}\ \mathrm dt , \nonumber \\
\end{equation}
with
\begin{eqnarray*}
p_n=p_n\left(   \left(  \mathrm{lg}N_{ch}, \mathrm{lg}N_{\mu}   \right)_i \ | \    \mathrm{lg}E    \right)  \;  .
\end{eqnarray*}
One has to sum over all $N_\mathrm n$ elements contributing to the all-particle cosmic ray spectrum, in this analysis the five representative primaries.  $T_{\mathrm m}$ is the measurement time,  $\mathit{\Omega}_{\mathrm{tot}}$ the total solid angle accessible for the experiment and used for the analysis, and $A_{\mathrm f}$ the chosen fiducial area.  The term $p_n$ represents the conditional probability to reconstruct a certain combination of charged particle and muon number, respectively to get an entry in the cell $\left( \mathrm{lg}(N_{ch}), \mathrm{lg}(N_{\mu}) \right)_{i}$, if the air shower inducing particle was of the type $n$ and had an energy of $E$. More precisely, $p_n$ itself is a convolution combining the intrinsic shower fluctuations occurring whilst the air shower development, the detection and reconstruction efficiency as well as the properties of the observables' reconstruction process. The cosine term in $\cos \theta\ \mathrm dA$ accomplishes the transformation from the horizontal surface element to the effective detection area.\\
Equation (\ref{eq:convolution_integral}) can mathematically be understood as a system of coupled integral equations referred to as Fredholm integral equation of first kind. There are various methods to solve such an integral equation, albeit a resolvability often doesn't \textit{per se} imply uniqueness. In some preliminary tests it was found, that the unfolding algorithm of Gold \cite{lit:gold_alg} yields appropriate and robust solutions. It is a iterative procedure and \textit{de facto} related to a minimization of a chi-square function. For countercheck purposes all results are validated by means of two additional algorithms, an also iterative method applying Bayes' theorem \cite{lit:bayes_alg} performing very stable, too, and an regularized unfolding based on a combination of the least-squares method with the principle of reduced cross-entropy \cite{lit:entropie_alg}, that yields slightly poorer results. All these solution strategies have in common that the response\footnote{Also named kernel or transfer function; and more precisely it is rather a matrix than a simple function.} function $p_n$ of Eq.(\ref{eq:convolution_integral}) has to be known \textit{a priori}. It is parametrized based on Monte Carlo simulations. The air shower development is simulated by means of CORSIKA~\cite{lit:corsika} 6.307 based on the interaction models \mbox{QGSJET-II-02}~\cite{lit:qgsjet-ii} and \mbox{FLUKA} 2002.4~\cite{lit:fluka}. The experiment's response is simulated using CRES\footnote{\underline{C}osmic \underline{R}ay \underline{E}vent \underline{S}imulation, a program package developed for the KASCADE~\cite{lit:kascade_allgemein_nimpaper} detector simulation.}~1.16/07, which bases on GEANT~3.21~\cite{lit:GEANT3_21} detector description and simulation tool. 
\section{Error analysis}
The determination of the elemental energy spectra will be subjected to influences of different error sources. They can roughly be classified in two categories: uncertainties induced, or at least appearing whilst the deconvolution process and those embedded in the computed response function caused by the limited Monte Carlo statistics. 
\subsection{Uncertainties whilst the deconvolution}
Firstly, the used data set is only a small sample based on a limited exposure, and hence suffering from statistical uncertainties. They are propagated through the unfolding algorithm and affect the quality of the solution. Furthermore, the used deconvolution method itself can introduce a systematic bias. The influences of both sources can be evaluated by means of a frequentist approach. Assuming appropriate\footnote{Spectral indices close to those estimated by KASCADE are used in order to have realistic ones. However, some unlikely spectra are tested, too.} spectral indices some trial elemental energy spectra are specified based on which a test data sample can be generated using Eq.(\ref{eq:convolution_integral}). Subsequently, these data samples are unfolded. Since the true solution is \textit{a priori} known, the deconvolution result can be compared to it to reveal statistical fluctuations induced by the limited measurement time and a possible systematic bias induced by the unfolding method.
\subsection{Influences of limited Monte Carlo statistics} 
\begin{figure}[!t]
  \vspace{5mm}
  \centering
  \includegraphics[width=0.8\columnwidth]{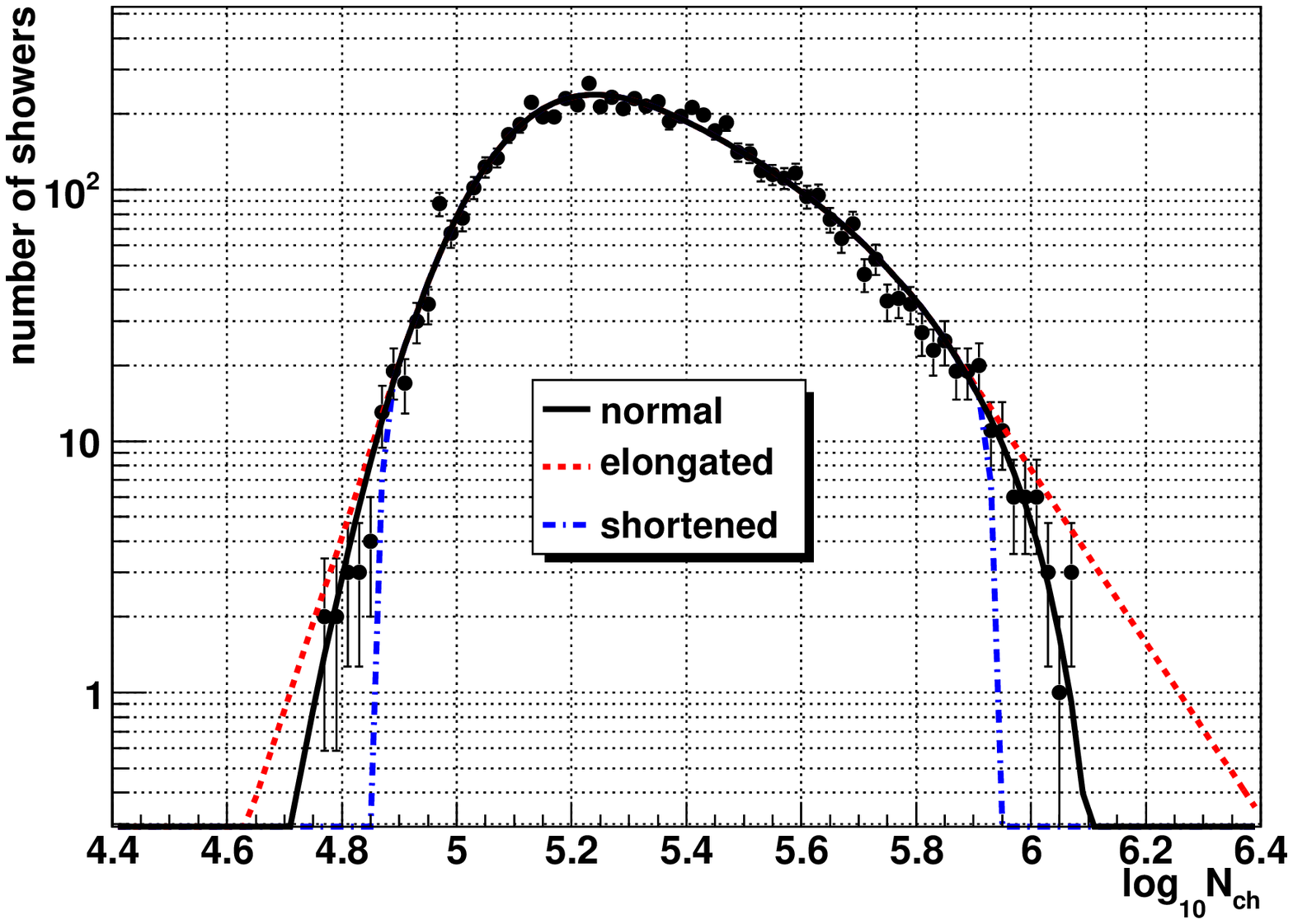}
  \caption{The simulated charged particle number distribution of $2\times10^{15}$~eV hydrogen induced air showers with angles of incidence less than $18^{\circ}$. The distribution is fitted based on different approaches (see text).}
  \label{fig:error_est}
 \end{figure}

The amount of simulated air showers is strongly limited by reason of computing time. Due to the limited Monte Carlo statistics, the computation of the response function, i.e. the parametrization of the intrinsic shower fluctuations as well as of the detector properties, will only be possible under certain uncertainties resulting in an systematic error of the finally unfolded solution. In Fig.~\ref{fig:error_est} exemplarily the simulated charged particle number distribution in case of hydrogen induced air showers with primary energy of $2\times10^{15}$~eV is shown. A scattering around the used parametrization (``normal'') can be observed. This statistical uncertainty will be treated conservatively: Considering the computed fit parameters and their errors some new sets of parameters are calculated by means of a random generator. Based on each set, new response functions can be computed and used to unfold the data. Comparing the results reveals the caused systematic uncertainty in the solution.\\
The distributions' tails have to be inspected in more detail. Because of the very low statistics, the tails can vary within a certain range without worsen the fit result. In particular the right tail describing the fluctuations in direction to higher energies can have an important impact on the unfolded solution due to the steeply falling flux of cosmic rays. The systematic influence of the tails will conservatively be estimated by computing two additional response functions assuming in contrast to the standard case, in agreement with the statistical uncertainties, either a very fast decreasing or an elongated tail (cf. Fig.~\ref{fig:error_est}). Using both for a deconvolution and comparing the results yields the maximal systematic error range caused by the uncertainty in the tails description. \\
\section{Results and conclusion}
\begin{figure*}[!t]
   \centerline{\includegraphics[clip, width=0.9\columnwidth, bb=0 0 567 356]{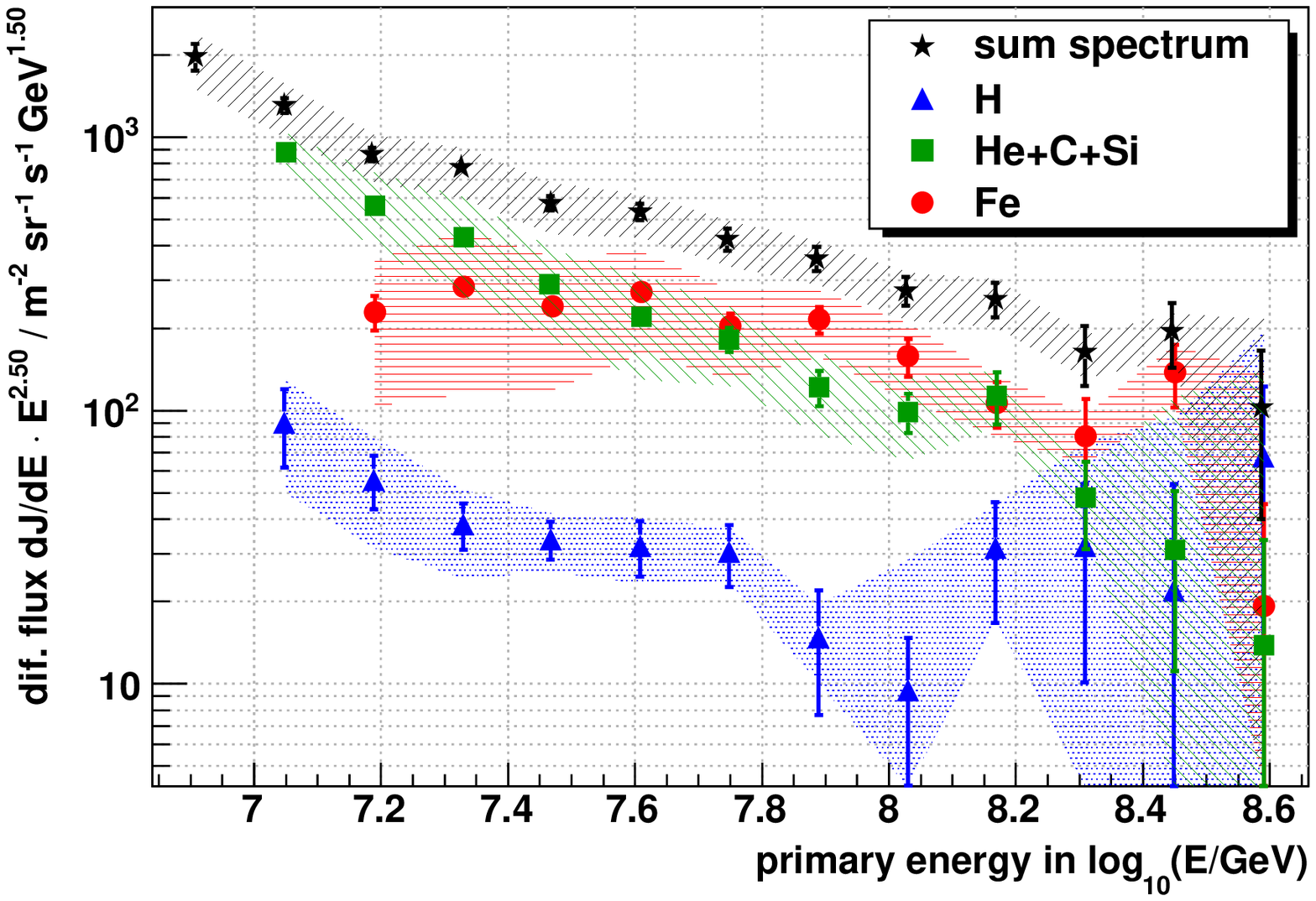}\label{fig2D}
              \hfil
              \includegraphics[clip, width=0.9\columnwidth,height=0.57\columnwidth, bb=0 0 567 336]{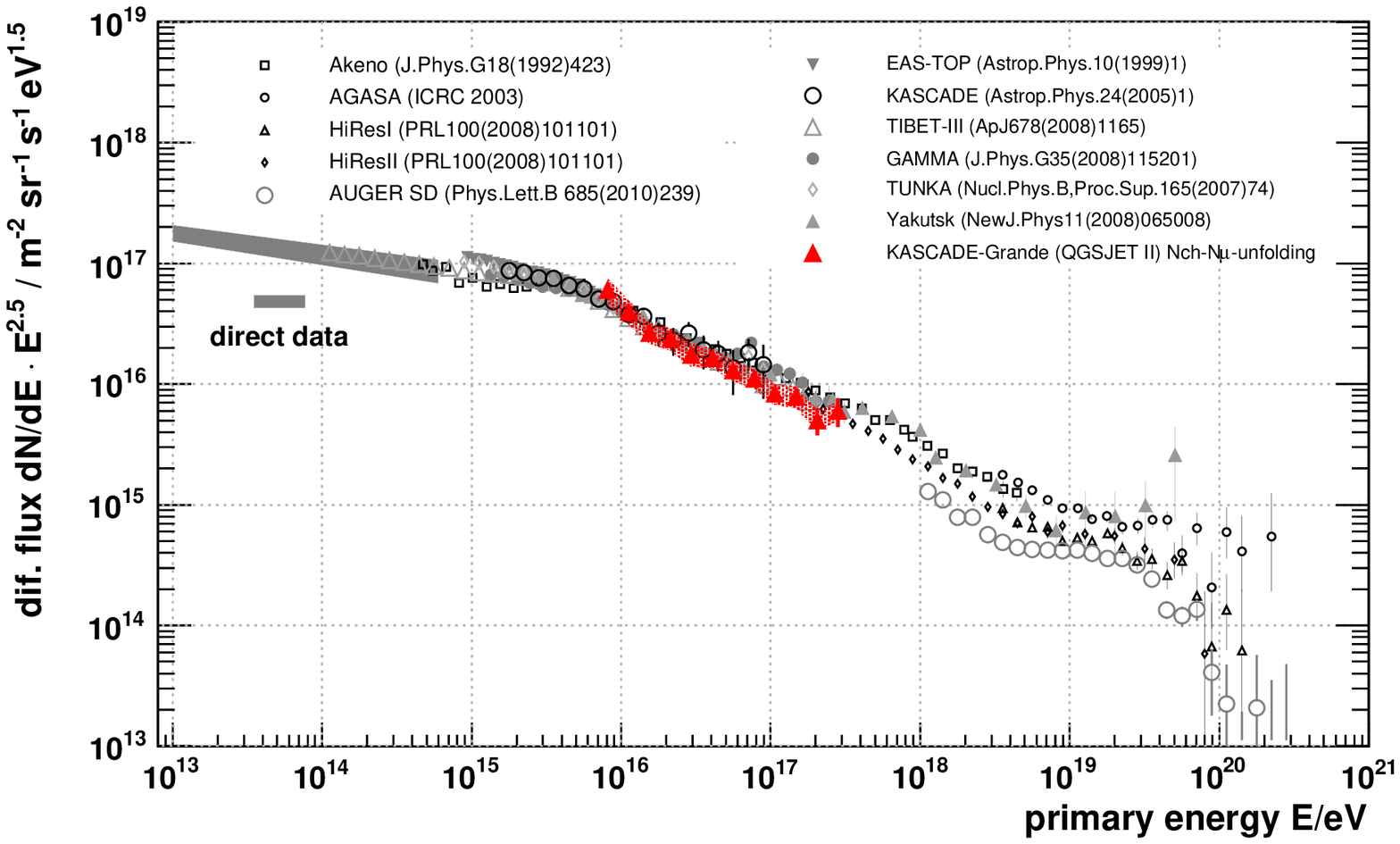}
 \label{fig3D}
             }
   \caption{Depicted are the unfolded energy spectra for H and Fe, a combined spectrum for He, C and Si, as well as the all-particle spectrum (left panel). The all-particle spectrum conforms well to those of other experiments (right panel).}
   \label{fig:unfolded_spectra}
 \end{figure*}

\begin{figure}[!b]
 \vspace{2mm}
  \centering
  \includegraphics[clip, width=0.9\columnwidth, bb=0 0 567 368]{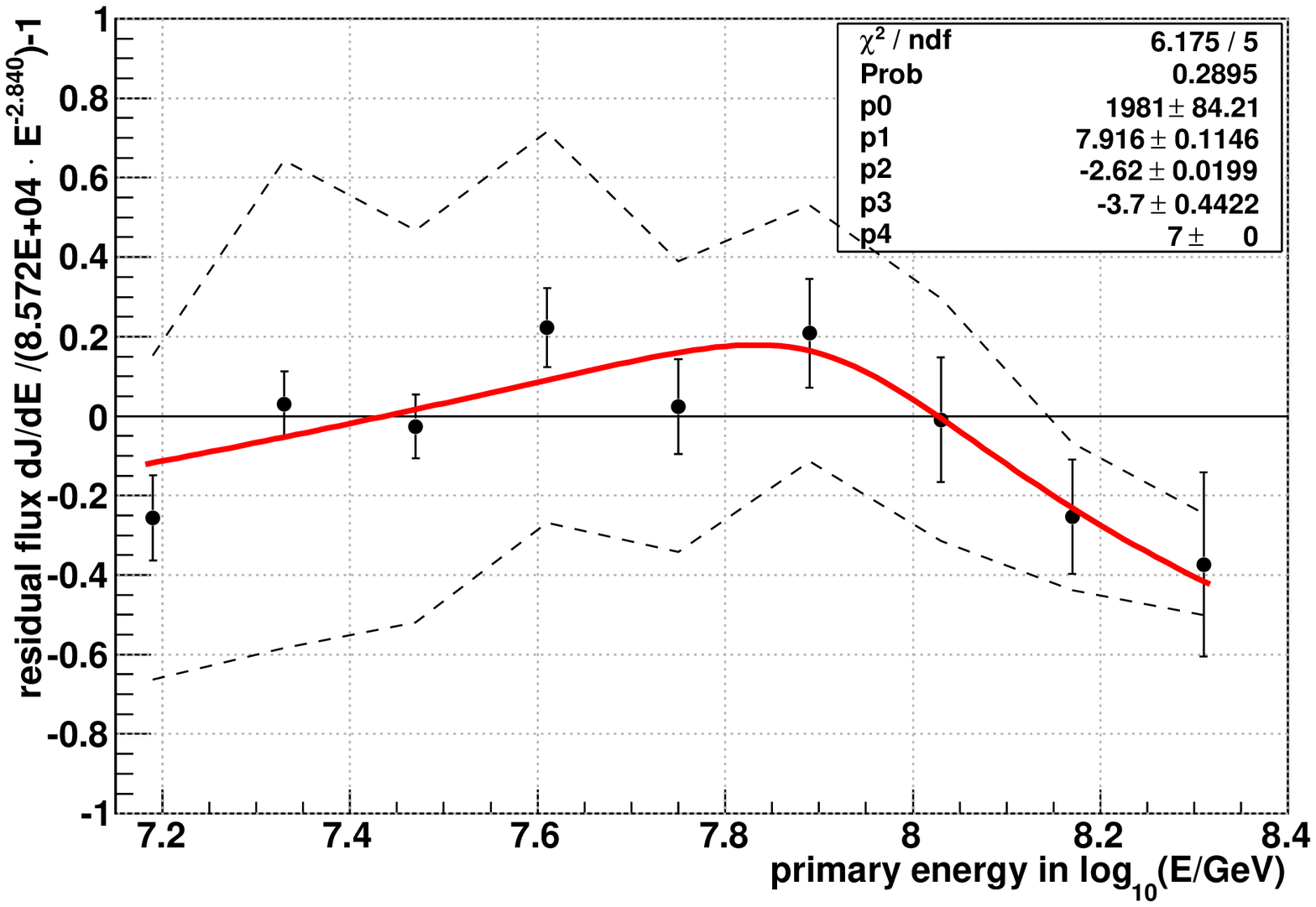}
  \caption{Residual flux between the iron spectrum shown in Fig.~\ref{fig:unfolded_spectra} and a spectrum that was derived by a single power law fit to that iron spectrum. Additionally, the iron spectrum is now fitted by a double power law.}
  \label{fig:residu}
 \end{figure}

In Fig.~\ref{fig:unfolded_spectra} left panel the unfolded differential energy spectra of hydrogen and iron as well as the sum\footnote{The intermediate component was combined because of the poorer results suffering from low statistics and making the plot confusing without giving further insights.} of the three single spectra of helium, carbon, and silicon are shown, representing respectively the fluxes of the light, heavy, and intermediate mass groups of cosmic rays. In addition, all five unfolded spectra are summed up to the all-particle flux, which is compatible to the results of other experiments (right panel) and agrees very well with the KASCADE-Grande all-particle spectrum published in \cite{lit:kascade-grande-all_particle}. The shaded band indicates the methodical uncertainties while the error bars represent the statistical error originating from the limited measurement time. 
With increasing energy the heavy component gets the dominant contributor to the cosmic ray composition. This agrees with the results of KASCADE \cite{lit:kascade-unfolding} where a reduction of the light component beyond the first knee was found.\\
Both in the all-particle and the iron spectrum there is by eye a slight bending discernible at around $10^{17}$~eV. However, the change in the all-particle spectrum reveals not to be significant. In this context, one should keep in mind that this spectrum is the sum of all five elemental spectra unfolded separately, and by this is affected by their uncertainties. For the determination of the all-particle spectrum with KASCADE-Grande, there are more precise methods available, e.g. that one introduced in \cite{lit:kascade-grande-all_particle} stating a high significance for a change in the spectral index of the all-particle spectrum at around $1\times 10^{17}$~eV.\\
In order to judge the possible structures in the unfolded iron spectrum, it was fitted preliminarily by a single power law. However, the resulting chi-square probability for such a featureless single power law was below 1\% ($\chi^2/\mathrm{\textit{ndf}}=18.9/7$). In Fig.~\ref{fig:residu}, the residual flux between the iron spectrum shown in Fig.~\ref{fig:unfolded_spectra}, left panel, and such a spectrum that was derived by a single power law fit is depicted in order to emphasize the deviations between the single power law and the unfolded spectrum.
Additionally, the iron spectrum is now fitted by a double power law:
\begin{equation}\label{eq:knee}
\frac{\mathrm d J(E)}{\mathrm d\, \mathrm{lg}E}=
p_0 \times E^{p_2} \times \left( 
1+
\left( 
\frac{E}{p_1}
\right) ^{p_4}
\right)^{\left(  p_3-p_2  \right)/p_4}\;  ,
\end{equation}
where $p_1=\mathrm{lg}(E_{\mathrm{knee}}/\mathrm{GeV})=7.9\pm 0.1$ corresponds to the knee position, while $p_2=-2.62\pm0.02$ and $p_3=-3.7\pm0.4$ are the spectral indices below and above the knee. The sharpness of the knee structure is encoded in $p_4=7.0$ and was fixed without worsening the fit's quality, while $p_0$ is a free normalization parameter. This fit describes the spectrum significantly better (chi-square probability at around 30\% with $\chi^2/\mathrm{\textit{ndf}}=6.2/5$), giving strong indications for a kink in the iron flux at around 80~PeV. \\
Comparing\footnote{And assuming that the mass groups represented by H and Fe actually consist only, or at least primarily, of that two primaries.} the position of this potential iron knee to that for hydrogen (at around 2~PeV to 4~PeV) gives indications for a scaling of the knee positions with the charge of the nuclei rather than with their atomic mass number. This would encourage the cosmic ray acceleration models based on magnetic fields. \\
To summarize, there is a strong indication for a kink in the iron-like spectrum at around 80~PeV as well as for a dependence of the cosmic ray acceleration process on the charge of the nuclei, both on the premise that especially the used model QGSJET-II-02 describes the physics of hadronic interactions at these energies with a high level of reliability. 
%\vspace{\baselineskip}

\clearpage

%% file: icrc0405.tex
%%
% 32nd International Cosmic Ray Conference 2011 Beijing China

%Class Required
%%% for classical LaTeX

%The paper title
\title{Primary energy reconstruction from the S(500) observable recorded with the KASCADE-Grande array}
%The short title will appear at the header of the even pages.

\shorttitle{G. Toma \etal Primary energy spectrum from S(500) - KASCADE-Grande}

%All paper authors
\authors{
G.~Toma$^{1}$,
W.D.~Apel$^{2}$,
J.C.~Arteaga-Vel\'azquez$^{3}$,
K.~Bekk$^{2}$,
M.~Bertaina$^{4}$,
J.~Bl\"umer$^{2,5}$,
H.~Bozdog$^{2}$,
I.M.~Brancus$^{1}$,
P.~Buchholz$^{6}$,
E.~Cantoni$^{4,7}$,
A.~Chiavassa$^{4}$,
F.~Cossavella$^{5,13}$,
K.~Daumiller$^{2}$,
V.~de Souza$^{8}$,
F.~Di~Pierro$^{4}$,
P.~Doll$^{2}$,
R.~Engel$^{2}$,
J.~Engler$^{2}$,
M. Finger$^{5}$, 
D.~Fuhrmann$^{9}$,
P.L.~Ghia$^{7}$, 
H.J.~Gils$^{2}$,
R.~Glasstetter$^{9}$,
C.~Grupen$^{6}$,
A.~Haungs$^{2}$,
D.~Heck$^{2}$,
J.R.~H\"orandel$^{10}$,
D.~Huber$^{5}$,
T.~Huege$^{2}$,
P.G.~Isar$^{2,14}$,
K.-H.~Kampert$^{9}$,
D.~Kang$^{5}$, 
H.O.~Klages$^{2}$,
K.~Link$^{5}$, 
P.~{\L}uczak$^{11}$,
M.~Ludwig$^{5}$,
H.J.~Mathes$^{2}$,
H.J.~Mayer$^{2}$,
M.~Melissas$^{5}$,
J.~Milke$^{2}$,
B.~Mitrica$^{1}$,
C.~Morello$^{7}$,
G.~Navarra$^{4,15}$,
J.~Oehlschl\"ager$^{2}$,
S.~Ostapchenko$^{2,16}$,
S.~Over$^{6}$,
N.~Palmieri$^{5}$,
M.~Petcu$^{1}$,
T.~Pierog$^{2}$,
H.~Rebel$^{2}$,
M.~Roth$^{2}$,
H.~Schieler$^{2}$,
F.G.~Schr\"oder$^{2}$,
O.~Sima$^{12}$,
G.C.~Trinchero$^{7}$,
H.~Ulrich$^{2}$,
A.~Weindl$^{2}$,
J.~Wochele$^{2}$,
M.~Wommer$^{2}$,
J.~Zabierowski$^{11}$
}
%All the affiliations.
\afiliations{
$^1$ National Institute of Physics and Nuclear Engineering, Bucharest, Romania\\
$^2$ Institut f\"ur Kernphysik, KIT - Karlsruher Institut f\"ur Technologie, Germany\\
$^3$ Universidad Michoacana, Instituto de F\'{\i}sica y Matem\'aticas, Morelia, Mexico\\
$^4$ Dipartimento di Fisica Generale dell' Universit\`a Torino, Italy\\
$^5$ Institut f\"ur Experimentelle Kernphysik, KIT - Karlsruher Institut f\"ur Technologie, Germany\\
$^6$ Fachbereich Physik, Universit\"at Siegen, Germany\\
$^7$ Istituto di Fisica dello Spazio Interplanetario, INAF Torino, Italy\\
$^8$ Universidade S$\tilde{a}$o Paulo, Instituto de F\'{\i}sica de S\~ao Carlos, Brasil\\
$^9$ Fachbereich Physik, Universit\"at Wuppertal, Germany\\
$^{10}$ Dept. of Astrophysics, Radboud University Nijmegen, The Netherlands\\
$^{11}$ Soltan Institute for Nuclear Studies, Lodz, Poland\\
$^{12}$ Department of Physics, University of Bucharest, Bucharest, Romania\\
\scriptsize{
$^{13}$ now at: Max-Planck-Institut Physik, M\"unchen, Germany; 
$^{14}$ now at: Institute Space Sciences, Bucharest, Romania; 
$^{15}$ deceased; 
$^{16}$ now at: Univ Trondheim, Norway
}
}
%email address of the contact person
\email{gabriel.toma@nipne.ro}

%The abstract.
\abstract{We present a method to reconstruct the primary energy spectrum of cosmic rays 
from the charged particle densities recorded with the KASCADE-Grande detector. 
The KASCADE-Grande is hosted by the Karlsruhe Institute for Technology (KIT), 
Germany and is operated by an international collaboration. It has been shown 
that the charged particle density becomes independent of the primary mass at 
certain fixed distances from the shower axis and that it can be used as an 
estimator of the primary energy. Such distance is a characteristic of the detector 
and in the case of the KASCADE-Grande experiment it was shown that it is 500 
m, hence the notation S(500). A relation is established by means of simulations 
between the primary energy of cosmic rays and the S(500). We account for the 
attenuation of inclined showers by applying the CIC method. By using the 
simulation derived calibration we build the primary energy spectrum from the 
recorded S(500). Several sources of systematic uncertainties are identified and 
their contribution to the final result is evaluated. Additionally we apply an 
unfolding to account for statistical fluctuations. The features of the obtained 
spectrum are discussed in relation to the result of another independent 
technique that is applied at KASCADE-Grande.}
%The keywords

% B E G I N   D O C U M E N T
\maketitle

\section{Introduction}
\begin{figure}[!t]
  \vspace{5mm}
  \centering
  \includegraphics[width=3.in]{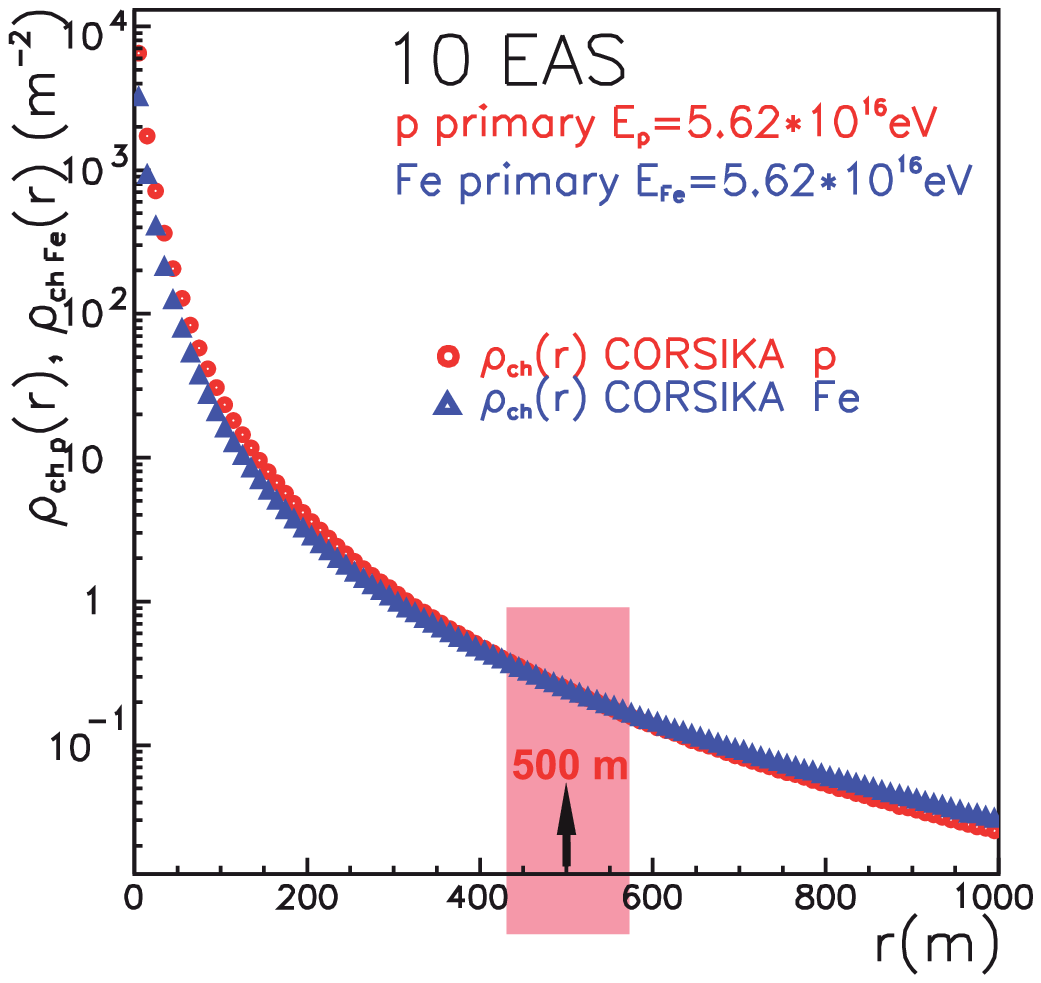}
  \caption{Simulations show that,  for the case of the KASCADE-Grande experimental layout, the particle density becomes independent of the primary mass at around 500 m distance from the shower core; this plot shows averaged simulated lateral distributions for different primary types with equal energy.}
  \label{fig1E}
 \end{figure}

Previous investigations have shown that the charged particle density in air showers becomes independent of the primary mass at large but fixed distances from the shower axis and that it can be used as an estimator for the primary energy \cite{Hillas}. Such a distance is characteristic for a given experiment. Based on this property a method was derived to reconstruct the primary energy spectrum from the particular value of the charged particle density, observed at such specific radial distances. While in the AGASA experiment the technique was applied for a distance of 600~m to the shower axis \cite{sx}, in the case of the KASCADE-Grande array, detailed simulations \cite{S500} have shown that the particular distance for which this effect takes place is about 500~m (see Fig.~\ref{fig1E} and Fig.~\ref{fig4E}). Hence the notation S(500) for the charged particle density at 500~m distance from the shower core. The distance is measured in a plane normal to the shower axis and containing the shower core. The data recorded in the detector plane is projected on the normal plane taking into account the attenuation effects characteristic to inclined events.\\
The study has been performed for both simulated (Fig.~\ref{fig5E}) and experimental (Fig.~\ref{fig6E}) events, using identical reconstruction procedures \cite{Sima}. The CORSIKA Monte Carlo EAS simulation tool \cite{corsika} is used to simulate air showers, with the QGSJETII model embedded for high energy interactions \cite{models}.\\
The energy deposits of particles in the KASCADE-Grande detector stations are recorded along with the associated temporal information (arrival times of particles). Using appropriate \textbf{L}ateral \textbf{E}nergy \textbf{C}orrection \textbf{F}unctions (LECF), the energy deposits are converted into particle densities. The LECF functions are dependent on the shower zenith angle \cite{Toma} and on the position of the station around the shower core (i.e. the LECF are dependent on the angle of incidence of particles in detectors). For every recorded event, a Linsley \cite{Linsley} \textbf{L}ateral \textbf{D}ensity \textbf{F}unction (LDF) is used to evaluate the particle density at the radial range of interest, 500~m.\\
The described reconstruction is performed independently from the standard reconstruction applied at KASCADE-Grande, based on the N$_{ch}$-N$_{\mu}$ approach and described in \cite{Mario}.
 
 \begin{figure}[!t]
  \vspace{5mm}
  \centering
  \includegraphics[width=3.in]{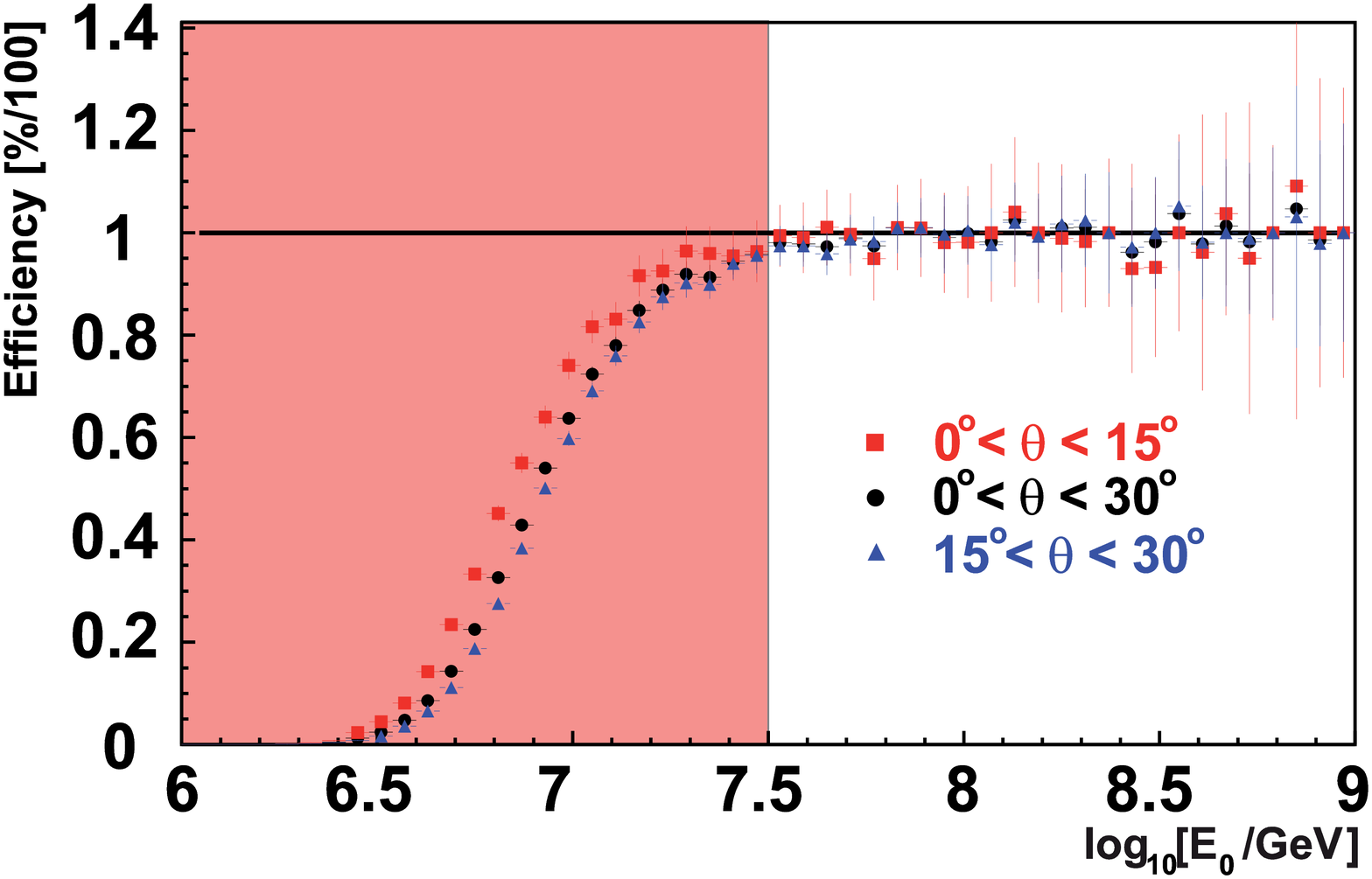}
  \caption{S(500) reconstruction efficiency for different zenith angular ranges and for the entire shower sample (all quality cuts applied); the reconstruction efficiency exceeds 95\% at log$_{10}$[$E_{0}$/GeV]\textgreater7.5}
  \label{fig2E}
 \end{figure}

\section{KASCADE-Grande}
In its general structure and operation, the KASCADE-Grande array \cite{kg} resembles other cosmic ray experiments.  Hosted by the Karlsruhe Institute of Technology (KIT), Campus North, Germany at 110~m a.s.l and operated by an international collaboration, it is composed of many detector stations that are distributed over a wide area and it has been designed to record charged particle densities, without disentangling the type of secondary that is interacting in the sensitive medium. The shape of the array is rectangular with a length of $\approx$700~m. The lateral particle density distribution is subsequently inferred by adjusting the data registered with the detector stations to an a-priori assumed lateral distribution function.\\
Historically, the KASCADE-Grande detector array is an extension of a smaller array (the KASCADE array, operated since 1996). KASCADE was designed to record air showers initiated by primaries with energies in the 10$^{14}$-10$^{17}$~eV range (including the knee range). The extension of the original KASCADE array was guided by the intention to extend the energy range for efficient EAS detection to 10$^{16}$-10$^{18}$~eV (Fig.~\ref{fig2E}). This extended energy range provides various interesting aspects: the expected transition from galactic to extragalactic cosmic rays and, in particular the question whether there exists a ''second knee'' in the energy spectrum.\\

\section{The constant intensity cut method (CIC)} 
\begin{figure}[!t]
  \vspace{5mm}
  \centering
  \includegraphics[width=3.in]{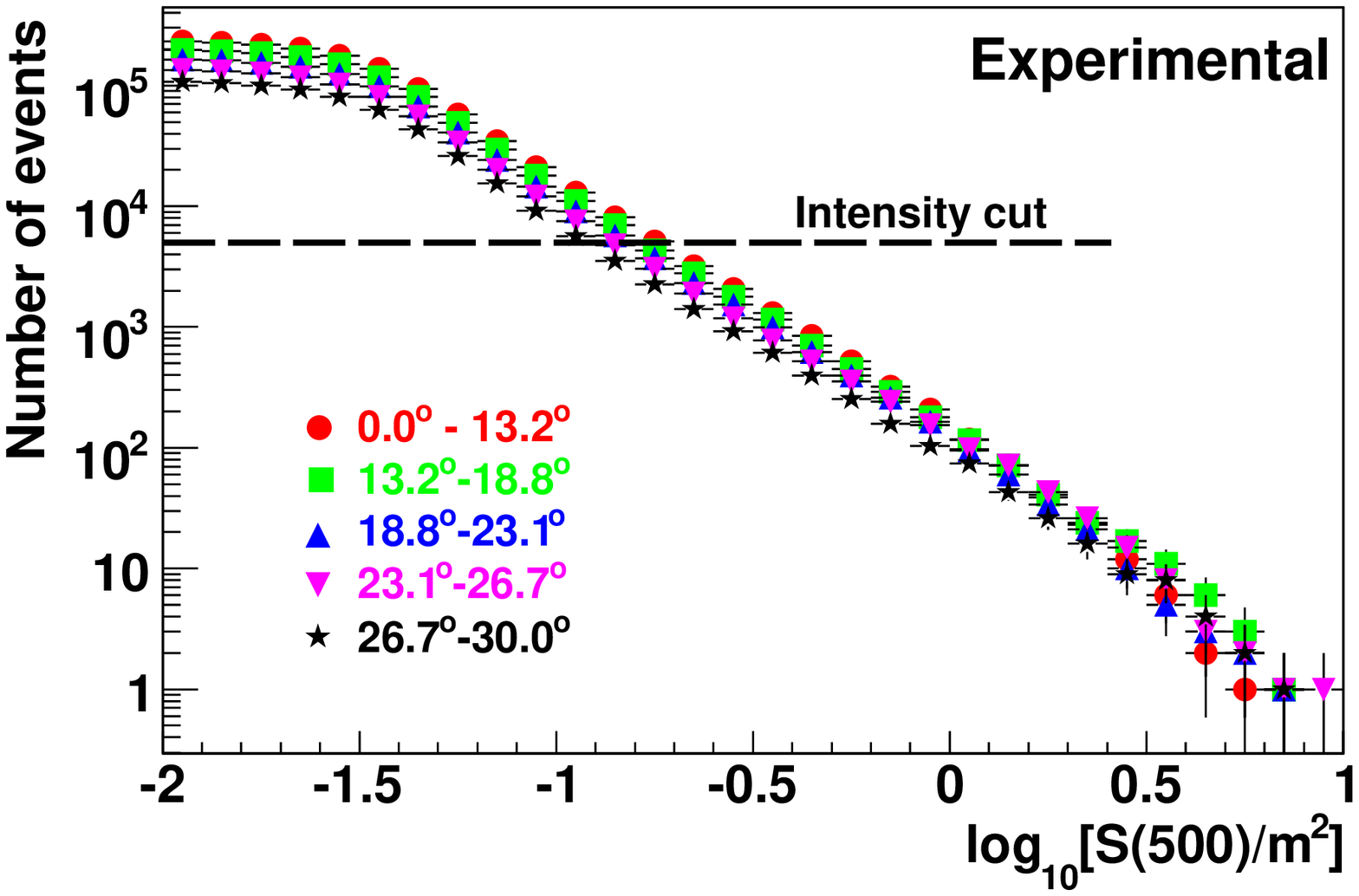}
  \caption{Integral S(500) spectra;  the horizontal line is a constant intensity cut at an arbitrarily chosen intensity; by assuming an exponential attenuation pattern the attenuation length of S(500) was evaluated to $\lambda$(500)=754~$\pm$~8~g$\cdot$cm$^{-2}$.}
  \label{fig3}
 \end{figure}

For a given event sample, an EAS observable could have different values for events induced by identical primaries ($E_{0}$,$A_{0}$), but arriving from different zenith angles (due to EAS attenuation through the atmosphere). This is also the case for the S(500). To reach the detectors at ground level, an inclined event will propagate along an extended path through the atmosphere compared to a vertical shower. The recorded particle densities at a given radial range for the two events will be different. One has to correct for this effect before performing an analysis simultaneously on all EAS events. This is achieved by applying the \textbf{C}onstant \textbf{I}ntensity \textbf{C}ut (CIC) method (Fig.~\ref{fig3}) \cite{cic}. The method is based on the assumption that for a given minimum primary energy threshold we should record the same flux of events (i.e. primaries) from all zenith angles. Therefore we divide our shower sample in sub-samples corresponding to different zenith angular bins that subtend equal solid angles. Then we look in the integral S(500) spectra of each sub-sample at a given arbitrary intensity. We establish a correlation between the S(500) corresponding to the given intensity in each spectrum and the corresponding zenith angle. This correlation we use as a correction function. All reconstructed S(500) values are corrected for attenuation by bringing them to the value they would have at a chosen reference angle. For the present study the reference angle is considered to be 21$^{\circ}$, since the zenith angular distribution for the recorded EAS sample peaks at this value. The CIC correction is derived entirely from recorded experimental data and is independent from simulated studies.

\section{Conversion to energy}
\begin{figure}[!t]
  \vspace{5mm}
  \centering
  \includegraphics[width=3.in]{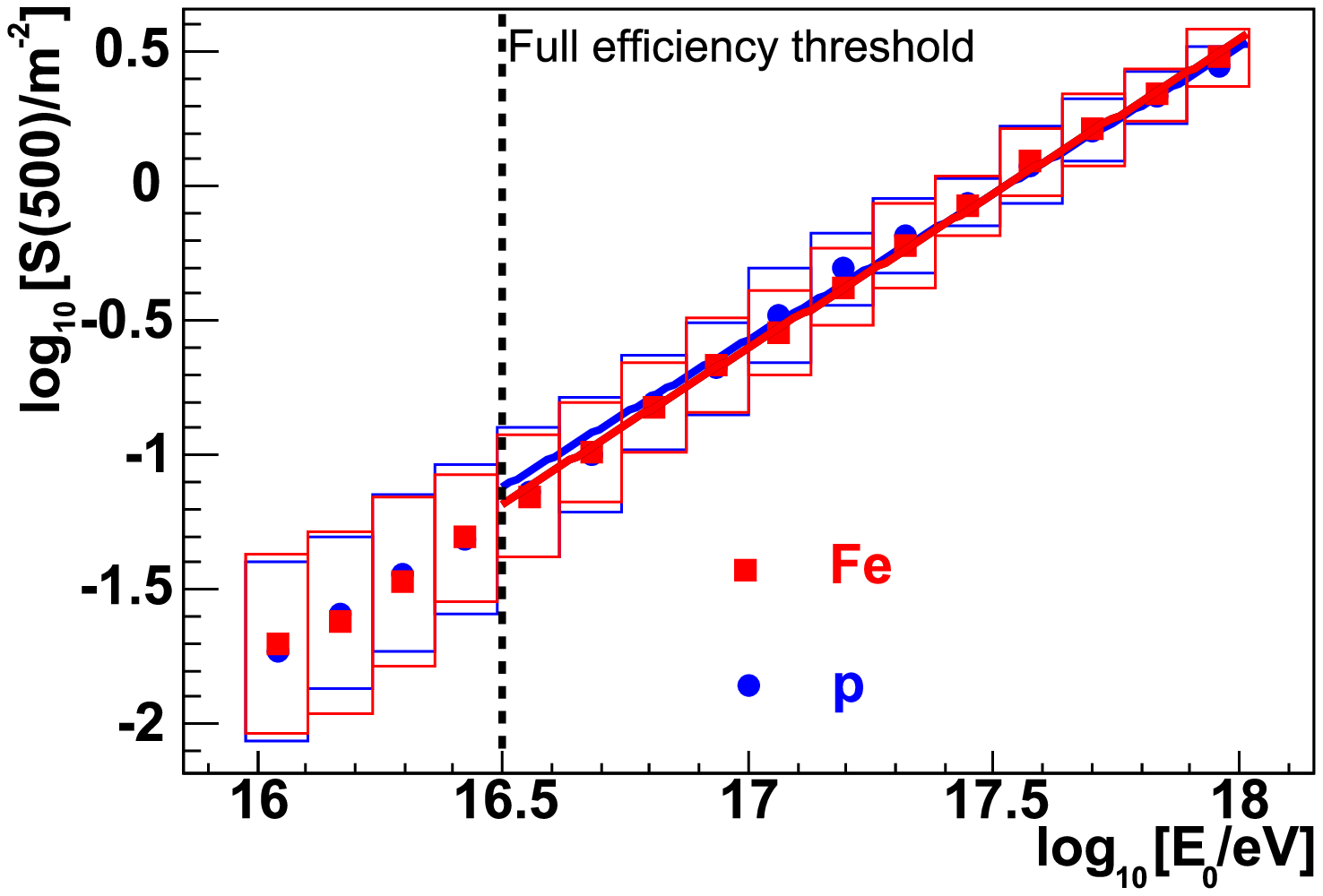}
  \caption{The dependence of S(500) on the primary energy E$_{0}$ for two different primaries (showers in fairly equal proportions for the two masses); the box-errors are the errors on the spread; the errors on the mean are represented with bars and are dot-sized; straight lines represent power law fits.}
  \label{fig4E}
 \end{figure}

For the experimental EAS sample, the total time of acquisition was 1173 days for a 500~$\times$~600~m$^{2}$ fiducial area. The same quality cuts were used for both simulated and experimental events. Only those events are accepted for which the zenith angle is below 30$^{\circ}$, the reconstructed shower core is positioned inside the detector array and not too close to the border, and the event is triggered by more than 24 Grande stations (to ensure that density data is recorded close to the 500~m radial range, i.e. the shower is large enough). A good quality of the fit to the Linsley distribution is a further important criterion.\\
A calibration of the primary energy E$_{0}$ with S(500) was derived from simulations (see Fig.~\ref{fig4E}). For the systematic contribution to the total error, several sources of systematic uncertainties have been identified: the spectral index of the simulated shower sample (which is different from the true one) is acting as a source of systematic uncertainty ($<$1\% contribution), the S(500)-E$_{0}$ calibration ($<$1\% contribution), the CIC method ($<$1\% contribution), the statistical fluctuations in the simulated shower sample (7\%) and the choise of a certain reference angle at which to perform the S(500) attenuation correction (7\% contribution).\\
The energy resolution has also been evaluated from simulations by calculating the difference between the true and the reconstructed primary energy (applying CIC to the simulated data) and was found to be 22\% for E$_{0}$=10$^{17}$ eV (for all primaries) with a slight decrease with increasing energy.

 \begin{figure}[!t]
  \vspace{5mm}
  \centering
  \includegraphics[width=3.in]{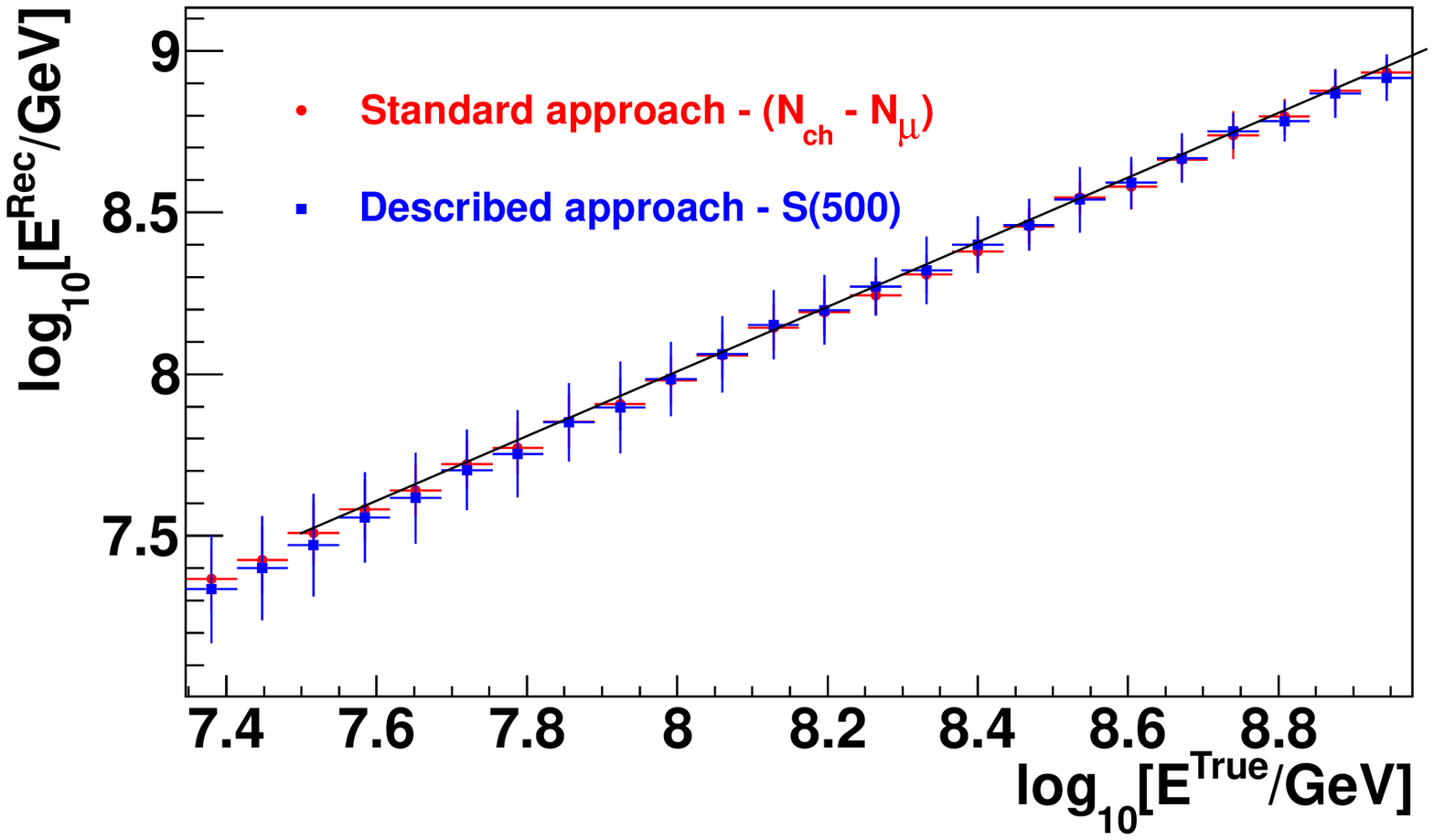}
  \caption{Comparison between the true (known) energy of primaries (five masses in fairly equal proportions) used in the simulated shower sample and the result of the reconstructions: the described reconstruction based on S(500) and the result of the standard KASCADE-Grande approach based on the N$_{ch}$~-~N$_{\mu}$ approach; the plot is a profile histogram, the error bars show the spread of data and the continuous line is the identity.}
  \label{fig5E}
 \end{figure}

 \begin{figure}[!t]
  \vspace{5mm}
  \centering
  \includegraphics[width=3.in]{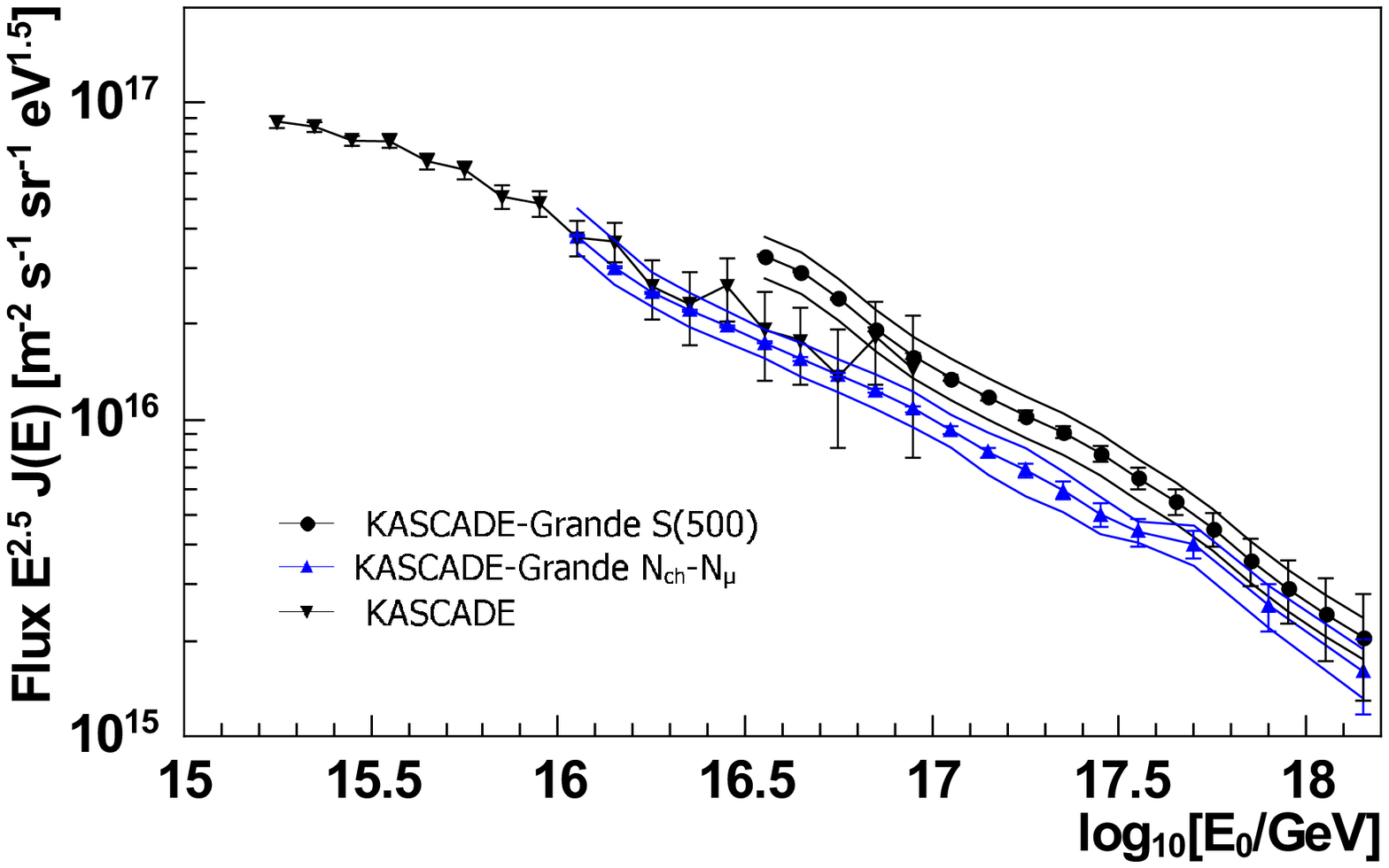}
  \caption{Reconstructed experimental energy spectrum by KASCADE-Grande from S(500)/CIC, multiplied by E$^{2.5}$ together with the result of the standard reconstruction procedure (based on the  N$_{ch}$ -N$_{\mu}$ correlation) and the results of KASCADE towards lower energies; the continuous lines above and below the spectrum show the systematic uncertainties.}
  \label{fig6E}
 \end{figure}

\section{The correction based on a response matrix}
In the reconstruction of an observable, the final result may be under- or over-estimating the true value. When representing the energy flux as a histogram a particular reconstructed event may be stored in the wrong (neighboring) energy bin. Thus in every energy bin of our spectrum we will have the data correctly belonging to that bin, but also data that was migrating from neighboring bins. As the energy spectrum is very steep (spectral index $\gamma\approx$-3) we expect that for a given energy bin, the mis-reconstructed events falling into it will be coming predominantly from lower energy bins. This induces a shift of the reconstructed spectrum and a change of the spectral index (due to the variable energy resolution of the method with the energy).\\
It is possible to account for the effect of fluctuations by calculating (from simulations) how many events migrate and where they add a contribution. Therefore a correction procedure based on a response matrix is derived and applied to the experimental data. The spectrum presented in Fig.~\ref{fig6E} includes the result of this correction.

\section{Conclusions}

The primary energy spectrum has been reconstructed from the particular case of charged particle densities recorded in the stations of the KASCADE-Grande array at 500 m distance from the shower axis, S(500). The study was performed for experimentally recorded events and also for simulated ones. The analysis of simulated events was needed in order to evaluate the reconstruction efficiency and quality and to derive a calibration curve $E_{0}$~-~S(500). The CIC method was applied on each shower in order to correct for attenuation effects. Using the simulation-derived calibration the S(500) values are converted into primary energy. The S(500)-derived KASCADE-Grande spectrum is composition independent. An evaluation of the various uncertainty sources has been done and a correction based on a response matrix has been employed to account for the effects of the fluctuations on the spectral index of the reconstructed energy spectrum. The S(500)-derived all-particle primary energy spectrum (Fig.~\ref{fig6E}) shows a shift and a slightly different spectral index compared to an independently performed reconstruction approach based on the total number of charged particles and the muon number. In an event by event comparison, the profile plot in fig.~\ref{fig5E} shows that when reconstructing simulated events with the two approaches there is good agreement between the methods. Present and future investigations are directed towards understanding the origin of the difference between the experimental results of the two methods. A possible source for this effect could be the shape of the lateral density distribution for simulated events that appears to be different from the shape of the experimental ones, although there could be additional sources for this shift.

\vspace*{0.5cm} \footnotesize{
{\bf Acknowledgement:} The KASCADE-Grande experiment is supported by the BMBF of Germany, the MIUR and INAF of Italy, the Polish Ministry of Science and Higher Education (grant 2009-2011), and the Romanian Authority for Scientific Research, UEFISCSU, grant PNII-IDEI no. 461/2009 and grant PN 09370105. Part of the investigation was funded in the frame of the DAAD doctoral scholarship A/06/09016 Ref. 322 and by KIT – Campus North in the frame of research visits.}

%\vspace{\baselineskip}
\

\clearpage

%% file: icrc0740.tex
%%
% 32nd International Cosmic Ray Conference 2011 Beijing China

%Class Required
%%% for classical LaTeX

%The paper title
\title{Tests of hadronic interaction models with the
KASCADE-Grande muon data}
%The short title will appear at the header of the even pages.

\shorttitle{J.C. Arteaga \etal Tests of hadronic models in KASCADE-Grande}

%All paper authors
\authors{
J.C.~Arteaga-Vel\'azquez$^{1}$,
W.D.~Apel$^{2}$,
K.~Bekk$^{2}$,
M.~Bertaina$^{3}$,
J.~Bl\"umer$^{2,4}$,
H.~Bozdog$^{2}$,
I.M.~Brancus$^{5}$,
P.~Buchholz$^{6}$,
E.~Cantoni$^{3,7}$,
A.~Chiavassa$^{3}$,
F.~Cossavella$^{4,13}$,
K.~Daumiller$^{2}$,
V.~de Souza$^{8}$,
F.~Di~Pierro$^{3}$,
P.~Doll$^{2}$,
R.~Engel$^{2}$,
J.~Engler$^{2}$,
M. Finger$^{4}$, 
D.~Fuhrmann$^{9}$,
P.L.~Ghia$^{7}$, 
H.J.~Gils$^{2}$,
R.~Glasstetter$^{9}$,
C.~Grupen$^{6}$,
A.~Haungs$^{2}$,
D.~Heck$^{2}$,
J.R.~H\"orandel$^{10}$,
D.~Huber$^{4}$,
T.~Huege$^{2}$,
P.G.~Isar$^{2,14}$,
K.-H.~Kampert$^{9}$,
D.~Kang$^{4}$, 
H.O.~Klages$^{2}$,
K.~Link$^{4}$, 
P.~{\L}uczak$^{11}$,
M.~Ludwig$^{4}$,
H.J.~Mathes$^{2}$,
H.J.~Mayer$^{2}$,
M.~Melissas$^{4}$,
J.~Milke$^{2}$,
B.~Mitrica$^{5}$,
C.~Morello$^{7}$,
G.~Navarra$^{3,15}$,
J.~Oehlschl\"ager$^{2}$,
S.~Ostapchenko$^{2,16}$,
S.~Over$^{6}$,
N.~Palmieri$^{4}$,
M.~Petcu$^{5}$,
T.~Pierog$^{2}$,
H.~Rebel$^{2}$,
M.~Roth$^{2}$,
H.~Schieler$^{2}$,
F.G.~Schr\"oder$^{2}$,
O.~Sima$^{12}$,
G.~Toma$^{5}$,
G.C.~Trinchero$^{7}$,
H.~Ulrich$^{2}$,
A.~Weindl$^{2}$,
J.~Wochele$^{2}$,
M.~Wommer$^{2}$,
J.~Zabierowski$^{11}$
}
%All the affiliations.
\afiliations{
$^1$ Universidad Michoacana, Instituto de F\'{\i}sica y Matem\'aticas, Morelia, Mexico\\
$^2$ Institut f\"ur Kernphysik, KIT - Karlsruher Institut f\"ur Technologie, Germany\\
$^3$ Dipartimento di Fisica Generale dell' Universit\`a Torino, Italy\\
$^4$ Institut f\"ur Experimentelle Kernphysik, KIT - Karlsruher Institut f\"ur Technologie, Germany\\
$^5$ National Institute of Physics and Nuclear Engineering, Bucharest, Romania\\
$^6$ Fachbereich Physik, Universit\"at Siegen, Germany\\
$^7$ Istituto di Fisica dello Spazio Interplanetario, INAF Torino, Italy\\
$^8$ Universidade S$\tilde{a}$o Paulo, Instituto de F\'{\i}sica de S\~ao Carlos, Brasil\\
$^9$ Fachbereich Physik, Universit\"at Wuppertal, Germany\\
$^{10}$ Dept. of Astrophysics, Radboud University Nijmegen, The Netherlands\\
$^{11}$ Soltan Institute for Nuclear Studies, Lodz, Poland\\
$^{12}$ Department of Physics, University of Bucharest, Bucharest, Romania\\
\scriptsize{
$^{13}$ now at: Max-Planck-Institut Physik, M\"unchen, Germany; 
$^{14}$ now at: Institute Space Sciences, Bucharest, Romania; 
$^{15}$ deceased; 
$^{16}$ now at: Univ Trondheim, Norway
}
}
%email address of the contact person
\email{arteaga@ifm.umich.mx}

%The abstract.     
\abstract
{
  The KASCADE-Grande experiment is an air-shower ground-based observatory
  designed to study in detail the energy spectrum and composition of primary
  cosmic rays in the region of $10^{16}-10^{18} \, \mbox{eV}$. These analyses are based
  on precise measurements of the charged, electron and muon numbers of the
  cosmic ray air-showers performed through different detector systems which
  come into play simultaneously in KASCADE-Grande during the data acquisition.
  Due to the quality of the data and the number of air-shower observables at
  disposal through the experiment the collected data proves to be also useful
  to test hadronic interaction models used for air-shower simulations. In this
  contribution, predictions of the QGSJET II-2, SIBYLL 2.1 and EPOS 1.99
  hadronic models are confronted with the KASCADE-Grande muon data.
  Besides, the influence of these models on the all-particle energy spectrum 
  derived from the muon size is also investigated.
}
%The keywords

% B E G I N   D O C U M E N T
\maketitle

%Begin the section.

 \begin{figure*}[!t]
 \centering
 \includegraphics[height=2.0in, width=3.2in]{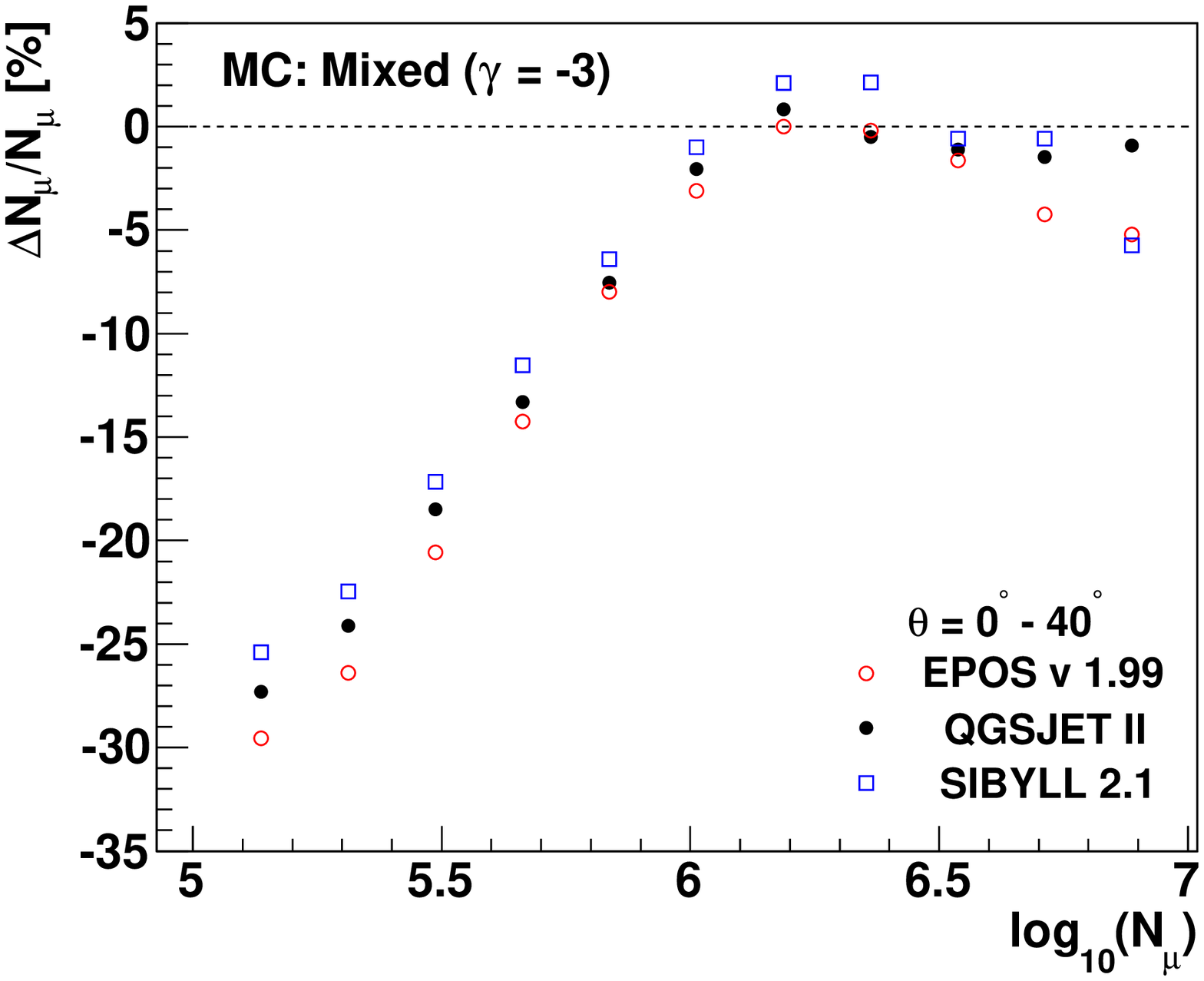}
 \includegraphics[height=2.0in, width=3.2in]{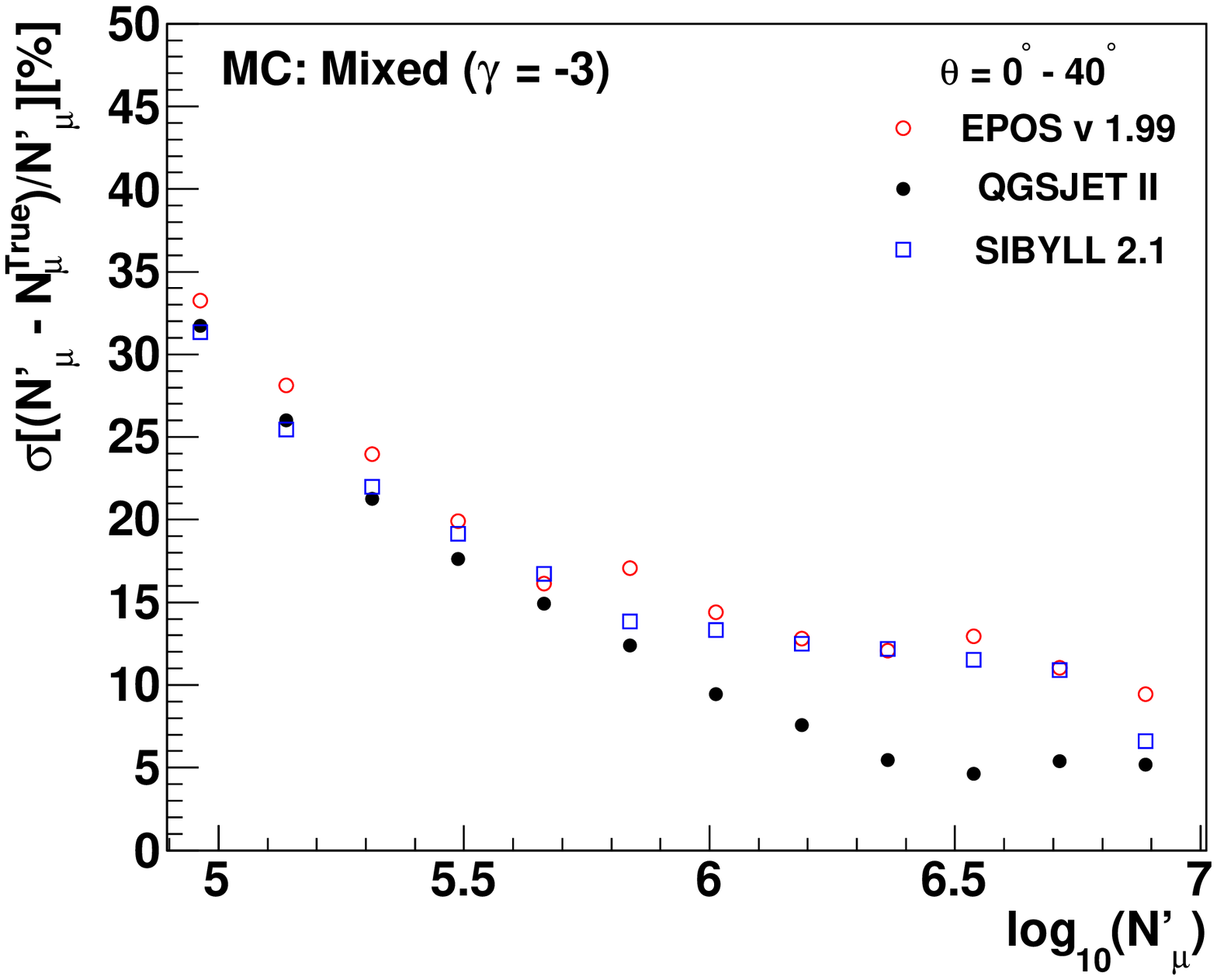}
  \caption{ Left: Mean values of the muon correction function shown against the 
  reconstructed muon number $N_\mu$. Right: Fluctuations for the muon number after 
  using the correction function versus the corrected muon number
  ($N^{\prime}_\mu$) assuming a mixed primary composition. MC results for
  different hadronic interaction models are presented.}
 \label{icrc0740_fig01}
\end{figure*}

\section{Introduction}

   The interpretation of high-energy cosmic ray data relies on the  extensive air 
 shower (EAS) simulations in which an important source of uncertainty is 
 the description of hadronic interactions. The main problem is that for the 
 physics of air showers the relevant processes lie in the kinematical region of 
 small transverse momenta where QCD cannot be applied perturbatively and, in most
 of the cases, where no data is available. In this way, phenomenological models and
 parametrizations of accelerator data at low energies must be invoked. These models 
 are later extrapolated to high energies where they are used as valuable tools to
 understand the EAS observations at cosmic ray observatories \cite{Knapp}. Actually, air shower
 experiments can also help as laboratories to test and improve the modern 
 hadronic interaction models in energy regions which at the moment are not reachable 
 in particle accelerators. This task requires the measurement of as many EAS properties
 as possible with enough precision, such as the ones performed with the KASCADE experiment \cite{kascadeF}. 
 In the energy interval of $10^{14}-10^{16}$, tests of early versions of the hadronic 
 interaction models QGSJET, EPOS, DPMJET and SIBYLL have been carried out with the 
 KASCADE observatory (see \cite{k-epos166} and references therein) through detailed 
 studies involving the hadron, electron and muon contents of air showers. None of 
 the above models was able to describe simultaneously all the KASCADE EAS data in the 
 energy region around 1 PeV. Now, with the enhanced detector KASCADE-Grande \cite{kg-NIM10F}, the 
 possibility  to extend these tests to the  $10^{16}-10^{18}$ eV interval is opened.
 
  First investigations on this subject have been already performed with KASCADE-Grande. 
 For example, in \cite{MTD} it has been shown 
 that predictions with QGSJET II about the muon production height ($H_\mu$) distributions 
 for EAS with zenith angles below $\theta < 18^{\circ}$ show a discrepancy
 compared to the measured data.
 On the other hand, in \cite{kang-xviisvhecri} measurements on the muon lateral 
 distributions ($\rho_\mu$) of EAS performed with KASCADE-Grande were confronted with
 the expected results of QGSJET II-2 and EPOS 1.99 finding an agreement between
 experiment and model predictions \cite{kang-xviisvhecri}. Some tests regarding the 
 sensitivity of the KASCADE-Grande results to the QGSJET II-2 and EPOS 1.99 hadronic 
 models have also been done, in particular, associated with the composition and the 
 all-particle energy spectrum of cosmic rays derived from the $\rho_\mu$ and the charged 
 particle number ($N_{ch}$) analyses, respectively \cite{kang-xviisvhecri}. EPOS 1.99 favors 
 an abundance of light primary particles in the data and predicts a higher flux 
 ($\sim 33 \%$) in comparison with QGSJET II-2 \cite{kang-xviisvhecri}. 

  In this work, the hadronic interaction models QGSJET II-2 \cite{qgsF}, SIBYLL 2.1 \cite{sibyll}
 and EPOS v1.99 \cite{eposF} are tested with the KASCADE-Grande muon data. The advantage of using 
 muons is that in an air shower these particles undergo less atmospheric interactions 
 than the electromagnetic component and, in consequence, they reflect directly the physics
 of the first hadronic interactions in EAS. Other tests and comparisons employing 
 the $\rho_\mu$ distributions and the all-particle energy spectrum, derived by
 combining the information from the $N_{ch} - N_\mu$ observables, are performed in 
 \cite{souza-icrc11, bertaina-icrc11}.
 
 \vspace{-0.2pc}
  In KASCADE-Grande,  measurements of the total muon number in EAS ($N_\mu$,
 number of muons greater than $230 \, \mbox{MeV}$) are performed 
 with an array of $192 \times 3.2 \, \mbox{m}^2$ shielded scintillator detectors
 belonging to the former KASCADE experiment \cite{kascadeF}. On the other
 hand, arrival times and charged particle densities, employed for estimations of the 
 EAS arrival direction, $N_{ch}$ content and core position, are measured with an 
 enhancement called the Grande array, which is composed by $37 \times 10 \, \mbox{m}^2$ 
 plastic scintillator detectors scattered on a surface of $0.5 \, \mbox{km}^2$ \cite{kg-NIM10F}.

 \section{Description of data sets}

 For the present analysis, all air shower simulations were performed with CORSIKA 
 \cite{corsF} using Fluka \cite{flukaF} to treat hadronic interactions in the low energy
 regime. At high energies, the hadronic interaction models QGSJET II-2, SIBYLL 2.1 and 
 EPOS v1.99, subjects of this investigation, were employed. The response of the detector 
 was simulated with a GEANT 3.21 based code. Sets with spectral indexes 
 $\gamma = -2.8, -3, -3.2$ were produced for several 
 primary particle assumptions: H, He, C, Si, Fe and a mixed composition
 scenario (all single primaries in equal abundances).

\begin{table*}[t]
\begin{small}
\begin{center}
\begin{tabular}{l|cccccccccc}
\hline
      & \multicolumn{9}{c}{$\Lambda_\mu$ $(\mbox{g/cm}^2)$}\\
      & \multicolumn{3}{c}{$\gamma = -2.8$} &  \multicolumn{3}{c}{$\gamma = -3.0$} &  \multicolumn{3}{c}{$\gamma = -3.2$}\\
Model & H & Mixed & Fe & H& Mixed & Fe& H& Mixed & Fe\\
\hline
EPOS 1.99  &$445 \pm 26$& $624 \pm 31$&$636 \pm 37$ 
           &$459 \pm 23$& $607 \pm 30$&$624 \pm 31$ 
           &$476 \pm 25 \,$& $614 \pm 30$&$604 \pm 30$ \\
QGSJET II  &$824 \pm 33$ & $832 \pm 31$& $690 \pm 43$ 
           &$900 \pm 40$ & $833 \pm 31$& $693 \pm 42$ 
           &$897 \pm 100$& $825 \pm 50$& $750 \pm 62$\\
SIBYLL 2.1 &$546 \pm 44$& $657 \pm 29$&$681 \pm 46$ 
           &$637 \pm 39$& $672 \pm 29$&$688 \pm 38$ 
           &$725 \pm 44 \,$& $681 \pm 29$&$699 \pm 40$\\
\hline
\end{tabular}
\caption{Muon attenuation lengths extracted from Monte Carlo 
      data. The first column represents the hadronic interaction model. The
      corresponding composition scenario
      and spectral index, $\gamma$, of the MC sample under study are specified
      at the upper lines of the table.}\label{table1}
\end{center}
\end{small}
\vspace{-1pc}
\end{table*}

 Selection cuts were applied to both experimental and MC data. They were
 chosen according to MC studies to avoid as much as possible the influence of
 systematic uncertainties in the measurements of the EAS parameters. The
 selected data were composed of events with more than 11 triggered stations in
 Grande, shower cores inside a central area of 
 $1.52 \times 10^{5} \, \mbox{m}^2$ and arrival directions confined to
 the zenith angle interval of $\Delta \theta = 0^\circ-40^\circ$. These events were registered
 during  stable periods of data acquisition and passed successfully the
 standard reconstruction procedure of KASCADE-Grande \cite{kg-NIM10F}. 
 Additionally, only showers with $\log N_\mu > 5.1$ were considered for this work.
 Both the experimental and simulated data were analyzed and reconstructed with 
 the same algorithms.  With the above quality cuts, the effective time of
 observation with KASCADE-Grande was equivalent to 1424 days. The threshold
 for full efficiency was found at $\log_{10} N_\mu \approx 5.4$.

 \section{Description of the analysis and results}

 To start with, all muon data was corrected for systematic uncertainties
 using muon correction functions derived from MC simulations for each
 hadronic interaction model assuming mixed composition and $\gamma = -3$. 
 The functions were parametrized with respect to core position, azimuthal and zenithal angles, 
 and muon size. In Fig. \ref{icrc0740_fig01} the mean value of the muon
 correction function for different hadronic interaction models is plotted against the 
 uncorrected $N_\mu$. In general, after correction the systematic error on the
 muon number above threshold is found to be almost independent of the
 corrected muon size, $N^{\prime}_\mu$, and smaller than $6 \%$. Fluctuations
 on $N^{\prime}_\mu$ were also investigated. They are shown in Fig.
 \ref{icrc0740_fig01} for a scenario with mixed composition and $\gamma =
 -3$. Note that although fluctuations exhibit the same tendency independently
 of the hadronic model under consideration, they present slight differences
 in magnitude, as it is the case for the correction functions. Therefore, some
 differences are expected when interpreting the same muon experimental
 data with the hadronic interaction models under consideration. 

\begin{figure}[!b]
  \centering
  \includegraphics[height=2.0in, width=3.2in]{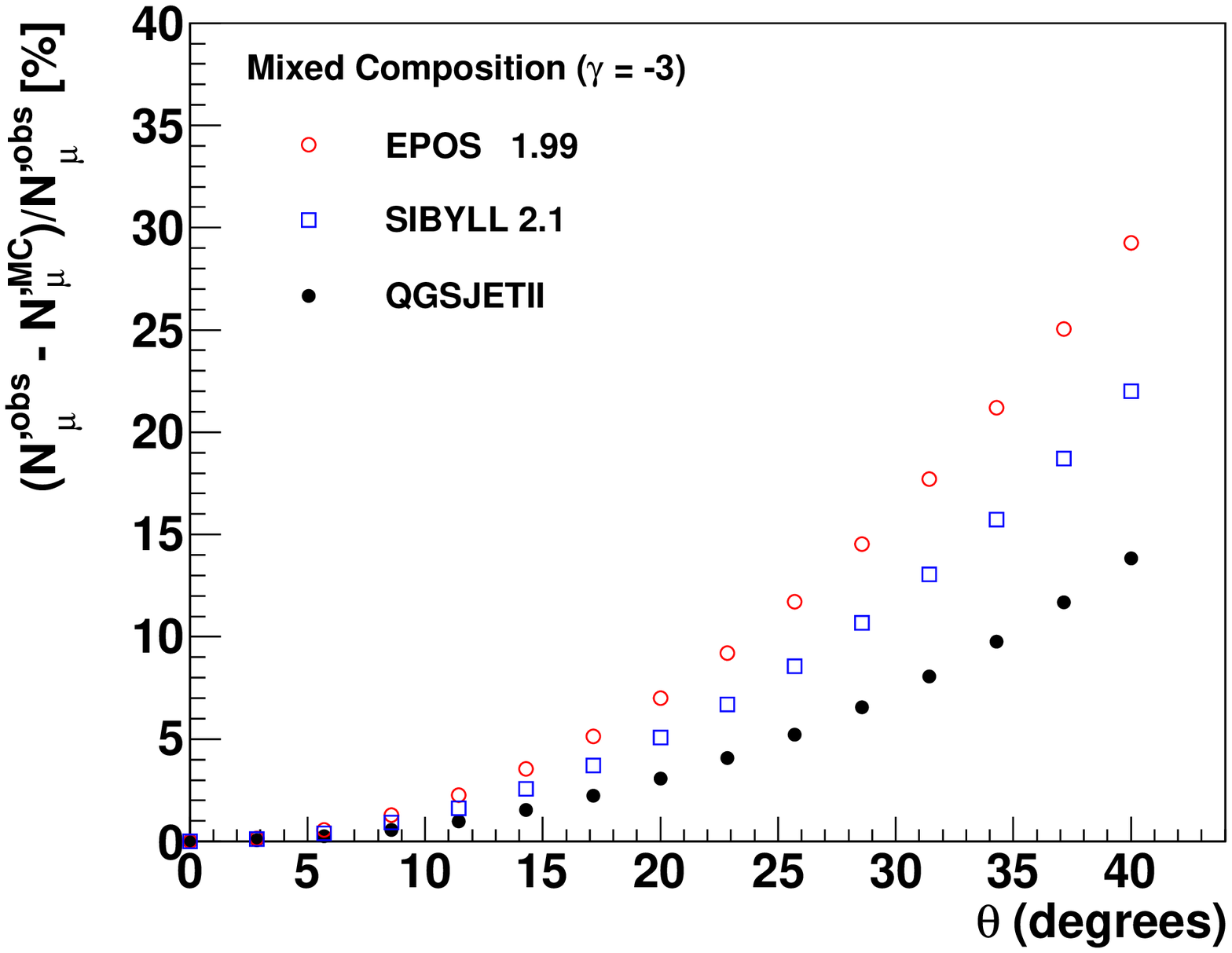}
  \caption{Relative deviation of measured data from model predictions for the
  dependence of the muon content with the zenith angle 
  in the framework of three hadronic interaction models.}
  \label{icrc0740_fig02}
 \end{figure}

 In a second step, to test the hadronic interaction models with the KASCADE-Grande
 muon data, predictions on the evolution of the muon content with the arrival
 zenith angle of the EAS were confronted with observations. The task was done
 comparing the expected and observed values of the muon attenuation length, $\Lambda_\mu$. 
 This quantity was extracted by applying the Constant Intensity Cut (CIC) method 
 to the data as described in reference \cite{arteagaF} but using a global fit
 to the attenuation curves, $\log_{10} N^{\prime}_\mu(\theta)$, with the known 
 formula
 \begin{equation}
   N^\prime_\mu = N^{\prime 0}_{\mu} \mbox{exp}[-X_0 \mbox{sec}(\theta)/\Lambda_\mu],
   \label{eqn1F}
 \end{equation}
 where $X_0 = 1023 \, \mbox{g/cm}^2$ is the average atmospheric depth for
 vertical showers and $N^{\prime 0}_\mu$ is a normalization parameter to
 be determined for each attenuation curve. The results for $\Lambda_\mu$ are 
 presented in Tables \ref{table1} and \ref{table2}. Discrepancies between the 
 experimental values and the simulation results can be observed for the studied models.
 The differences do not disappear when modifying the primary composition or
 spectral index. As a consequence, the predicted evolution of the muon
 component with the zenith angle, $N^{\prime}_\mu(\theta)$ (see equation 1) 
 shows also a disagreement with the observations. The percentage of
 deviation for $N^\prime_\mu (\theta)$ between experiment and simulations 
 (mixed composition scenario with $\gamma = -3$) inside the frameworks of QGSJET
 II-2, EPOS 1.99 and SIBYLL 2.1 is presented in
 Fig. \ref{icrc0740_fig02}. Formula  \ref{eqn1F} was employed to calculate the curves
 of Fig. \ref{icrc0740_fig02} using the values for $\Lambda_\mu$ of Tables
 \ref{table1} and \ref{table2}  and normalizing data in such a way that the 
 predicted $N^\prime_\mu$ at $\theta = 0^\circ$ agrees with the measured
 value. Several factors, which are model dependent, may come into play in the 
 observed differences: from the predicted muon correction function (see graphs 
 in Fig. \ref{icrc0740_fig01}), up to the description of the production,
 evolution and fluctuations of the shower, therefore one should be cautious when
 extracting conclusions from these differences.

\begin{table}[!t]
\begin{small}
\begin{center}
\begin{tabular}{l|ccc}
\hline
       &\multicolumn{3}{c}{\hspace{2pc} $\Lambda_\mu$ $(\mbox{g/cm}^2)$}\\
Model  &&&\\
\hline
EPOS 1.99    &&&$1851 \pm 142$ \\                
QGSJET II    &&&$1383 \pm 84 \,$\\
SIBYLL 2.1   &&&$1443 \pm 86 \,$\\
\hline
\end{tabular}
\caption{Muon attenuation lengths extracted from KASCADE-Grande data under the
       framework of different hadronic interaction models. When comparing MC and experimental
      data it should be understood that the same $N_\mu$ correction function was employed.}\label{table2}
\end{center}
\end{small}
\vspace{-1pc}
\end{table}

\begin{figure}[!t]
  \centering
  \includegraphics[height=2.0in, width=3.2in]{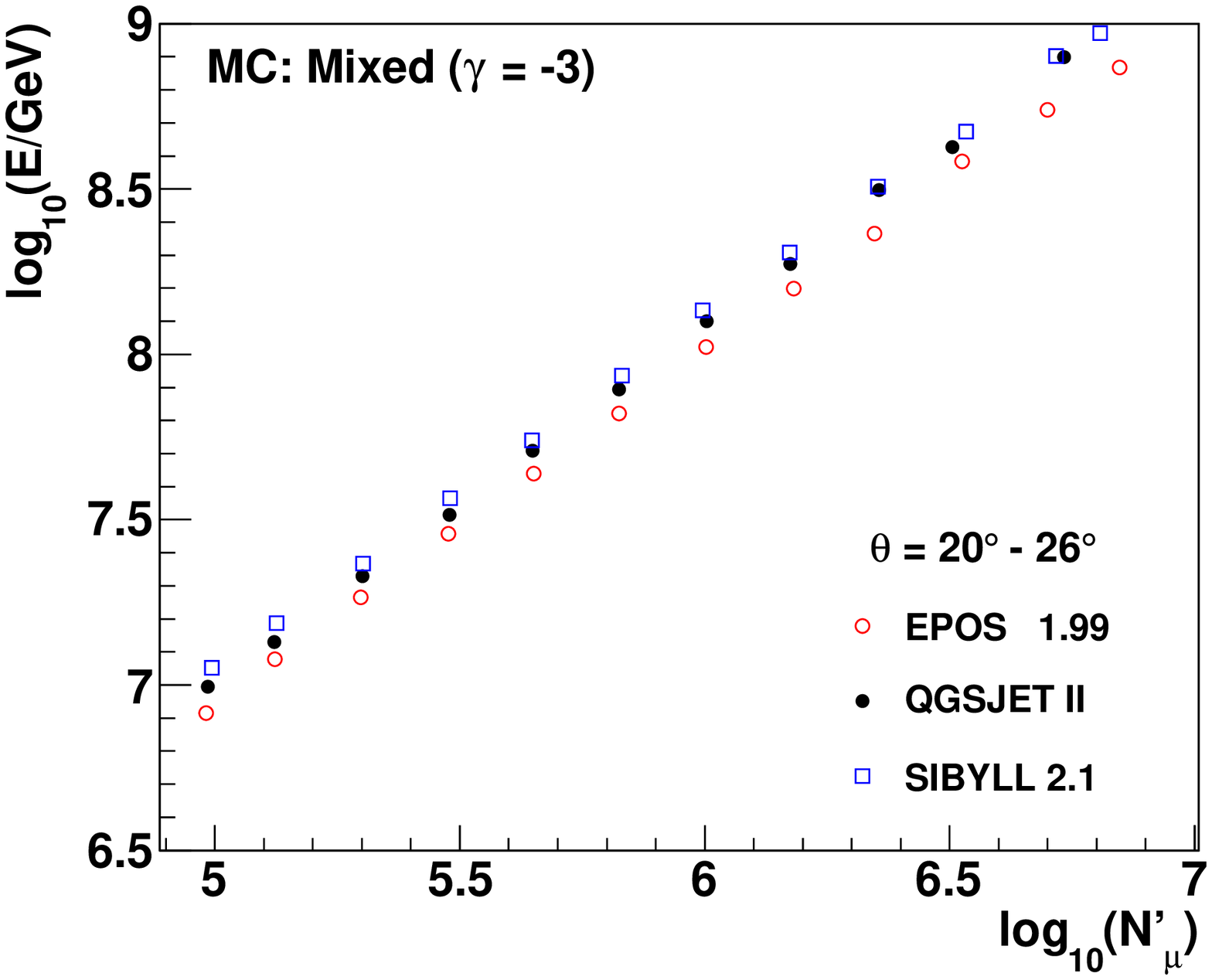}
  \caption{Primary energy as a function of the corrected muon number estimated
  from MC simulations for a mixed composition scenario with $\gamma = -3$
  using QGSJET II-2, EPOS 1.99 and SIBYLL 2.1.}
  \label{icrc0740_fig03}
 \end{figure}

\begin{figure}[!t]
  \centering
  \includegraphics[height=2.0in, width=3.2in]{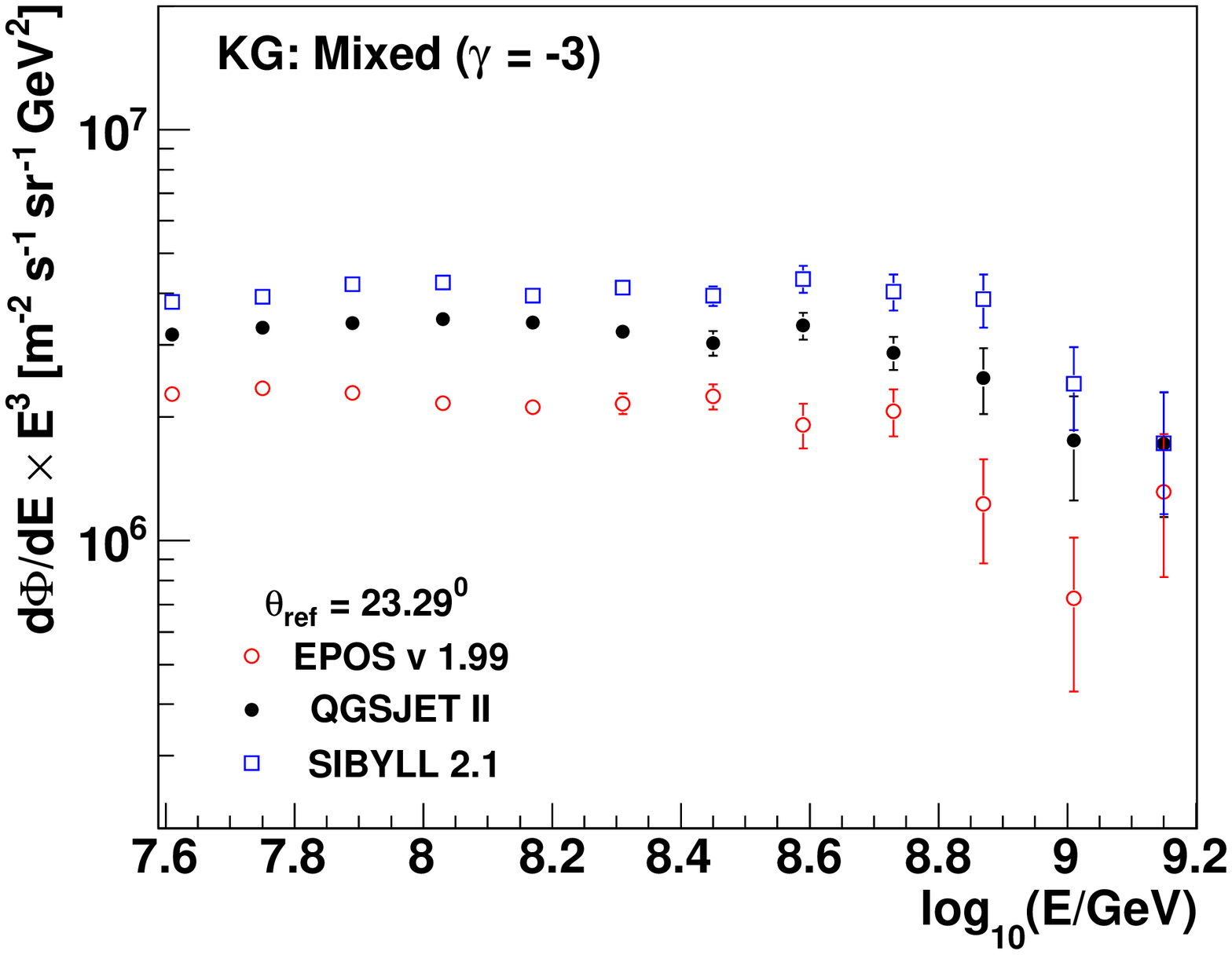}
  \caption{Energy spectra derived from KASCADE-Grande muon data using
  several hadronic models and assuming a mixed composition scenario with $\gamma = -3$.}
  \label{icrc0740_fig04}
 \end{figure}

 To finalize this analysis, the KASCADE-Grande energy spectra were reconstructed
 using QGSJET II-2, EPOS 1.99 and SIBYLL 2.1,  under the assumption of a mixed composition
 scenario and a spectral index $\gamma = -3$, and were compared with each other to 
 test the influence of the hadronic interaction model on the 
 KASCADE-Grande muon data. Details about the method of reconstruction can be
 found in \cite{arteagaF}. One important difference with respect to reference \cite{arteagaF}
 is that in the present work, the energy spectra were unfolded using the
 Gold algorithm \cite{gold}. As it was the case in \cite{arteagaF} the primary energy 
 was calculated using a calibration function of the form
 $E = \alpha_\mu [N^{\prime}_\mu (\theta_{ref})]^{\beta_\mu}$, where
 $N^{\prime}_\mu (\theta_{ref})$ is the corrected muon number of the EAS that 
 is expected at $\theta_{ref}$ according to the CIC method. $\theta_{ref} = 23.3^\circ$ is a
 zenith angle of reference. On the other hand,
 $\alpha_\mu$ and $\beta_\mu$ are model dependent parameters, which are determined 
 from a fit to the MC data presented in Fig. \ref{icrc0740_fig03}. Note that  
 at a fixed energy, the predictions of EPOS 1.99 for  
 the mean number of muons in the interval $\theta = 20^\circ - 26^\circ$ give 
 values higher by $14 \%$ compared to  QGSJET II-2, and $21 \%$ to SIBYLL
 2.1.  The reconstructed energy spectra are shown in Fig. \ref{icrc0740_fig04} 
 for the full efficiency region. 

  As seen in Fig. \ref{icrc0740_fig04}, the all-particle energy spectrum 
 derived with QGSJET II-2 is lower than the spectra calculated with SIBYLL
 2.1 but larger than the flux estimated with EPOS 1.99. In particular, at $E =
 10^8 \, \mbox{GeV}$, the above differences are of the order of $23 \%$ and
 $36 \%$, respectively. These values are comparable to those derived from 
 reference \cite{kang-xviisvhecri} ($\sim 30\%$) where the 
 charged particle number is used instead to reconstruct the energy spectra
 within the QGSJET II-2 and the EPOS 1.99 frameworks. In the latter case, however,
 the EPOS 1.99 fluxes are shifted to higher energies in comparison with those 
 estimated with QGSJET II-2.

 \section{Conclusions}

  Three hadronic interaction models: QGSJET II-2, EPOS 1.99 and SIBYLL 2.1
  were tested in this paper by comparing their predictions for the 
  attenuation of the muon content of EAS in the atmosphere with the
  observations of KASCADE-Grande. It was found that the above
  hadronic interaction models do not describe this aspect of the measured muon data. The sensitivity of
  the all-particle energy spectrum derived from the $N_\mu$ measurements to the hadronic
  interaction models was investigated. The spectrum reconstructed using 
  EPOS 1.99 is lower ($\sim 36\%$) in comparison with that from QGSJET II-2. The latter being 
  smaller ($\sim 23\%$) than the spectrum derived with SIBYLL 2.1

\vspace*{0.5cm} \footnotesize{
{\bf Acknowledgment:} KASCADE-Grande is supported by the BMBF of Germany, the MIUR and INAF of Italy, the
Polish Ministry of Science and Higher Education (in part by grant for
2009-2011) and the Romanian Authority for Scientific Research. This study was
partly supported by the DAAD-Proalmex program (2009-2010) and CONACYT. }

%\vspace{\baselineskip}
\

\clearpage

%% file: icrc0953.tex
%%
% 32nd International Cosmic Ray Conference 2011 Beijing China

%Class Required
%%% for classical LaTeX
%The paper title
\title{A direct measurement of the muon component of air showers 
 by the KASCADE-Grande Experiment}
%The short title will appear at the header of the even pages.

\shorttitle{V. de Souza \etal Muon Density - KASCADE-Grande}

%All paper authors
\authors{
V.~de Souza$^{8}$,
W.D.~Apel$^{1}$,
J.C.~Arteaga-Vel\'azquez$^{2}$,
K.~Bekk$^{1}$,
M.~Bertaina$^{3}$,
J.~Bl\"umer$^{1,4}$,
H.~Bozdog$^{1}$,
I.M.~Brancus$^{5}$,
P.~Buchholz$^{6}$,
E.~Cantoni$^{3,7}$,
A.~Chiavassa$^{3}$,
F.~Cossavella$^{4,13}$,
K.~Daumiller$^{1}$,
F.~Di~Pierro$^{3}$,
P.~Doll$^{1}$,
R.~Engel$^{1}$,
J.~Engler$^{1}$,
M. Finger$^{4}$, 
D.~Fuhrmann$^{9}$,
P.L.~Ghia$^{7}$, 
H.J.~Gils$^{1}$,
R.~Glasstetter$^{9}$,
C.~Grupen$^{6}$,
A.~Haungs$^{1}$,
D.~Heck$^{1}$,
J.R.~H\"orandel$^{10}$,
D.~Huber$^{4}$,
T.~Huege$^{1}$,
P.G.~Isar$^{1,14}$,
K.-H.~Kampert$^{9}$,
D.~Kang$^{4}$, 
H.O.~Klages$^{1}$,
K.~Link$^{4}$, 
P.~{\L}uczak$^{11}$,
M.~Ludwig$^{4}$,
H.J.~Mathes$^{1}$,
H.J.~Mayer$^{1}$,
M.~Melissas$^{4}$,
J.~Milke$^{1}$,
B.~Mitrica$^{5}$,
C.~Morello$^{7}$,
G.~Navarra$^{3,15}$,
J.~Oehlschl\"ager$^{1}$,
S.~Ostapchenko$^{1,16}$,
S.~Over$^{6}$,
N.~Palmieri$^{4}$,
M.~Petcu$^{5}$,
T.~Pierog$^{1}$,
H.~Rebel$^{1}$,
M.~Roth$^{1}$,
H.~Schieler$^{1}$,
F.G.~Schr\"oder$^{1}$,
O.~Sima$^{12}$,
G.~Toma$^{5}$,
G.C.~Trinchero$^{7}$,
H.~Ulrich$^{1}$,
A.~Weindl$^{1}$,
J.~Wochele$^{1}$,
M.~Wommer$^{1}$,
J.~Zabierowski$^{11}$
}
%All the affiliations.
\afiliations{
$^1$ Institut f\"ur Kernphysik, KIT - Karlsruher Institut f\"ur Technologie, Germany\\
$^2$ Universidad Michoacana, Instituto de F\'{\i}sica y Matem\'aticas, Morelia, Mexico\\
$^3$ Dipartimento di Fisica Generale dell' Universit\`a Torino, Italy\\
$^4$ Institut f\"ur Experimentelle Kernphysik, KIT - Karlsruher Institut f\"ur Technologie, Germany\\
$^5$ National Institute of Physics and Nuclear Engineering, Bucharest, Romania\\
$^6$ Fachbereich Physik, Universit\"at Siegen, Germany\\
$^7$ Istituto di Fisica dello Spazio Interplanetario, INAF Torino, Italy\\
$^8$ Universidade S$\tilde{a}$o Paulo, Instituto de F\'{\i}sica de S\~ao Carlos, Brasil\\
$^9$ Fachbereich Physik, Universit\"at Wuppertal, Germany\\
$^{10}$ Dept. of Astrophysics, Radboud University Nijmegen, The Netherlands\\
$^{11}$ Soltan Institute for Nuclear Studies, Lodz, Poland\\
$^{12}$ Department of Physics, University of Bucharest, Bucharest, Romania\\
\scriptsize{
$^{13}$ now at: Max-Planck-Institut Physik, M\"unchen, Germany; 
$^{14}$ now at: Institute Space Sciences, Bucharest, Romania; 
$^{15}$ deceased; 
$^{16}$ now at: Univ Trondheim, Norway
}
}
%email address of the contact person
\email{vitor@ifsc.usp.br}

%The abstract.
\abstract{The muon component of atmospheric air showers is a very important
information in astroparticle physics because it is related to
the primary particle mass and also because it depends on the
hadronic interactions happening in the shower development. 
In this paper, we study the muon densities measured by the
KASCADE-Grande experiment and illustrate its importance in
composition studies and testing of hadronic interaction models.
The data analysed here was measured by the KASCADE-Grande detector and
lies in the $10^{16}-10^{18}$ eV energy range.
The measured muon density is compared to predictions
of EPOS 1.99, QGSJet II and Sibyll 2.1 hadronic interaction
models. This paper is an update of the results presented in
\cite{bib:vitor:icrc2009}. The paper extends its scope by testing two
extra hadronic models Sibyll 2.1 and EPOS 1.99.}


\maketitle

\section{Introduction}

\begin{figure}[t]
\begin{center}
  \includegraphics
  [width=0.4\textwidth]{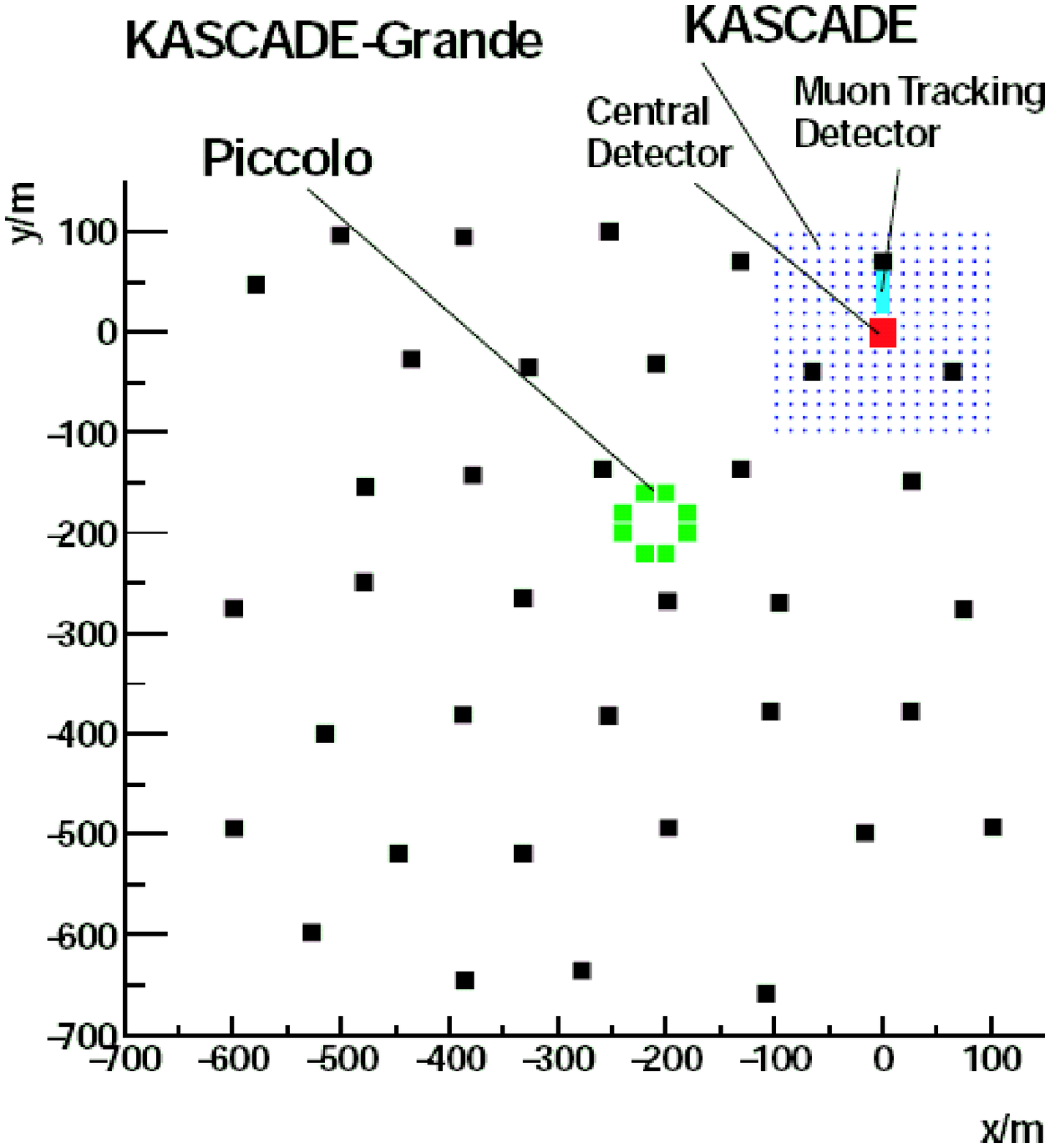}
\end{center}
%\vspace{-0.7cm}
\caption{Representation of the KASCADE-Grande detectors.}
\label{fig:kascade}
\end{figure}

Cosmic rays with energy range
between $10^{16}$ and $10^{18}$ eV have the potential to reveal
interesting astrophysical phenomena occurring in the Universe.
In this energy range the transition between galactic to extragalactic 
predominance of the flux might occur. This transition is expected to
cause an evolution of the mean mass of primary particles. 

The KASCADE-Grande experiment (see figure ~\ref{fig:kascade}) has been set up to
measure primary cosmic rays in this energy range in order to help in the
understanding of these questions.
The experiment is located at the Karlsruhe Institue of Technology,
Campus North, Germany,
where, besides the  existing KASCADE~\cite{bib:KA2} array, two new
detector set ups (Grande and Piccolo) have been installed. The
experiment is able to sample different components of extensive air
showers (electromagnetic, muonic and hadronic) with high accuracy and
covering a surface of 0.5 km$^2$. For an overview of the actual setup
of the KASCADE-Grande Experiment see ref.~\cite{bib:andreas}.

Muons are the
messengers of the hadronic interactions of the particles in the shower
and therefore are a powerful tool to determine the primary particle
mass and to study the hadronic interaction models.

In this article we present studies of the muon component of the
showers. This paper is an update of a previous publication
\cite{bib:vitor:icrc2009} in which the KASCADE-Grande Collaboration has
shown the muon densities for the first time. In this contribution, we
analyse two more years of data representing an substancial increase
in the number of events. In this publication two extra hadronic
interation models are also tested: EPOS 1.99 and Sibyll 2.1.

\begin{figure}[t]
\begin{center}
  \includegraphics
  [width=0.4\textwidth]{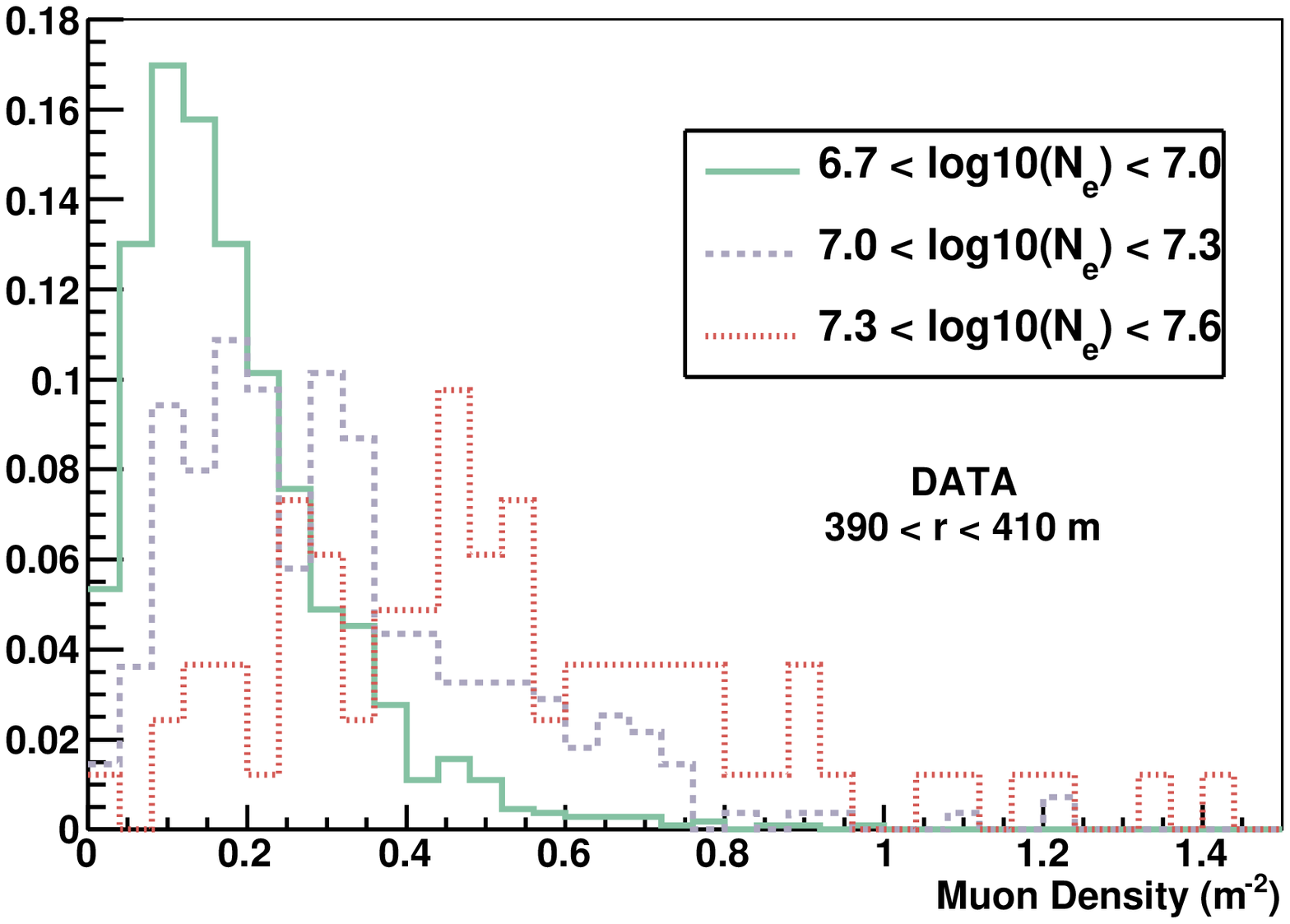}
\end{center}
\vspace{-0.7cm}
\caption{Distribution of the density of muons for three cuts in the
  total number of electrons.}
\label{fig:hist:ldf}
\end{figure}

\section{Reconstruction}

As explained in \cite{bib:vitor:icrc2009} and repeated here, the main
parameters used in this study are the density of muons and 
the total number of electrons in the shower for which the
reconstruction accuracy is going to be discussed below. For the
reconstruction accuracy of the shower geometry see
ref.~\cite{bib:federico}.

The density of muons is directly measured by the KASCADE
622 $m^2$ scintillators. These detectors are shielded by 10 cm of lead and 4 cm
of iron, corresponding to 20 radiation lengths and a threshold of 230
MeV for vertical muons. The error in the measurement of the energy deposit was
experimentally determined to be smaller than 10\%~\cite{bib:KA2}.

For each shower, the density of
muons is calculated as follows. The muon stations are grouped in rings
of 20 m distance from the shower axis. The sum of the signals measured
by all muon stations inside each ring is divided by the effective detection
area of the stations. Therefore the muon density as a function of the
distance from the shower axis is measured in a very direct way. No
fitting of lateral distributions is needed in these calculations. 

The total number of electrons in the shower is reconstructed in a
combined way using KASCADE and KASCADE-Grande stations. A lateral
distribution function (LDF) of the Lagutin type can be fitted to the
density of muons measured by the KASCADE
detector~\cite{bib:fuhrmann}. After that, using the fitted function, the
number of muons at any distance from the shower axis can be
estimated. The KASCADE-Grande stations measure the number of charged
particles. The number of electrons at each KASCADE-Grande stations is
determined by subtracting from the measured number of charged
particles the number of muons estimated with the LDF fitted to the
KASCADE stations.

\begin{figure}[t]
\begin{center}
  \includegraphics
  [width=0.4\textwidth]{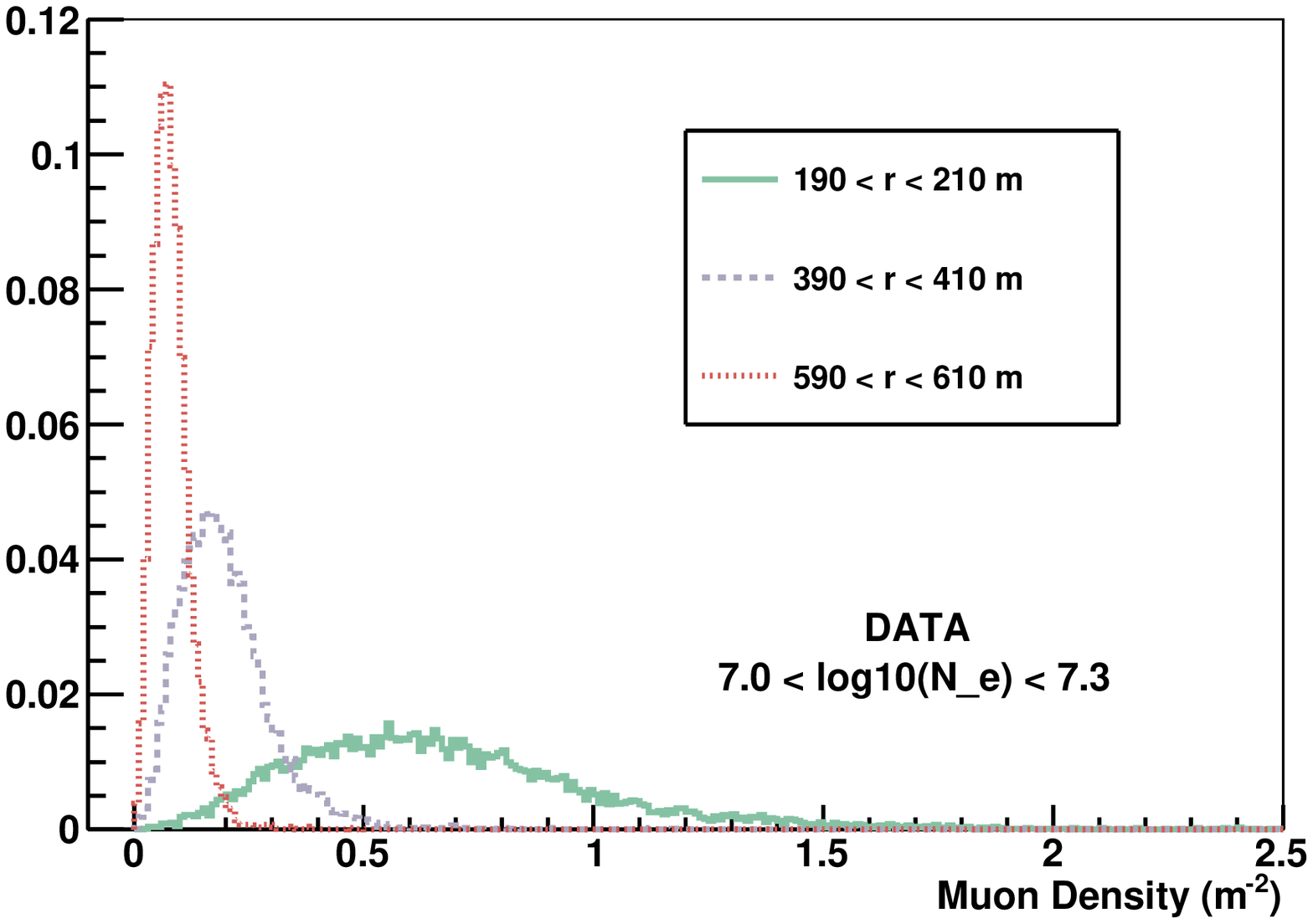}
\end{center}
\vspace{-0.7cm}
\caption{Distribution of the density of muons for three distances from the shower axis.}
\label{fig:hist:elect}
\end{figure}

At this stage, the number of electrons at each KASCADE-Grande station
is known. Finally, a modified NKG~\cite{bib:nkg} function is fitted to this data and
the total number of electrons is determined in the fit.

Quality cuts
have been applied to the events in this analysis procedure. We have
required
more than 19 KASCADE-Grande stations with signal. 
The showers used in all analyses along this paper were reconstructed with zenith angle
between 0 and 42 degrees.
The same quality
cuts were applied to the simulated events used for reconstruction
studies and to the data
presented in the following section.  After the quality cuts,
the total number of electrons can be estimated with a systematic shift
smaller than 10\% and a statistical uncertainty smaller than 20\%
along the entire range considered in this paper~\cite{bib:federico}.

Figure~\ref{fig:hist:ldf} shows the measured density of muons at three distances
from the shower axis for events with a total number of 
electrons ($N_e$) in the range $7.0 < Log10(N_e) < 7.3$ ($\approx 10^{17}$ eV).
Similar plots were obtained for other $N_e$ ranges.

Figure~\ref{fig:hist:elect} shows the density of muons at 400 m from
the shower axis for events with total number of 
electrons ($N_e$) in the range  $6.7 < Log10(N_e) < 7.0$, $7.0 <
Log10(N_e) < 7.3$ and  $7.3 < Log10(N_e) < 7.6$. Similar plots were
obtained for other distances from the shower axis.

Figure~\ref{fig:hist:ldf} and Figure~\ref{fig:hist:elect} show the
general expected trend: a) decrease of the muon density with
increasing distance from the shower axis and b) increase of the
muon density with increasing total number of electrons. 
In the next sections we explore these relations in order to show the
capabilities of the KASCADE-Grande experiment for a composition study
and for tests of the hadronic interaction models.

We present data for  $7.0 <
Log10(N_e) < 7.3$  and $390 < r < 410 \ m$, these cuts have been chosen
in order to minimize the fluctuation of the signal and the
reconstruction inaccuracy and to maximize the
number of showers for which we have data, however the same
conclusions would be 
drawn for all parameter cuts.

\section{Simulation}

\begin{figure}[t]
\begin{center}
  \includegraphics
  [width=0.4\textwidth]{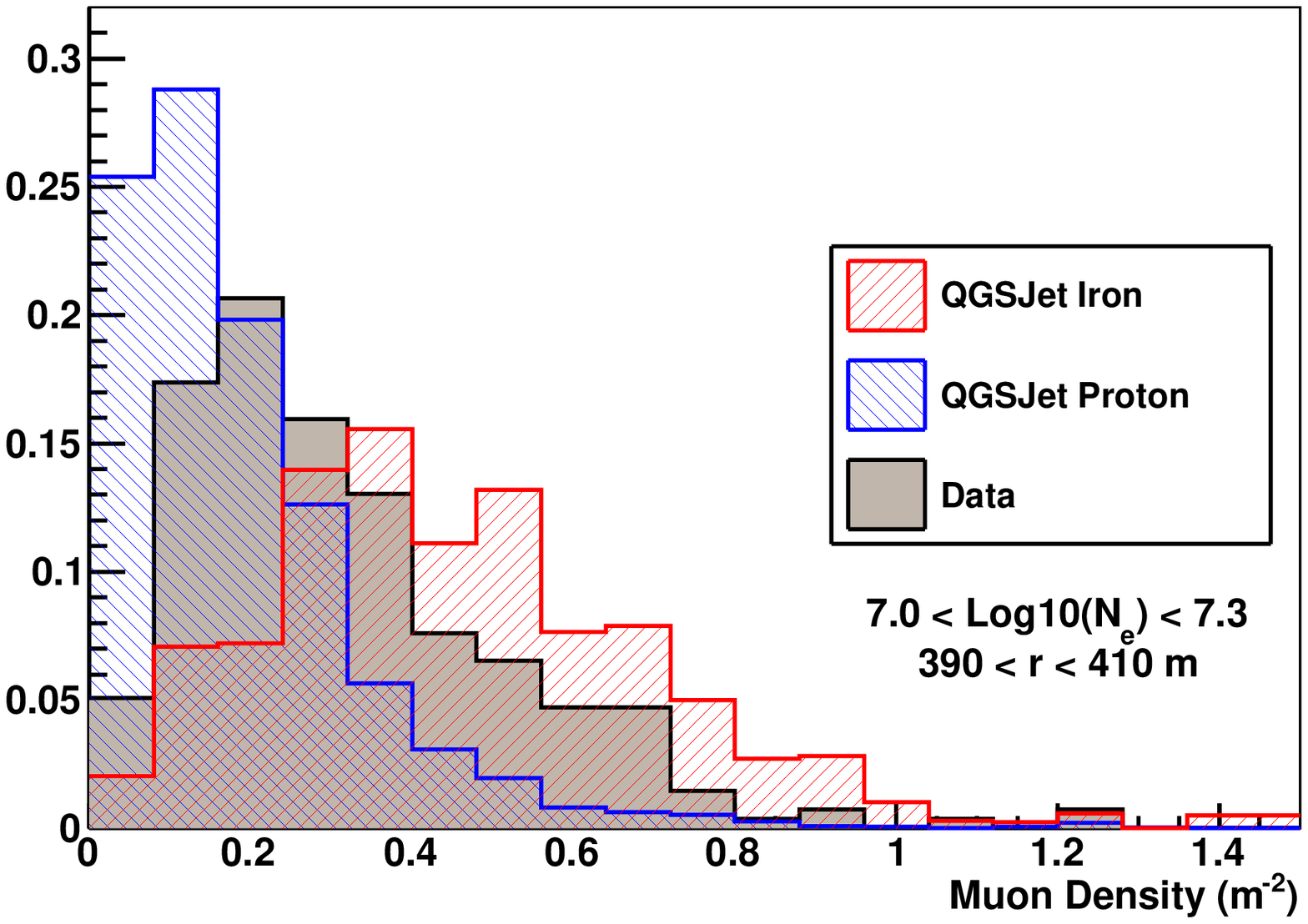}
\end{center}
\vspace{-0.7cm}
\caption{Measured distribution of the density of muons at 400 m
  compared to the predictions of QGSJet II.}
\label{fig:hist:qgsjet}
\end{figure}

\begin{figure}[t]
\begin{center}
  \includegraphics
  [width=0.4\textwidth]{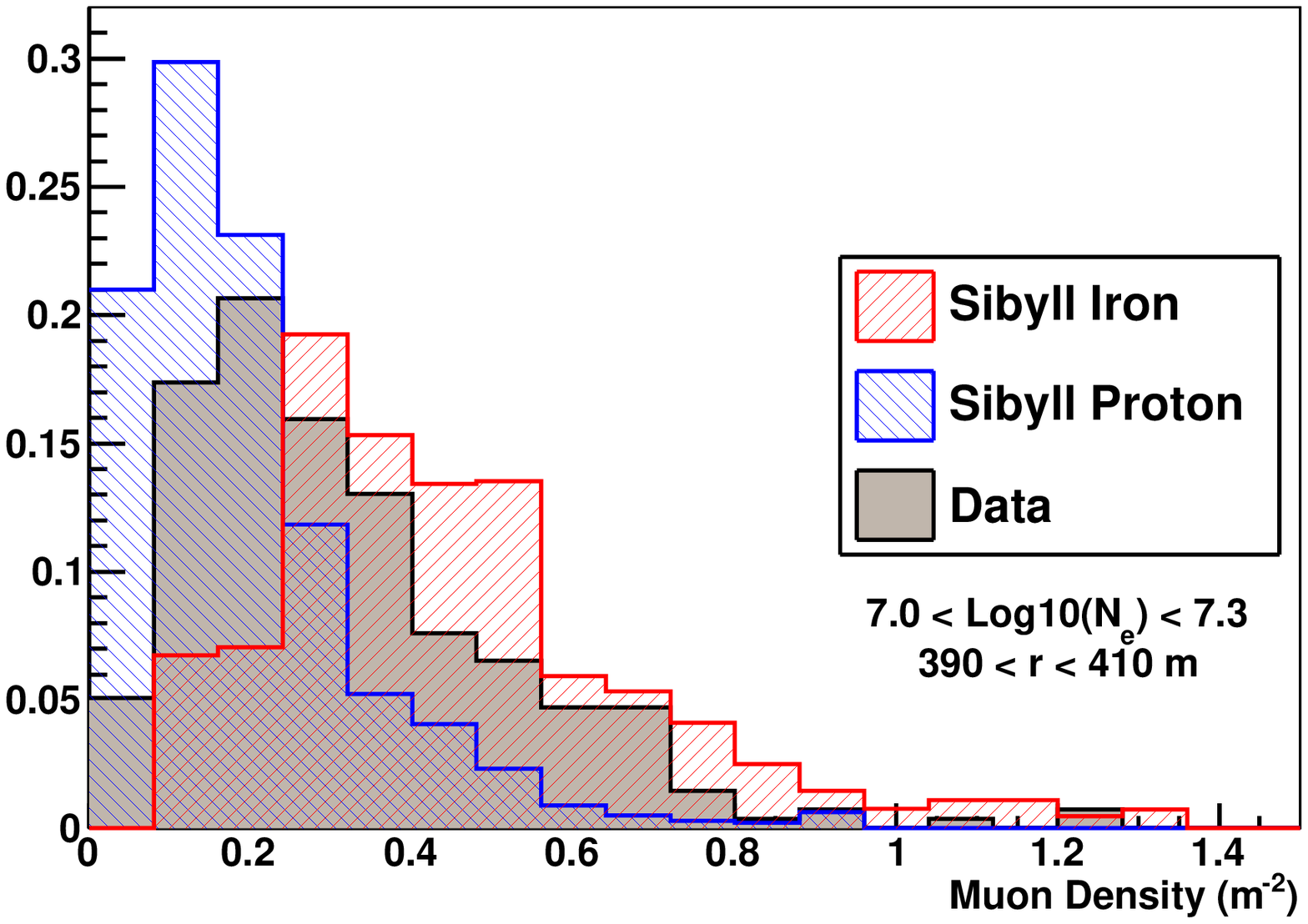}
\end{center}
\vspace{-0.7cm}
\caption{Measured distribution of the density of muons at 400 m
  compared to the predictions of Sibyll 2.1.}
\label{fig:hist:sibyll}
\end{figure}

For all studies in this paper we have used the CORSIKA
~\cite{bib:corsika} simulation program with the
FLUKA~\cite{bib:fluka} option for low energy hadronic interactions.
Three high energy hadronic interaction models
were used EPOS 1.99~\cite{bib:epos}, QGSJet II 
~\cite{bib:qgsjet} and Sibyll 2.1~\cite{bib:sibyll}. No thinning is used~\cite{bib:corsika}.

CORSIKA showers are
simulated through the detectors and reconstructed in the same way as
the measured data, such that a direct comparison between data and
simulation is possible.

\begin{figure}[t]
\begin{center}
  \includegraphics
  [width=0.4\textwidth]{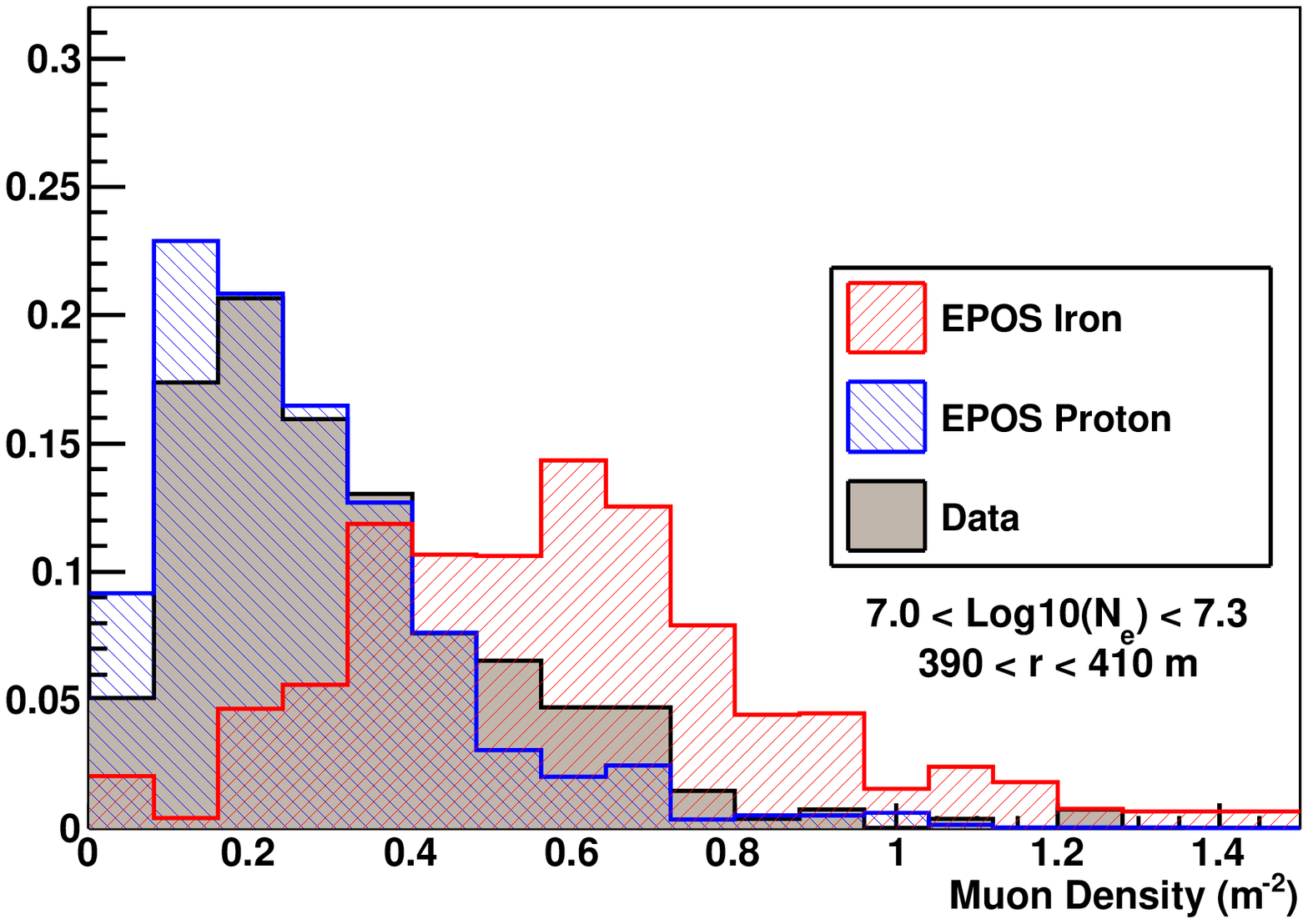}
\end{center}
\vspace{-0.7cm}
\caption{Measured distribution of the density of muons at 400 m
  compared to the predictions of EPOS 1.99.}
\label{fig:hist:epos}
\end{figure}

Figures \ref{fig:hist:qgsjet}, \ref{fig:hist:epos} and \ref{fig:hist:sibyll}  show the
comparison of the measured density of muons to values predicted by
QGSJet II, EPOS 1.99 and Sibyll 2.1, respectively. For the three hadronic interactions models we show
the limiting cases of proton and iron nuclei as primary particles. It
can be seen in figures \ref{fig:hist:qgsjet}, \ref{fig:hist:epos} and \ref{fig:hist:sibyll} that the
data lie well 
within the proton and iron limits for QGSJet II, EPOS
1.99 and Sibyll 2.1. These graphics are
going to be further discussed in the next sections.

\section{Analysis}

\begin{figure}[t]
\begin{center}
  \includegraphics
  [width=0.5\textwidth]{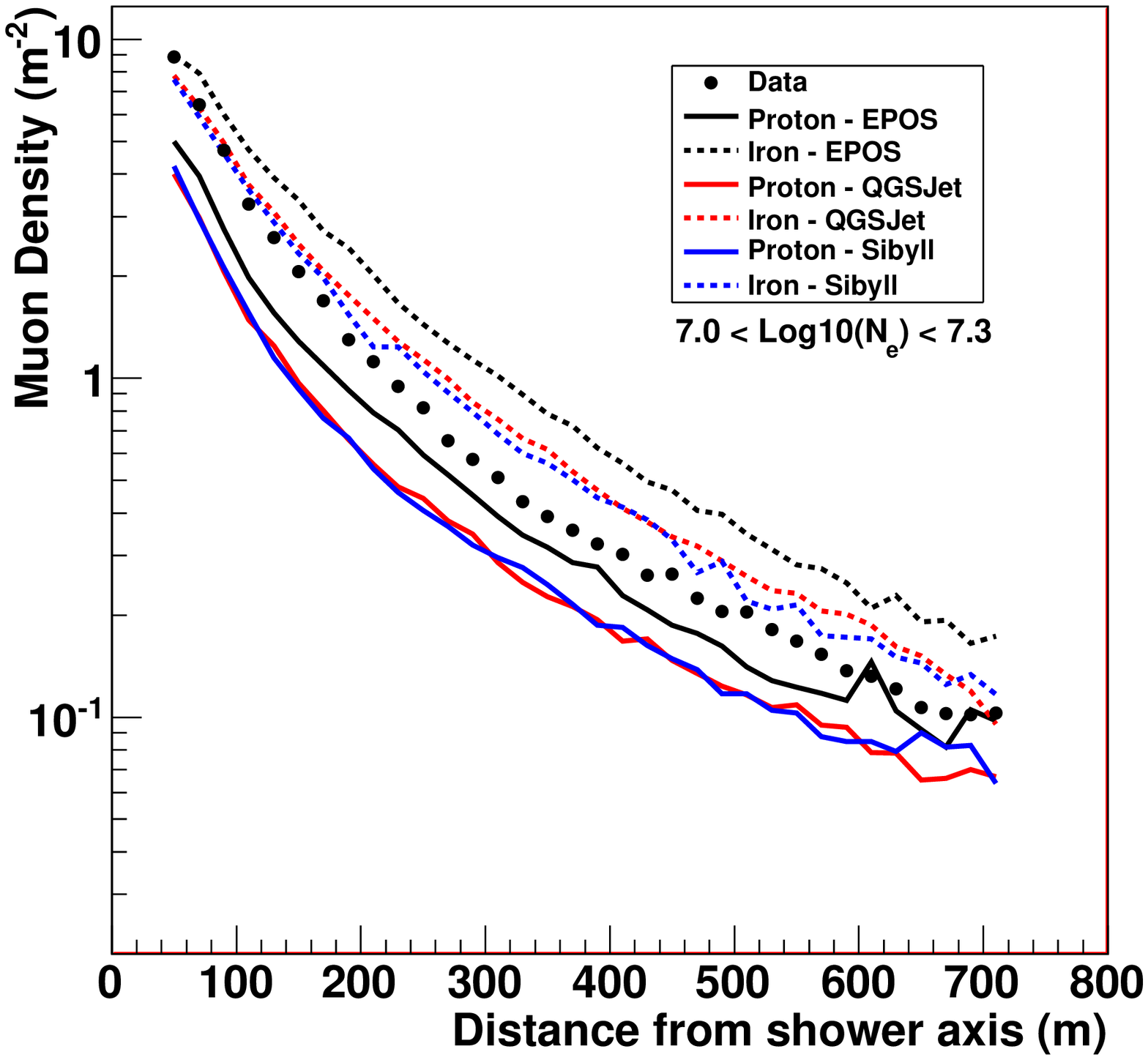}
\end{center}
\vspace{-0.7cm}
\caption{Lateral distribution of muons compared to the predictions of
  QGSJet II, EPOS 1.99 and Sibyll 2.1.}
\label{fig:ldf}
\end{figure}

Figure \ref{fig:ldf} shows the mean muon density as a function of the
distance from the shower axis compared to the predictions of QGSJet
II, EPOS 1.99 and Sibyll 2.1. The three hadronic interaction models bracket the data
within the proton and iron limits for the entire range of distances
from 100 to 720 meters. For distances further than 720 meters the
statistics is not enough for a conclusion. 

Interesting to note is also
the slope of the LDF. Considering an equal probability trigger for
protons and iron primaries as a function of distance from the shower
axis, one can analyse the slope of the LDF as a test of the hadronic
interaction models or as a composition parameters.

The intermediate distance from 200 to 600 m is the most
significant range. For larger distance the statistical fluctuations
dominates and at short distances the accuracy of the core position
reconstruction 
impacts the measurements of the LDF resulting in a steeply falling
curve as seen in figure \ref{fig:ldf}.

It is clear that the measured LDF is not parallel to the QGSJet, nor
EPOS 1.99 nor Sibyll 2.1 curves. That shows that the slope of the LDF cannot
be well described by neither models if the composition is supposed to
be dominated by only one light or only one heavy primary. If the
composition is mixed the slope varies with distance and the
interpretation depends on the exact composition.

\begin{figure}[h]
\begin{center}
  \includegraphics
  [width=0.5\textwidth]{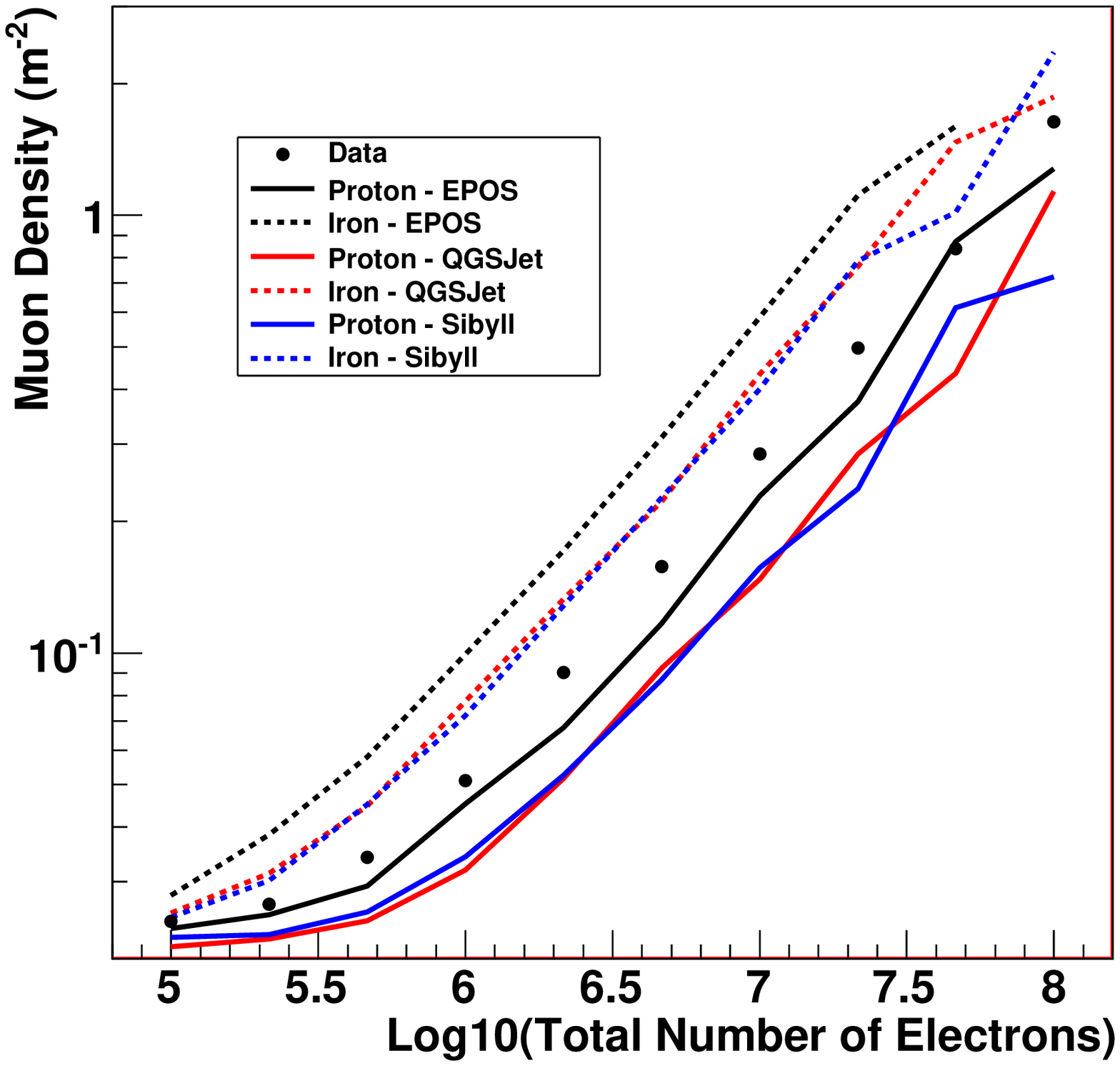}
\end{center}
\vspace{-0.7cm}
\caption{Muon density as a function of the total number of electrons
  compared to the predictions of QGSJet II, EPOS 1.99 and Sibyll 2.1.} 
\label{fig:elect}
\end{figure}

Figure~\ref{fig:elect} shows the evolution of the mean muon density
as a function 
of $N_e$. The calculations done with QGSJet II, EPOS 1.99  and Sibyll
2.1 using  proton and iron nuclei as primary particles bracket the data in
the entire range of $5 < Log10(N_{e}) < 8$.

Nevertheless, both figures \ref{fig:ldf} and  ~\ref{fig:elect} show that EPOS
1.99 would require a very light primary composition in order to fit the
data. On the other hand, QGSJet II and Sibyll 2.1 could fit the data with an intermediate
primary abundance between proton and iron nuclei.

Besides that, in figure~\ref{fig:elect} it is possible to analyse a
possible transition of     
the primary component with increasing total number of
electrons. The analysis done with both models show no abrupt change
in the compositon in the entire energy range.

\section{Conclusions}

The Grande array is in continuous and stable data taking since December
2003. The quality of the detector can be illustrated by the smooth
data curve and small fluctuations in figures~\ref{fig:ldf} and \ref{fig:elect}.

In this article, we have briefly described the procedure used to
measure the density of muons with the KASCADE array and we have studied its
correlation with the distance from the shower axis and the total
number of electrons in the shower. 

The density of muons in the shower is measured directly by
the KASCADE detectors. We have used this data to study the hadronic
interaction models QGSJet II, EPOS 1.99 and Sibyll 2.1. The plots
shown here are an update of the previous publication \cite{bib:vitor:icrc2009}

Figure~\ref{fig:elect} shows no abrupt change with increasing total
number of electrons up to $Log10( N_e )= 7.5 \approx 5 \times 10^{17}$ eV. The mean primary mass estimation
would depend on the hadronic interaction model used.

\clearpage

%% file: icrc0273.tex
%%
% 32nd International Cosmic Ray Conference 2011 Beijing China

%Class Required
%%% for classical LaTeX
%The paper title
\title{On the primary mass sensitivity of muon pseudorapidities measured
               with KASCADE-Grande}
%The short title will appear at the header of the even pages.

\shorttitle{J.~Zabierowski \etal Mass sensitivity of muon pseudorapidity}

%All paper authors
\authors{
J.~Zabierowski$^{1}$
P.~{\L}uczak$^{1}$,
P.~Doll$^{2}$,
W.D.~Apel$^{2}$,
J.C.~Arteaga-Vel\'azquez$^{3}$,
K.~Bekk$^{2}$,
M.~Bertaina$^{4}$,
J.~Bl\"umer$^{2,5}$,
H.~Bozdog$^{2}$,
I.M.~Brancus$^{6}$,
P.~Buchholz$^{7}$,
E.~Cantoni$^{4,8}$,
A.~Chiavassa$^{4}$,
F.~Cossavella$^{5,13}$,
K.~Daumiller$^{2}$,
V.~de Souza$^{9}$,
F.~Di~Pierro$^{4}$,
R.~Engel$^{2}$,
J.~Engler$^{2}$,
M. Finger$^{5}$, 
D.~Fuhrmann$^{10}$,
P.L.~Ghia$^{8}$, 
H.J.~Gils$^{2}$,
R.~Glasstetter$^{10}$,
C.~Grupen$^{7}$,
A.~Haungs$^{2}$,
D.~Heck$^{2}$,
J.R.~H\"orandel$^{11}$,
D.~Huber$^{5}$,
T.~Huege$^{2}$,
P.G.~Isar$^{2,14}$,
K.-H.~Kampert$^{10}$,
D.~Kang$^{5}$, 
H.O.~Klages$^{2}$,
K.~Link$^{5}$, 
M.~Ludwig$^{5}$,
H.J.~Mathes$^{2}$,
H.J.~Mayer$^{2}$,
M.~Melissas$^{5}$,
J.~Milke$^{2}$,
B.~Mitrica$^{6}$,
C.~Morello$^{8}$,
G.~Navarra$^{4,15}$,
J.~Oehlschl\"ager$^{2}$,
S.~Ostapchenko$^{2,16}$,
S.~Over$^{7}$,
N.~Palmieri$^{5}$,
M.~Petcu$^{6}$,
T.~Pierog$^{2}$,
H.~Rebel$^{2}$,
M.~Roth$^{2}$,
H.~Schieler$^{2}$,
F.G.~Schr\"oder$^{2}$,
O.~Sima$^{12}$,
G.~Toma$^{6}$,
G.C.~Trinchero$^{8}$,
H.~Ulrich$^{2}$,
A.~Weindl$^{2}$,
J.~Wochele$^{2}$,
M.~Wommer$^{2}$,
}
%All the affiliations.
\afiliations{
$^1$ Soltan Institute for Nuclear Studies, Lodz, Poland\\
$^2$ Institut f\"ur Kernphysik, KIT - Karlsruher Institut f\"ur Technologie, Germany\\
$^3$ Universidad Michoacana, Instituto de F\'{\i}sica y Matem\'aticas, Morelia, Mexico\\
$^4$ Dipartimento di Fisica Generale dell' Universit\`a Torino, Italy\\
$^5$ Institut f\"ur Experimentelle Kernphysik, KIT - Karlsruher Institut f\"ur Technologie, Germany\\
$^6$ National Institute of Physics and Nuclear Engineering, Bucharest, Romania\\
$^7$ Fachbereich Physik, Universit\"at Siegen, Germany\\
$^8$ Istituto di Fisica dello Spazio Interplanetario, INAF Torino, Italy\\
$^9$ Universidade S$\tilde{a}$o Paulo, Instituto de F\'{\i}sica de S\~ao Carlos, Brasil\\
$^{10}$ Fachbereich Physik, Universit\"at Wuppertal, Germany\\
$^{11}$ Dept. of Astrophysics, Radboud University Nijmegen, The Netherlands\\
$^{12}$ Department of Physics, University of Bucharest, Bucharest, Romania\\
\scriptsize{
$^{13}$ now at: Max-Planck-Institut Physik, M\"unchen, Germany; 
$^{14}$ now at: Institute Space Sciences, Bucharest, Romania; 
$^{15}$ deceased; 
$^{16}$ now at: Univ Trondheim, Norway
}
}
%email address of the contact person
\email{janzab@zpk.u.lodz.pl}

%The abstract.
\abstract{With the Muon Tracking Detector in the KASCADE-Grande experiment mean EAS 
muon pseudorapidities are investigated. Here we report on the results of 
studying the sensitivity of this quantity to the mass of primary cosmic 
ray particles. Obtained values of the mean logarithmic mass in the 
$10^{16}$ eV - $10^{17}$ eV range of primary energies, based on the 
QGSJetII - FLUKA interaction model combination, are compared with 
the results of other experiments. 
The validity of the model in reproducing experimentally measured 
pseudorapidity values and its comparison with the EPOS 1.99 is discussed.}
%The keywords

% B E G I N   D O C U M E N T
\maketitle

%Begin the section.
\section{Introduction}
\begin{figure*}[th]
\begin{center}
%\centering
\includegraphics*[width=20cm,height=22cm]{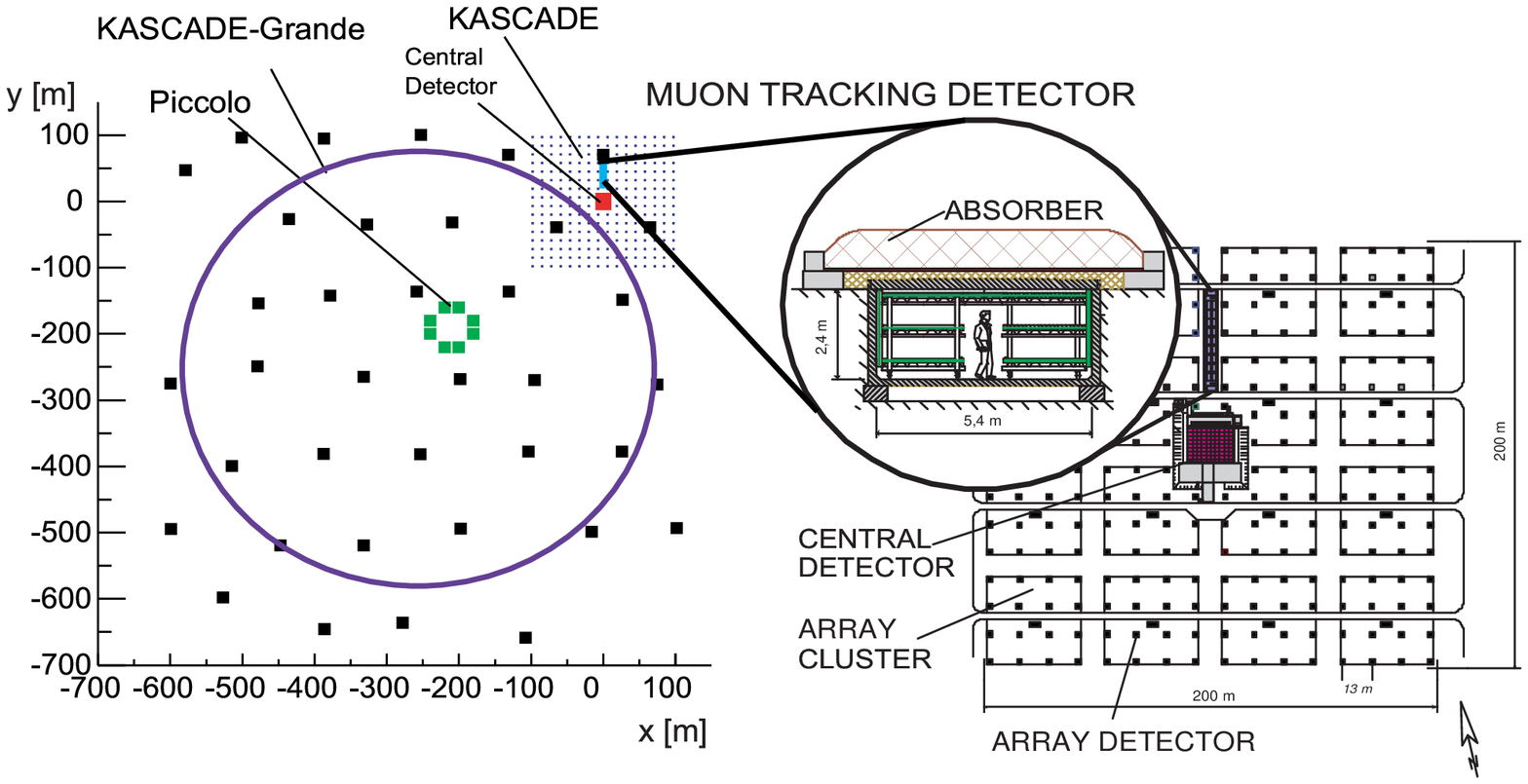}
\vspace{-14cm}
\caption{Layout of the KASCADE-Grande experiment distributed over the KIT - Campus North area.
KASCADE is situated in the North-East corner of the Campus; note the position of the Muon Tracking Detector.}
\label{fig1i}
\end{center}
\end{figure*}

The Muon Tracking Detector (MTD) \cite{MTDi} is one of the detector components in the KASCADE-Grande
EAS experiment \cite{KASG} (see Fig.\ref{fig1i}), operated at the Karlsruhe Institute of 
Technology (KIT) - Campus North, in Germany, by an 
international collaboration. The MTD measures directions of muon tracks in EAS
with excellent angular resolution of $\approx$~ $0.35^{\circ}$. 
These directional data allow to investigate the longitudinal development of the muonic component in showers which
is a signature of the development of the hadronic EAS core, being in turn dependent on the mass of the primary cosmic 
ray particle initiating a shower. Such studies can be done either by the determination of a mean
muon production height \cite{Hmu} or by using the mean pseudorapidity ($\eta$) of EAS muons,  
expressed in terms of their tangential ($\tau$) and radial ($\rho$) angles (quantities
reconstructed in the experiment) \cite{eta}, \cite{pylos}. In this work we investigate to what extent
one can use the muon pseudorapidity for the determination of primary mass.

\section{Muon pseudorapidities in KASCADE-Grande}
\begin{figure}[hb]
\begin{center}
\includegraphics[width=60mm]{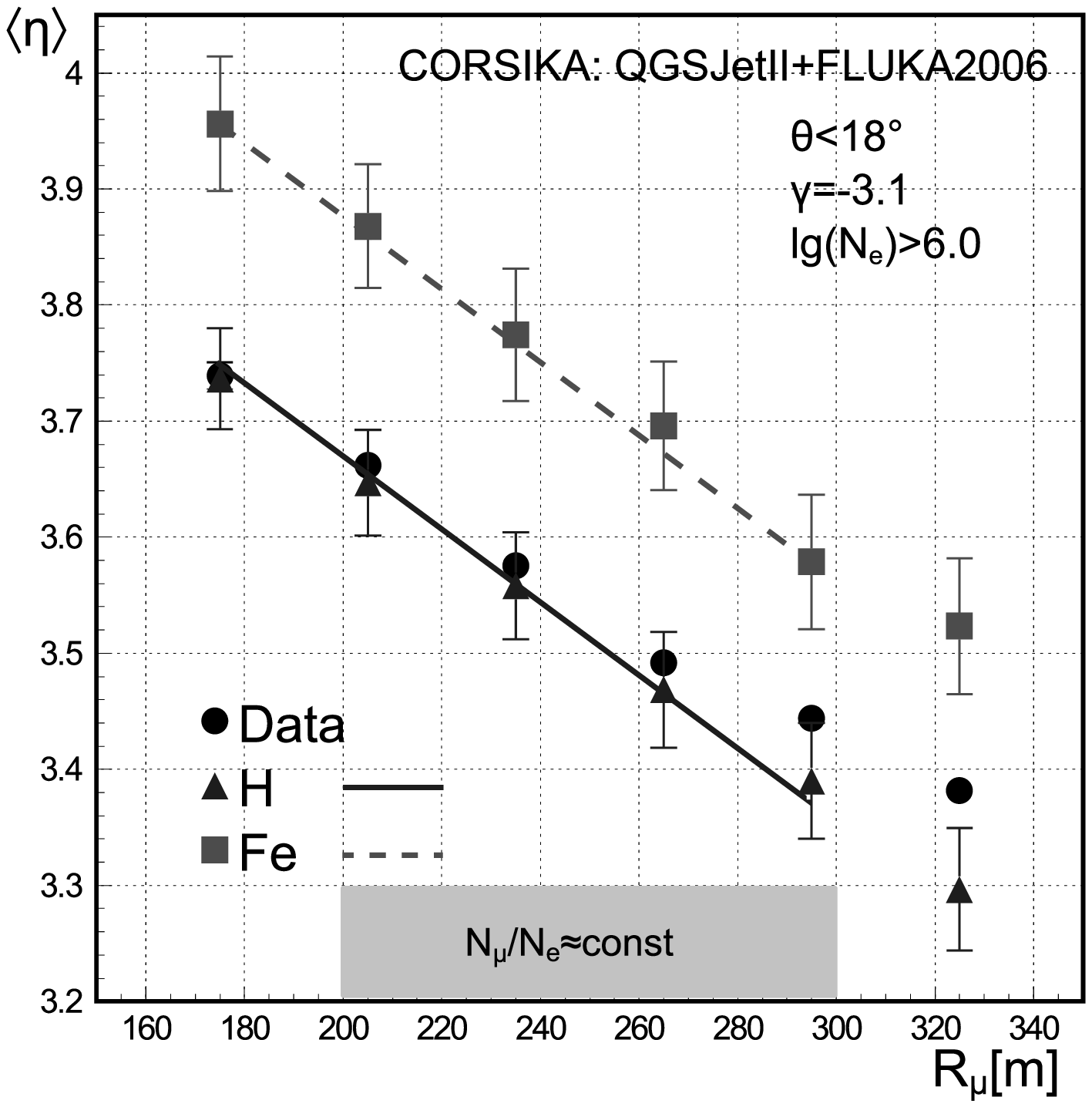}
\vspace{-.5cm}
\caption{Lateral distribution of mean muon pseudorapidity in the limited distance range (see text)
 measured in KASCADE-Grande and
compared with the CORSIKA simulation values for proton and iron primaries.}
\label{etalat}
\end{center}
\end{figure}

In KASCADE-Grande muons can be
registered up to 700~m from the shower core,
but normally different analyses are carried out in
specific distance ranges. 
%and in every distance ring 
%a certain "window" in the longitudinal shower profile is sensed. 
In Fig. \ref{etalat}
lateral distribution of mean EAS muon pseudorapidities measured with the MTD in KASCADE-Grande are
shown, together with the results of CORSIKA \cite{heck} Monte-Carlo simulations, employing a QGSJetII \cite{qgsi} and FLUKA2006
\cite{fluk} model combination. Requirement of shower size $log{N_e}>$ 6 ensures full experiment trigger efficiency
for registration, both, 
proton- and iron-initiated showers. 
Muons from iron showers are, per average,
produced higher than those from proton
showers, and this leads to the observed difference in the
mean pseudorapidities on ground, i.e., to the primary mass sensitivity of muon pseudorapidities.
In Fig.~\ref{etalat} one can notice in the experimental data, below 200~m, a bias towards
proton mass, resulting (for showers above $10^{16}$~eV) from some saturation effects in the MTD close to the shower core. 
Above 300~m, an increasing bias towards iron mass is observed, caused by
non-equal MTD trigger efficiency for all types of primaries there. Therefore, all subsequent investigations
are done in the distance range 200~m - 300~m, where, as it was checked with the $N_{\mu}/N_e$ ratio of showers
used in the analysis, registration efficiency of the MTD is
primary mass independent.

\section{Mean logarithmic mass calculated with muon pseudorapidities}

\begin{figure}[ht]
 \vspace{-2cm} 
  \centering
  \includegraphics[width=60mm]{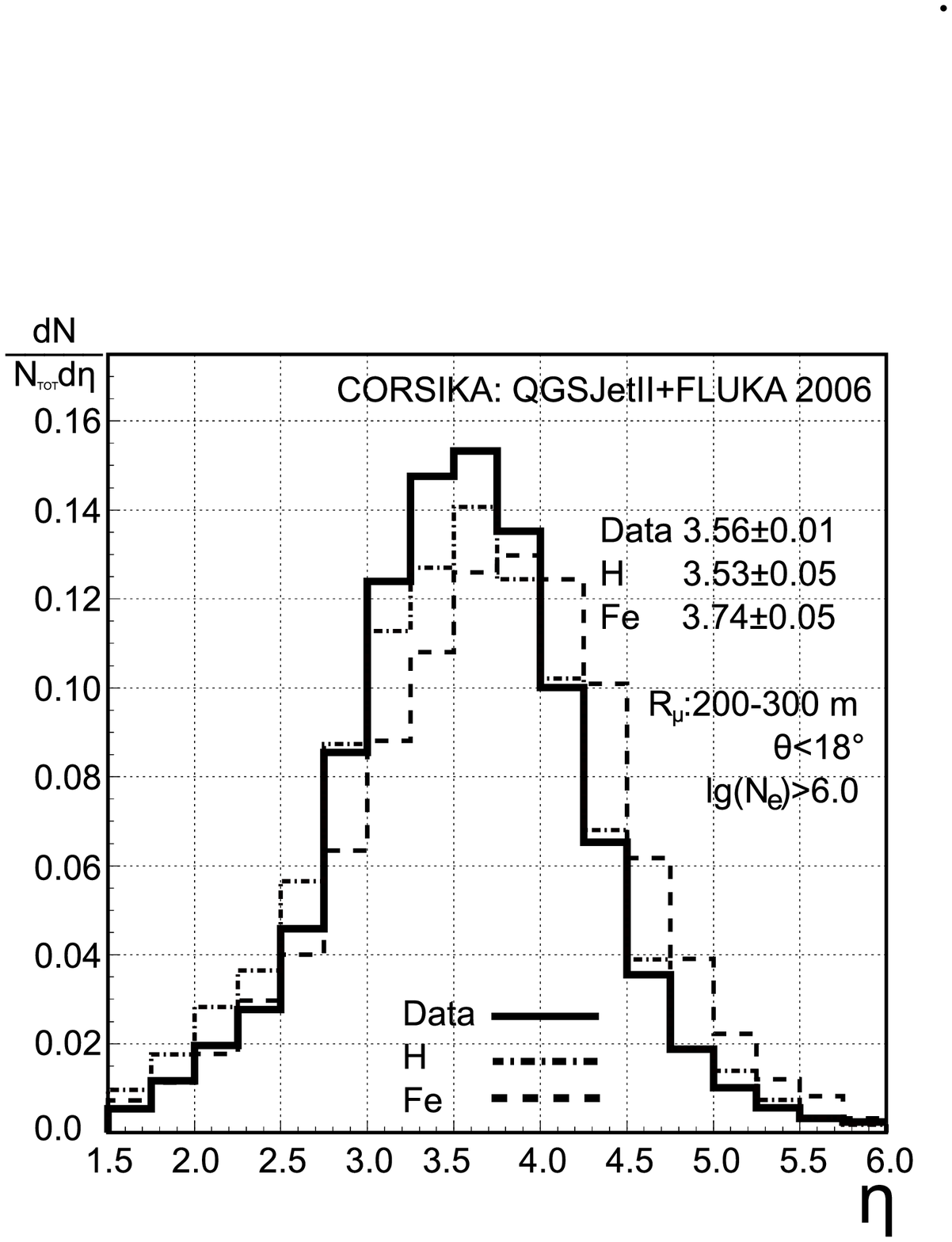}
  \caption{Distribution of measured muon pseudorapidity in the range 200-300~m from the shower core
  together with the simulation results for proton and iron initiated showers.}
  \label{distr}
 \end{figure}

The distribution of measured muon pseudorapidity in the range 200-300~m from the shower core shows, that
its mean value is contained between the simulated values for proton and iron primaries (Fig. \ref{distr}).

\begin{figure}[!t]
\begin{center}
\includegraphics[width=55mm]{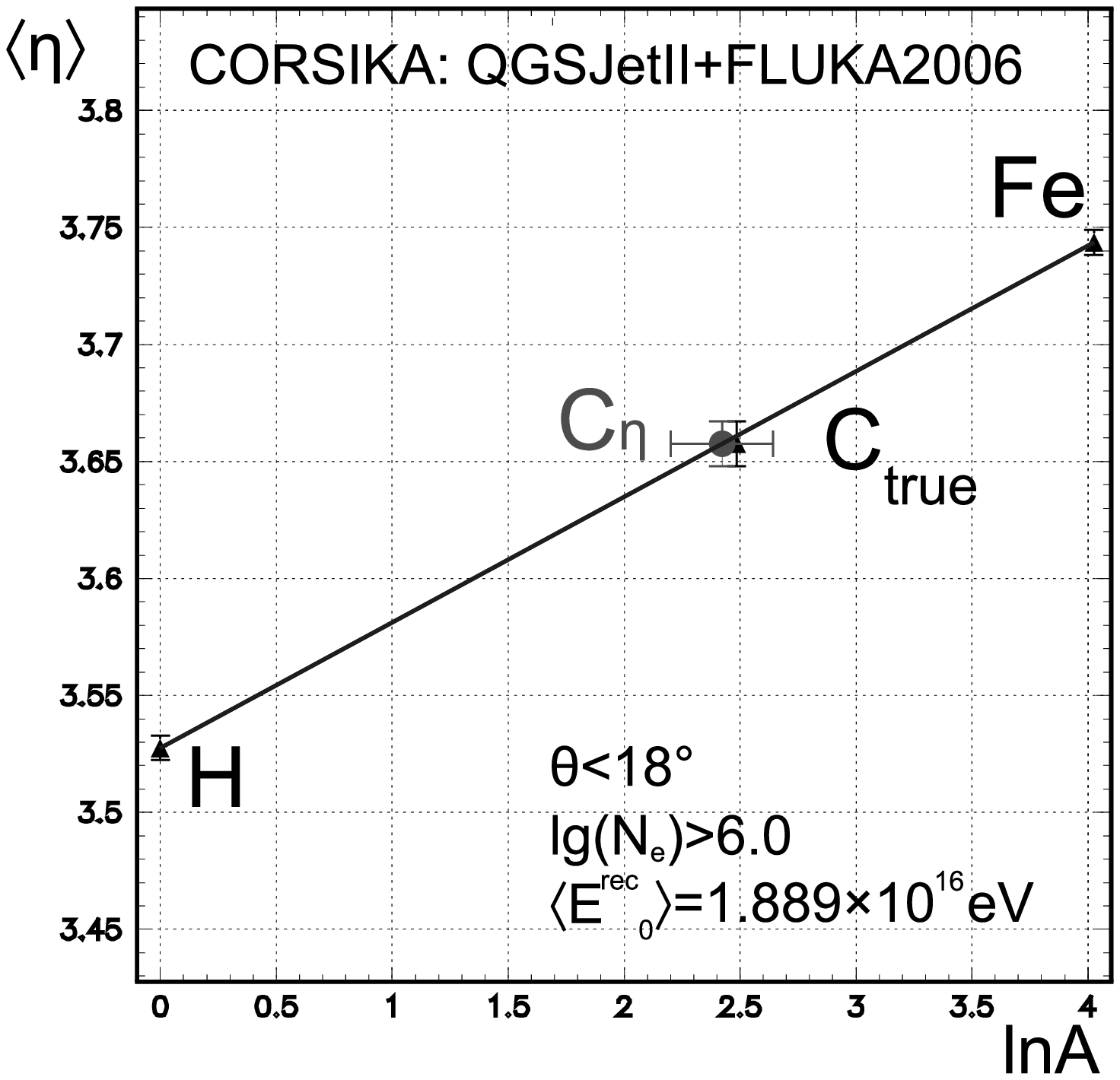}
\caption{Test of the linear dependence of mean EAS muon pseudorapidity on the logarithm
of primary mass. Triangles - simulated $<\eta>$ values for H,C and Fe versus true lnA on the x-axis.
A full circle - $<lnA>$ calculated for carbon, using its $<\eta>$ value.}
\label{lintest}
\end{center}
\end{figure}

In order to assess the average mass composition of cosmic rays with the standard $<lnA>$ 
procedure one has to be
sure, that the mean muon pseudorapidity linearly depends on primary mass. The results of the linearity check are 
shown in Fig. \ref{lintest}. Here, the $<lnA>$ was calculated from $\eta$ distribution for simulated carbon showers
and compared with the known value ln12=2.49. The mean value of calculated $lnA$ differs here by less than $2\%$
from the true value for carbon primary, justifying the use of muon pseudorapidity for the determination of $<lnA>$
of cosmic rays above $10^{16}~eV$. 

\begin{table*}[t]
\begin{center}
\begin{tabular}{l|ccc}
$log[E_{0}/GeV]$  & $<E_0>$ [GeV] & $<lnA>$  & $<lnA>$ \\
 range & &(QGSJetII-FLUKA) & EPOS1.99-FLUKA \\
\hline
$7.0 - 7.3$&(1.38$\pm$0.01)$\times10^7$  & 0.4$\pm$ 0.3 &  1.0$\pm$ 0.5\\
$>7.0$&(2.23$\pm$0.01)$\times10^7$   & 1.1$\pm$ 0.2  &   1.2$\pm$ 0.3\\
$7.3 - 7.6$&(2.69$\pm$0.01)$\times10^7$  & 1.6$\pm$ 0.3 &         -    \\
 $>7.3$ & (3.92$\pm$0.03)$\times10^7$ & 2.0$\pm$ 0.2  &  1.5$\pm$ 0.5\\
 $7.6 - 7.9$ & (5.34$\pm$0.02)$\times10^7$ &  2.2$\pm$ 0.5   &  2.4$\pm$ 1.3\\
$>7.6$  & (7.12$\pm$0.08)$\times10^7$ &  2.7$\pm$ 0.4  &  3.5$\pm$ 1.1\\
$>7.9$  & (13.20$\pm$0.26)$\times10^7$  &  2.5$\pm$ 1.4  &      -    \\
\hline
\end{tabular}
\caption{Results of the $<lnA>$ calculated in the $10^{16}-10^{17}$~eV 
primary energy range for two high-energy interaction models.}
\label{table}
\end{center}
\end{table*}

In the following the results for the primary energy decade $10^{16} - 10^{17}~eV$ will be given.
The primary energy was calculated using total number of charged particles and total number of muons in showers,
with a procedure described in \cite{marioz}. 
The energy ranges in which the $<lnA>$ was calculated are shown in the first column of Table \ref{table}.
In the second column mean energy values, taking into account the spectrum with an index -3.1, are given.
Columns 3 and 4 contain calculated $<lnA>$ values for QGSJetII and EPOS 1.99 \cite{eposi}, respectively. Due to the very limited
statistics of available simulations with EPOS not in all energy bins the calculation was possible. In those, where it was
possible, the errors (only statistical were considered) are, anyway, much larger than in case of the QGSJetII.

\begin{figure*}[th]
  \centering
  \includegraphics[width=5.3in,height=3.2in]{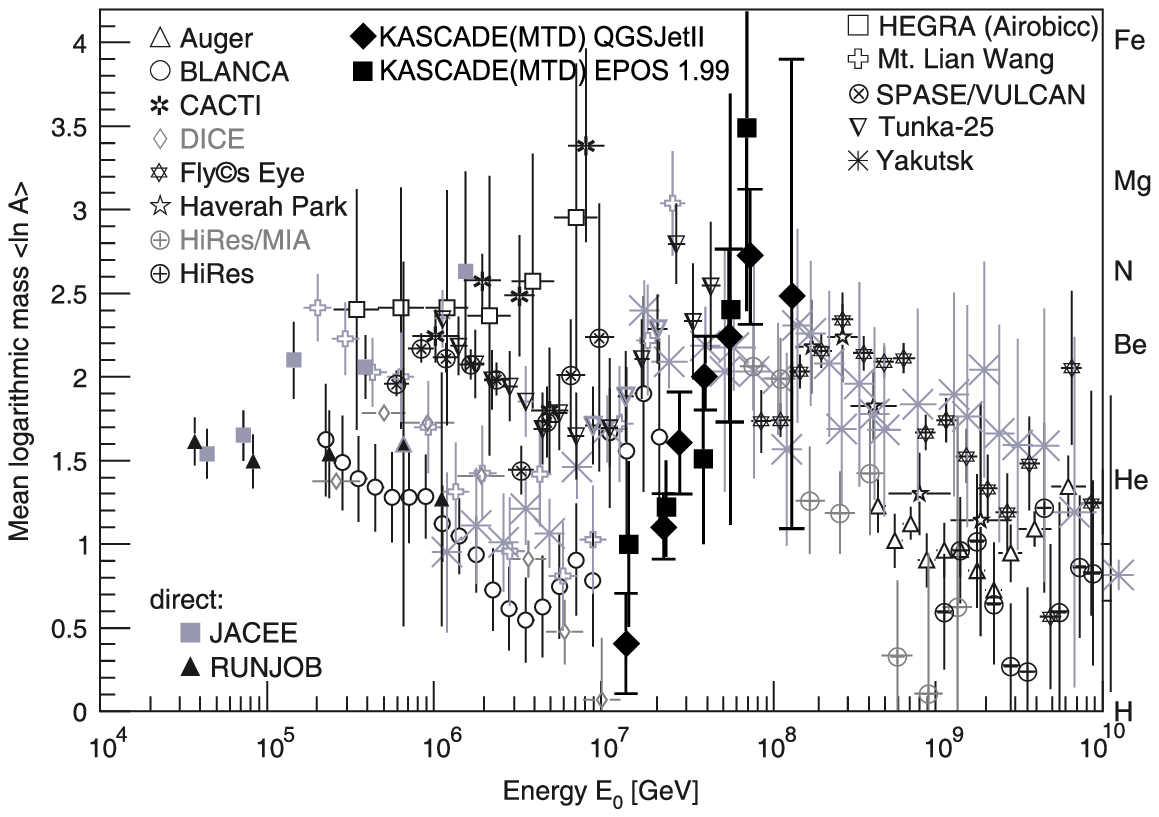}
  \caption{Results of mean logarithmic mass values obtained with mean muon pseudorapidities, shown together
  with the values derived from the average depth of the shower maximum by different experiments,
  from Fig. 14 (top) of Ref. \cite{hor}.
    }
  \label{lnasummary}
 \end{figure*}

The results from columns 3 and 4 of Table \ref{table} are presented in Fig. \ref{lnasummary},
together with a collection of $<lnA>$ values (taken from Fig. 14 (top) in Ref. \cite{hor}) derived from the average 
depth of the shower maximum by various experiments.

\section{Discussion and conclusions}

As seen in Fig. \ref{lnasummary} our results, obtained with EAS muons pseudorapidity measured on ground level,
are compatible with the collection of mean logarithmic mass values derived from the average depth of the shower maximum.
They show similar rise of $<lnA>$ with energy in the investigated primary energy range. This compatibility, rather with the
results obtained from measurement of the light generated during the shower development than the ones measured on
ground (e.g. electron/muon ratio) is because the mean muon pseudorapidity is also a signature of shower development.

Obtained values are generally lower than those from other experiments. The main reason for this can be, that
the QGSJetII model, despite improvement with respect to QGSJet01, gives still too high mean pseudorapidity values 
compared with the measurements. 

It was found in previous studies (e.g. \cite{jzparis}), that in the investigated distance range from the 
core about 40 $\%$ 
of the registered muons is produced in the hadronic interactions above 200~GeV, i.e., they are modeled in the 
simulations by QGSJetII. And this model creates more muons at the heights above 4-5~km, than observed experimentally
\cite{Hmu}. Such muons have higher mean muon pseudorapidity than the ones created deeper in the atmosphere 
\cite{lodz}, thus affecting the determination of the $<lnA>$.
 
 The mean pseudorapidity values obtained in simulations suffer from the insufficient shower statistics in the simulations.
 A standard approach is to use the same simulated shower 5 or 10 times over the detector area in order to reduce the time needed
 for simulations to acceptable value. In this particular investigation such an approach creates additional, systematic error
 in the determination of $<lnA>$, impossible to be quantified without using each simulated shower once, what would 
 require very significant increase of the simulation time. 
 
 In Fig. \ref{lnasummary} some $<lnA>$ values obtained with the EPOS 1.99 model are shown for the comparison. Due to the
 much smaller number of simulated data the statistical errors are much larger than in case of QGSJetII (see also Table \ref{table}). 
 Moreover, the above mentioned effect of shower statistics is influencing the results even more.
 
 One can say  that the general behaviour
 showing rise of the primary mass with the energy is also seen with this model, however it is not possible to compare 
 the values of the logarithmic mass 
 predicted by the two models in question.
 
 In conclusion, the mean muon pseudorapidity in EAS is a primary mass sensitive parameter and, providing the
 large enough number of simulated showers is available for the analysis, the $<lnA>$ of primay cosmic rays can be derived.
 However, the small difference between $<\eta>$ values for proton and iron initiated showers ($\approx$ 0.2) would require
 a significant increase in the number of simulated showers and, even then, the remaining uncertainties would prohibit
 to make a convincing mass composition studies with, at least, 3 mass groups.

\vspace*{0.75cm} \footnotesize{
{\bf Acknowledgement:} KASCADE-Grande is supported by the BMBF of Germany, the MIUR and INAF of Italy, the
Polish Ministry of Science and Higher Education (this work by grant for 2009-2011) and the Romanian Authority 
for Scientific Research.
}

%\vspace{\baselineskip}

\clearpage

%% file: icrc0274.tex
%%
% 32nd International Cosmic Ray Conference 2011 Beijing China

%Class Required
%%% for classical LaTeX
%The paper title
\title{Gamma-Ray Source Studies using a Muon Tracking Detector (MTD)}
%The short title will appear at the header of the even pages.

\shorttitle{P. Doll \etal Gamma-Ray Source Studies.. }

%All paper authors

\authors{
P.~Doll$^{1}$, K.~Daumiller$^{1}$, J.~Zabierowski$^{11}$,
W.D.~Apel$^{1}$, J.C.~Arteaga-Vel\'azquez$^{2}$, K.~Bekk$^{1}$,
M.~Bertaina$^{3}$, J.~Bl\"umer$^{1,4}$, H.~Bozdog$^{1}$,
I.M.~Brancus$^{5}$, P.~Buchholz$^{6}$, E.~Cantoni$^{3,7}$,
A.~Chiavassa$^{3}$, F.~Cossavella$^{4,13}$, V.~de Souza$^{8}$,
F.~Di~Pierro$^{3}$, R.~Engel$^{1}$, J.~Engler$^{1}$, M.
Finger$^{4}$, D.~Fuhrmann$^{9}$, P.L.~Ghia$^{7}$, H.J.~Gils$^{1}$,
R.~Glasstetter$^{9}$, C.~Grupen$^{6}$, A.~Haungs$^{1}$,
D.~Heck$^{1}$, J.R.~H\"orandel$^{10}$, D.~Huber$^{4}$,
T.~Huege$^{1}$, P.G.~Isar$^{1,14}$, K.-H.~Kampert$^{9}$,
D.~Kang$^{4}$, H.O.~Klages$^{1}$, K.~Link$^{4}$,
P.~{\L}uczak$^{11}$, M.~Ludwig$^{4}$, H.J.~Mathes$^{1}$,
H.J.~Mayer$^{1}$, M.~Melissas$^{4}$, J.~Milke$^{1}$,
B.~Mitrica$^{5}$, C.~Morello$^{7}$, G.~Navarra$^{3,15}$,
J.~Oehlschl\"ager$^{1}$, S.~Ostapchenko$^{1,16}$, S.~Over$^{6}$,
N.~Palmieri$^{4}$, M.~Petcu$^{5}$, T.~Pierog$^{1}$,
H.~Rebel$^{1}$, M.~Roth$^{1}$, H.~Schieler$^{1}$,
F.G.~Schr\"oder$^{1}$, O.~Sima$^{12}$, G.~Toma$^{5}$,
G.C.~Trinchero$^{7}$, H.~Ulrich$^{1}$, A.~Weindl$^{1}$,
J.~Wochele$^{1}$, M.~Wommer$^{1}$ } \afiliations{
$^1$ Institut f\"ur Kernphysik, KIT - Karlsruher Institut f\"ur Technologie, Germany\\
$^2$ Universidad Michoacana, Instituto de F\'{\i}sica y Matem\'aticas, Morelia, Mexico\\
$^3$ Dipartimento di Fisica Generale dell' Universit\`a Torino, Italy\\
$^4$ Institut f\"ur Experimentelle Kernphysik, KIT - Karlsruher Institut f\"ur Technologie, Germany\\
$^5$ National Institute of Physics and Nuclear Engineering, Bucharest, Romania\\
$^6$ Fachbereich Physik, Universit\"at Siegen, Germany\\
$^7$ Istituto di Fisica dello Spazio Interplanetario, INAF Torino, Italy\\
$^8$ Universidade S$\tilde{a}$o Paulo, Instituto de F\'{\i}sica de S\~ao Carlos, Brasil\\
$^9$ Fachbereich Physik, Universit\"at Wuppertal, Germany\\
$^{10}$ Dept. of Astrophysics, Radboud University Nijmegen, The Netherlands\\
$^{11}$ Soltan Institute for Nuclear Studies, Lodz, Poland\\
$^{12}$ Department of Physics, University of Bucharest, Bucharest, Romania\\
\scriptsize{ $^{13}$ now at: Max-Planck-Institut Physik,
M\"unchen, Germany; $^{14}$ now at: Institute Space Sciences,
Bucharest, Romania; $^{15}$ deceased; $^{16}$ now at: Univ
Trondheim, Norway }}

%email address of the contact person
\email{paul.doll@kit.edu}

%The abstract.
\abstract{A large area ($128m^2$) streamer tube detector, located
within the KASCADE-Grande experiment, has been built with the aim
to identify muons and their directions from extensive air showers
by track measurements. We discuss the possibility of observation
of Gamma-Ray sources by means of single isolated muons above the
background of cosmic-ray muons using a muon tracking detector
(MTD) exhibiting good angular resolution. Properties of the pion
photo-production process and of the MTD which support the
identification of Gammas are discussed. Preliminary Gamma spectrum
accumulated from Crab and the Mkn421 flux correlation with X-ray
(RXTE/PCA) are presented.}
%The keywords

% B E G I N   D O C U M E N T
\maketitle

%Begin the section.
\section{Introduction} \label{int}

A reliable understanding of the muon production by primary
gamma-rays is mandatory to gauge the sensitivity of the experiment
to gamma primaries. The Vector Meson Dominance Model is usually
employed for photo-nuclear interactions at gamma energies above a
few GeV. To illustrate the critical role played by the event
generators in predicting the muon content of showers, the
Feynman-x distribution of charged pions (in the laboratory frame)
as calculated by the FLUKA code demonstrates ~\cite{poirier1} a
basic difference between gamma and proton-induced collisions: The
gamma primaries lead to a much larger fraction of high-x
secondaries than the proton primaries, therefore, focussing the
pions to very forward direction, and, therefore, compensating for
the much smaller pion photo-production cross section.

High energy gamma rays produce muons in the Earth's atmosphere
that can be detected and reconstructed in relative shallow
underground muon detectors. Such detectors are sensitive to muon
energies of a few GeV. Although muons of such low energy compete
with a large background of cosmic ray muons, they can be
identified provided the detector has sufficient effective area and
resolution. Unlike air-Cherenkov telescopes
~\cite{{veritas},{magic},{hess}} muon detectors cover a large
fraction of the sky with a large duty cycle. The advantage is
considerable in studying the emission from highly variable
sources. Moreover, background multi-muon bundles can be
conveniently rejected without suppression of the predominantly
single-muon gamma signal.

Detected muons originate in gamma induced pion production with
some 20 times higher gamma energies ~\cite{l3+c}. A low energy
threshold for muon detection provides for the correspondingly low
energy gammas a deep view into the Universe. Muons of
$E_{\mu}>1~GeV$ associated with primary photons of several GeV
provide an almost attenuation-free window into the depth of the
universe suffering little from flux losses due to collisions with
IR and CMB radiation via pair production.

\section{Muons from Gammas}

Gamma rays initiate atmospheric cascades of mostly electrons and
photons, but also some muons. Muons originate from the decay of
charged pions which are also photo-produced  by high energy shower
photons ~\cite{gaisser} albeit with much smaller cross section.

The signal-to-noise ratio, defined as the number of events divided
by the square root of the number of background events in a
resolution pixel of ${\sigma^{\circ}}\times{\sigma^{\circ}}$,
depends on the detector area $A$ and the zenith angle $\Theta$ as
~\cite{gaisser}:
%\begin{equation}
$S/N^{1/2}= {{A^{1/2}} / {cos{\Theta}^{0.9}{\sigma}}}$.
%\end{equation}
The formula simply expresses that the signal-to-noise ratio is
improved for increased area $A$, better resolution $\sigma$ and
sources observed at larger zenith angle $\Theta$, where the cosmic
ray background muon rate is reduced.

The MTD ~\cite{doll01} is sensitive to the gamma energy region
above $\sim 10~GeV$ while the muon energy cut equals to $0.8~GeV$.

As mentioned above the background includes some fraction of
multi-muon events. Rejecting multi-muon events not only improves
the signal-to-noise ratio, it also improves the angular resolution
which may be degraded by less reliable reconstruction of complex
muon bundles initiated by high energy cosmic ray particles which
are accompanied by other shower particles. MTD with its detection
area of $A_{MTD}=128 m^{2}$ has below the shielding an average
rate of $2.5 kHz$ which results above an energy of $0.8~GeV$ in
$\sim 10^{7}$ tracks from background muons per year in a
$1^{o}\times 1^{o}$ pixel. Assuming a further reduction because of
the strong focusing of the gamma induced muons by about 10, this
rate may lead for Crab ~\cite{gaisser} to $S/N^{1/2}=40$ for one
year of running.

\section{Muon Tracking Detector (MTD)}
\label{detector}

The MTD ~\cite{doll01} is built out of Streamer Tube (ST) chambers
of 4 m length and located in the KASCADE-Grande experiment
~\cite{grande}. The ST chambers are grouped in, so called, {\it
modules}. Four modules, three positioned on horizontal planes
(top, middle, bottom) and one arranged vertically (wall), form a
muon telescope. The whole detector comprises 16 telescopes
arranged in two rows. The low mass structure of the detector
design reduces secondary interactions in the sensitive detector
part.

To deal with clean muon tracks an efficient filter absorbing a
large fraction of the low energy electromagnetic component of
proton induced showers is mandatory. The larger Bethe-Heitler
cross-section induced showers for gammas leading to pair
production and subsequent electromagnetic cascade are readily
absorbed by the filter. The energy threshold ($0.8~GeV$) is
suitable for background reduction. Following the characteristic
energy ${\epsilon}_{\mu}=m_{\mu}c^{2} 6.4km/{\tau}_{\mu}c\sim
0.8~GeV$ muons from proton induced showers do not survive over the
path length of 6.4 km and with energies close to the MTD threshold
but decay to electrons which are readily absorbed. Photo-nuclear
produced muons survive more easily because of comparatively lower
production height. This situation improves the S/N ratio for
photon detection. The stability of the MTD ~\cite{doll01} is very
good. The detector gas system follows precisely the atmospheric
gas pressure.

The first-level $\gamma$/proton discrimination is mostly achieved
by the characteristics of the MTD. The ratio of the number of hits
in the top module to the track number $N_{hit}^{top}/N_{track}$ in
each tower is a powerful tool in discriminating against hadron
induced air showers. Only tracks with one hit in each module are
accepted. Only hits with small cluster size 'cls' (clusters of
readout wires or readout strips) are accepted. The background of
high energy electrons or hadrons is further reduced by cuts in the
track quality $Q^{2}$ (see ~\cite{doll01}). Showering electrons or
hadrons lead to larger 'cls' and, therefore, smaller $Q^{2}$.

For the subsequent analysis the following cuts in the data have to
be employed: 1) Only data over one full sidereal day (86163 solar
seconds) are considered. 2) Only sidereal days where all 16
telescopes are functioning are used. 3) The daily rate should
follow a Gaussian distribution, and only days with rate within
$\pm 4\sigma$ are accepted. 4) Only data corrected for pressure
and temperature in the atmosphere and in the detector are
included.

\section{Gamma-Source Search}
\label{method}

To identify sources which emit high energy gammas and which are
with small probability converted to single muons deep in the
atmosphere, we have the possibility to restrict the arrival
direction and arrival time. The restriction in time and direction
has the task to reduce other disturbing sources like the Sun. We
have to cope with about 3 orders of magnitude larger background
from cosmic ray muons in a canonical $1^{o}\times1^{o}$ window in
the sky.

The optimal square bin size for the search of a point gamma-ray
source with the MTD is $1^{\circ}$ on a side, corresponding to a
Gaussian angular resolution of about $0.3^{\circ}$.

For the search of point sources, the fluctuation of events in a
fixed direction of the sky is investigated. When examining the
fluctuation of the number of muon events bin by bin and searching
for possible sources with a muon event excess, we employ the
numbers $N_{on}$, $T_{on}$, $N_{off}$, $T_{off}$. We use
Li-Ma~\cite{lima} formula  to examine the fluctuation of the
number of muon events bin by bin and to search for possible muon
event excess. In the current analysis this procedure is used with
bin size $\Delta\delta=0.25^{\circ}$ and
$\Delta\alpha=\Delta\delta/cos\delta$, confined to specific
regions on the sky to investigate the MTD sensitivity to specific
gamma-ray sources. Only specific track directions in $\Theta$ and
$\Phi$ and only specific arrival time intervals, where the
corresponding source is high (winter months for Crab)
($\Theta_{source}< 35^{\circ}$) are chosen to accumulate clean
single muon events in the $\delta$ versus $\alpha$ (declination
versus right ascension) plane.
% Crab appears in Fig.~3 at
%  $21.9^{\circ}, ~5.58~h$ which is
%$0.1^{\circ}$ further from the zenith and $0.15^{\circ}$ earlier
%in time compared to the known position ~\cite{magic1} of the Crab
%pulsar.

The analysis shows that the strong requirements of only 1 hit in
each detector module and all $cls<3$ give good profiles for Crab
and Mkn421.

Observation of Mkn421 with the ARGO-YBJ experiment is reported by
S.Vernetto et al.~\cite{vernetto}. The FERMI view of the TeV
blazar Mkn421 is given in ~\cite{paneque} and the MAGIC
observations of Mkn421 and related optical/X-ray/TeV
multiwavelength studies are reported in ~\cite{bonnoli}. Mkn421 is
a very active Galaxy nucleus showing strong variation in the X-ray
and TeV Gamma-ray fluxes and appears at $38.3^{\circ}$, $11.08~h$
which is $0.1^{\circ}$ closer to the zenith and $0.15^{\circ}$
later in time than the nominal position. The muon intensity is
sampled and weighted with the actual MTD efficiency and smoothed
afterwards. Again, only data epochs are considered when Mkn421 is
high. Interestingly, the source profile is improved when employing
days (properly matched in time) for which appreciable X-ray flux
($>2.5~photons/day$) is reported ~\cite{rxte}. For the Mkn421
source a correlation analysis with the X-ray flux measurements by
the All-Sky-Monitor (ASM) onboard the RXTE-satellite~\cite{rxte}
was performed providing the correlation as shown in Fig.~1. The
picture exhibits a clear correlation between the time averaged (1
day) high energy muon flux and the time averaged X-ray flux above
$2~keV$. A narrow window around the nominal Mkn421 position
improves the correlation in Fig. 1. Also a correlation between the
muon flux from the MTD in the region of Mkn421 and the gamma flux
in the GeV range recorded by FERMI satellite ~\cite{fermi} is
observed. The dashed line in Fig.~1 represents a fit to the
correlation and deserves further studies. Employing days with
almost constant X-ray activity from Mkn421, the conversion of GeV
gammas into muons in the atmosphere above the MTD can be
investigated. Fig.~2 shows the correlation of the muon rate with
the atmospheric pressure overburden. It shows the strong gain of
photo-produced pions with pressure especially after correction
(solid points in Fig.~2) for the pressure dependent track
detection efficieny of the MTD ~\cite{doll01}. The strong
dependence of the muon rate on the atmospheric pressure suggests
the pion photo-production to occur in the lower atmosphere.

Fig. 3 shows for Crab a two-dimensional multi-quadric
interpolation ~\cite{allison} of the data points in the $\delta$
versus $\alpha$ (declination versus right ascension) plane (in the
boundaries of a MAGIC presentation). The contour scale is in units
of excess counts per smoothing radius. The position of the source
as quoted by the MAGIC group ~\cite{magic1} is given by the cross.
The displacement of the source to larger zenith $\Theta$, and
smaller azimuth $\Phi$, and possible deformation is due to the
geomagnetic field $B_{geo}$ and is expected to depend on the
momentum of the muons and their orientation with respect to the
geomagnetic field components.

Gamma-ray induced muons in the deep atmosphere we expect to be
almost charge symmetric. Therefore, the spread of the source
profile is reproducing the muon momentum distribution. Muons are
considered to exhibit negative charge if they are deflected
towards the East from the line connecting the actual Crab position
with the geomagnetic field direction.  Mean deflection of about
$0.6^{\circ}$ is expected for 1~GeV muons created 1~km above
ground. Only negatively charged muons to the East side of the
source are considered for accumulating an angle-distance spectrum.
The frequency of muon tracks as function of $1/distance$ to the
nominal Crab source position is plotted in the Fig.~4. resulting
in steep falling spectrum. The inverse angle-distance is expressed
in muon momentum, assuming a $\sim p_{\mu} \times B_{geo}$
dependence. Assuming further that muons at threshold ($0.8~GeV$)
stem from gammas about 10 times higher in momentum would provide a
preliminary gamma energy scale. The preliminary flux normalization
considers the gamma flux attenuation in the air, the small
$\sigma_{\gamma\pi}$ cross section on air, the mass density of the
atmosphere and  the momentum dependence of the $\pi \Rightarrow
\mu$ conversion rate and the exposure of the MTD. The dashed line
represents a $flux_{\mu} \sim p_{\mu}^{-2}$ flux dependence. A
more detailed unfolding of the gamma source spectrum has still to
be carried out.
\begin{figure}
\begin{center}
\includegraphics [width=0.4\textwidth]{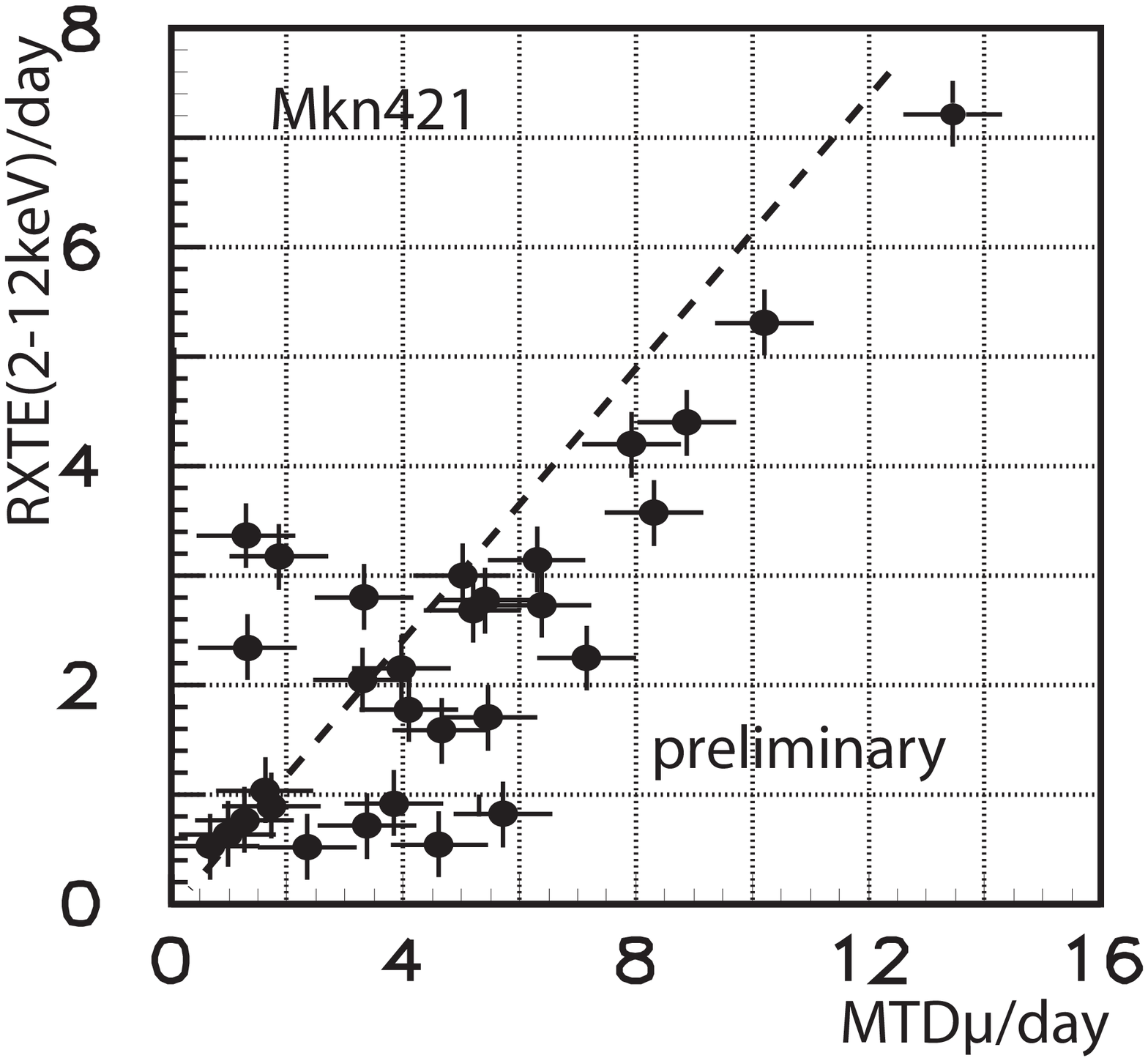}
\end{center}
\vspace{0.05 cm} \caption{Correlation of the X-ray flux with the
muon rates in a narrow window around the nominal Mkn421 position.
Data were taken in a high flux period during February 2010. The
dashed line emphasizes a linear dependence between the X-ray flux
and the GeV gamma ray flux. So far, only fraction of all data are
considered.} \label{fig1z}
\end{figure}
\vspace{0.05 cm}
\begin{figure}
\begin{center}
\includegraphics [width=0.4\textwidth]{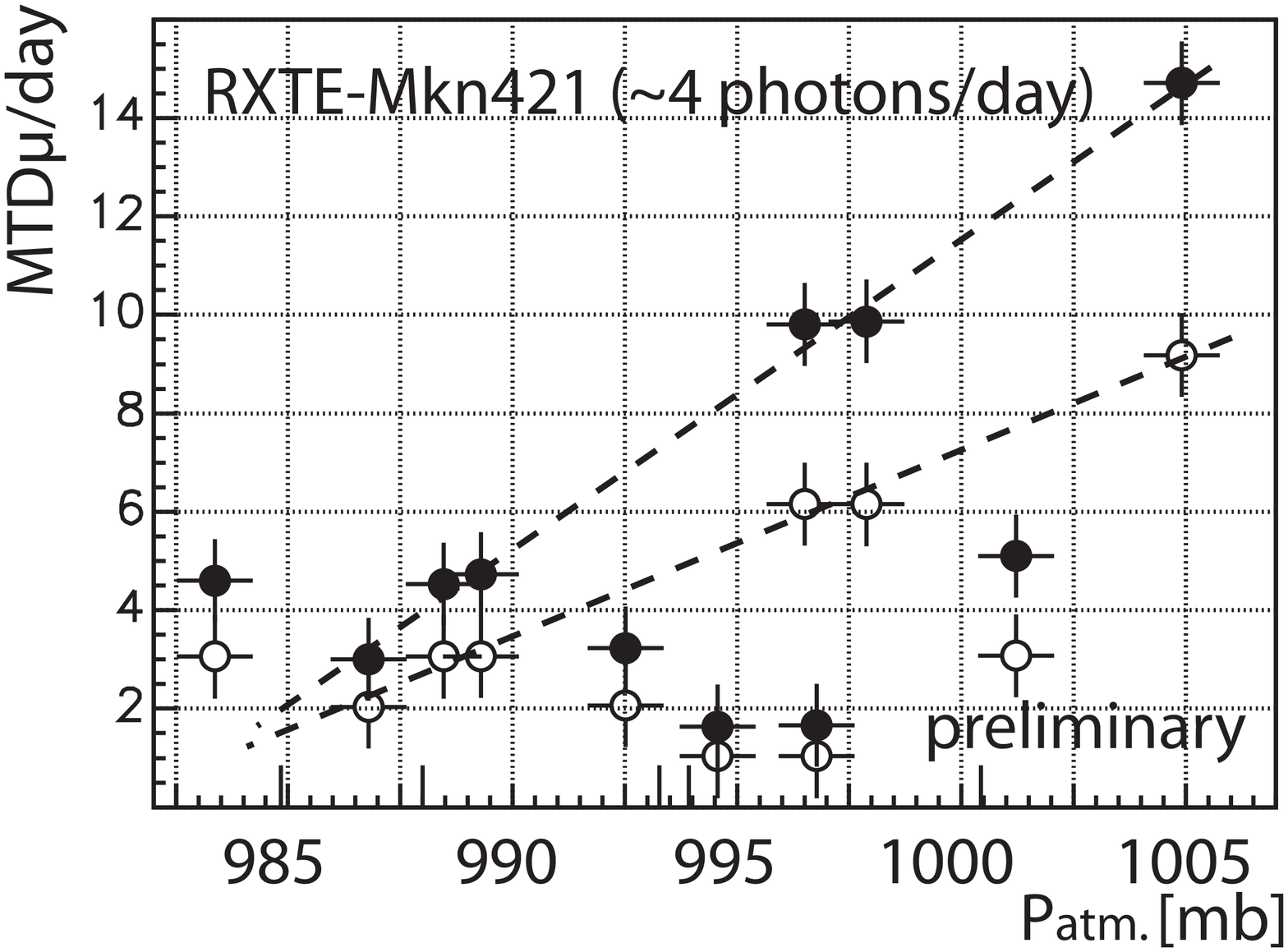}
\end{center}
\vspace{0.05 cm} \caption{Variation of the photo-pion production
yielding muons with atmospheric pressure for a window of almost
constant X-ray flux from Mkn421 (see Fig. 1, dashed region). The
full symbols consider the effect of the MTD track detection
efficiency.} \label{fig2j}
\end{figure}
%\vspace{0.1 cm}
\begin{figure}
\begin{center}
\includegraphics [width=0.3\textwidth]{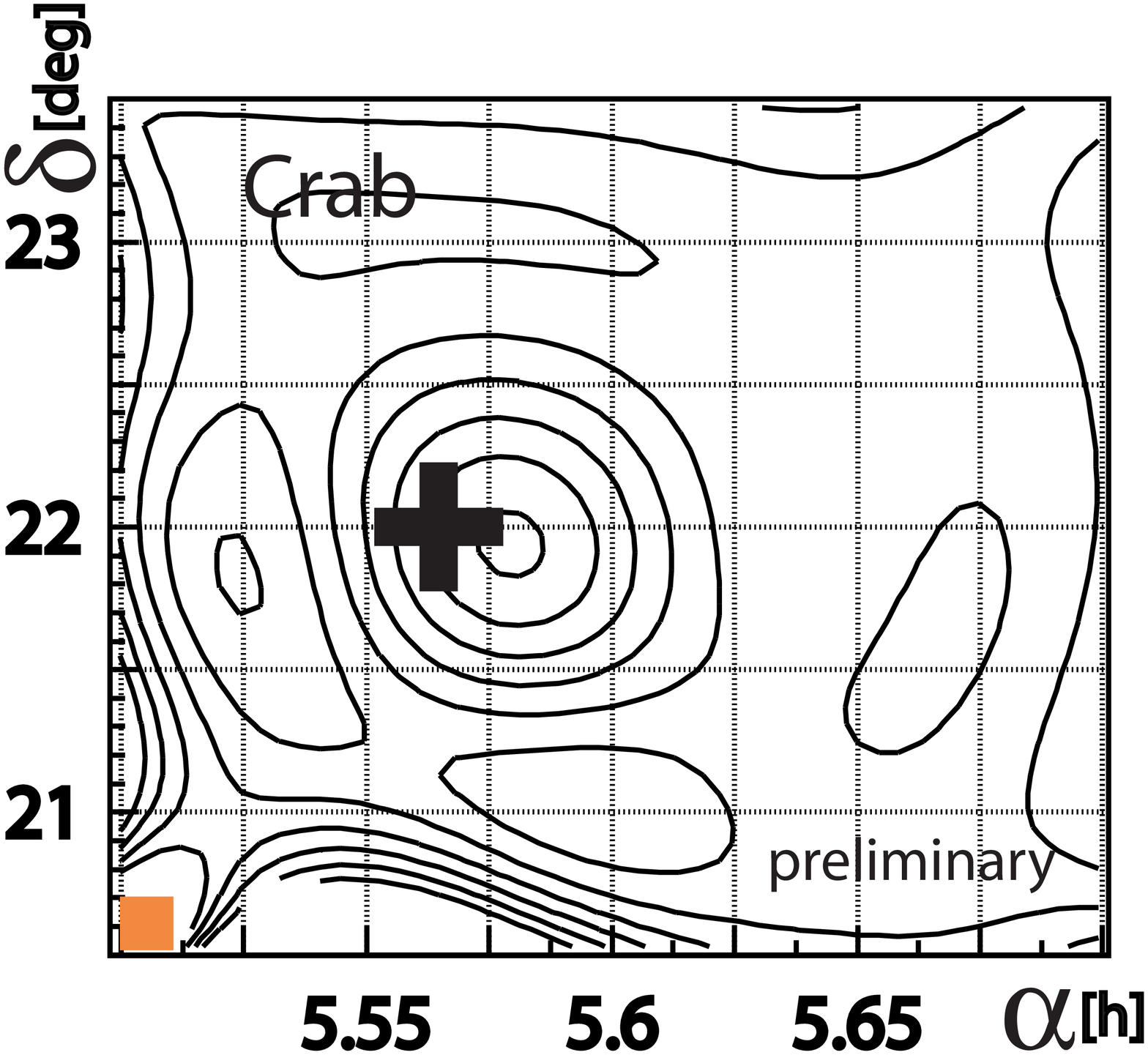}
\end{center}
\vspace{0.05 cm} \caption{Muon yield close to the nominal Crab
position in a $3^{\circ}\times 3^{\circ}$ field of view. The cross
represents the Crab position after MAGIC ~\cite{magic1}. The
contour lines vary in $25\%$ steps. So far, only fraction of all
data are considered. } \label{fig3j}
\end{figure}
\begin{figure}
\begin{center}
\includegraphics [width=0.4\textwidth]{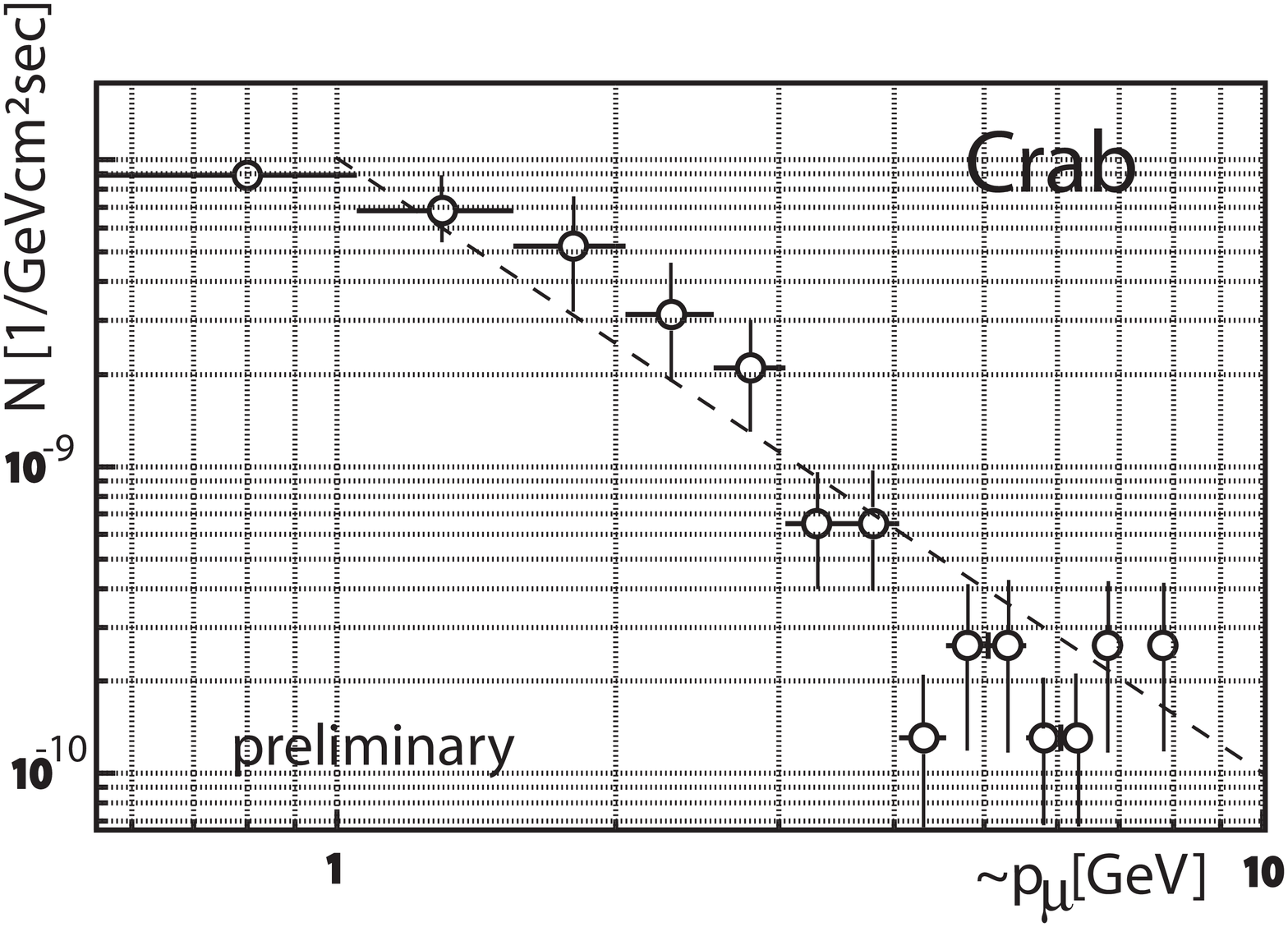}
\end{center}
\vspace{0.05 cm} \caption{Muon flux variation derived as function
of the inverse distance from the nominal source position and
expressed in momentum units above the muon momentum threshold of
the MTD~($0.8~GeV$). So far, only fraction of all data are
considered. } \label{fig4i}
\end{figure}

\section{Outlook}
The high resolution muon tracking detector MTD in the
KASCADE-Grande experiment demonstrates to be capable to identify
gamma point sources in the sensitivity range of Crab fluxes. The
highly variable gamma source Mkn421 provides in its 'high' state a
test beam to study the efficiency of $\gamma \Rightarrow \mu$
conversion in the atmosphere depending on the atmospheric
parameters, and further tuning of the MTD response to gammas.
Future analysis of a larger data sample will provide more detailed
information on the nature of high energy gamma source muons. There
is a common understanding that the high energy gamma source muons
serve as sensitive probes to investigate the high energy photon
interactions in the atmosphere, providing a gamma detector in the
multi-GeV range.

\vspace*{0.5cm} \footnotesize{ {\bf Acknowledgement:}
KASCADE-Grande is supported by the BMBF of Germany, the MIUR and
INAF of Italy, the Polish Ministry of Science and Higher Education
(this work in part by grant for 2009-2011) and the Romanian
Authority for Scientific Research.}

%\clearpage

\clearpage